\pdfoutput=1
\newcommand*{\ATLASLATEXPATH}{./}
\documentclass[texlive=2016,txfonts,USenglish,cernpreprint]{atlasdoc}

\usepackage[subfigure,block=none]{atlaspackage}

\usepackage{multirow}
\usepackage{bm}
\usepackage{slashed}
\usepackage{mathrsfs}
\usepackage{xspace} 
\usepackage{units}
\usepackage{rotating}
\usepackage{verbatim}
\usepackage{float}
\usepackage{hyperref}      
\usepackage{cleveref}
\usepackage{arydshln}
\usepackage{float,lscape}
\usepackage{color}

\usepackage{dcolumn}
\newcolumntype{d}[1]{D{.}{.}{#1}}
\newcolumntype{P}[1]{D{,}{\,\pm\,}{#1}}
\usepackage{atlasbiblatex}

\usepackage{atlasphysics}
\usepackage{atlasbsm}

\graphicspath{{./}{./}}

\usepackage{SUSY-2017-03-PAPER-defs}

\hypersetup{pdftitle={ATLAS draft},pdfauthor={The ATLAS Collaboration}}

\AtlasTitle{
Search for chargino--neutralino production using recursive jigsaw reconstruction in final states with two or three charged leptons in proton--proton collisions at $\sqrt{s}$ = 13 TeV with the ATLAS detector}

\date{\today}

\AtlasRefCode{SUSY-2017-03}

\AtlasNote{SUSY-2017-03}

\PreprintIdNumber{CERN-EP-2018-113}

\arXivId{1806.02293}

\AtlasJournal{PRD}
\AtlasJournalRef{\PRD 98 (2018) 092012}
\AtlasDOI{DOI:10.1103/PhysRevD.98.092012}

\AtlasAbstract{
A search for electroweak production of supersymmetric particles is performed in two-lepton and three-lepton final states using recursive jigsaw reconstruction, a technique that assigns reconstructed objects to the most probable hemispheres of the decay trees, allowing to construct tailored kinematic variables to separate the signal and background. The search uses data collected in 2015 and 2016 by the ATLAS experiment in $\sqrt{s}$ = 13 TeV proton--proton collisions at the CERN Large Hadron Collider corresponding to an integrated luminosity of 36.1 fb$^{-1}$. 
Chargino--neutralino pair production, with decays via W/Z bosons, is studied in
final states involving leptons and jets and missing transverse momentum for scenarios with large and intermediate mass-splittings between the parent particle and lightest supersymmetric particle, as well as for the scenario where this mass splitting is close to the mass of the $Z$ boson. The latter case is challenging since the vector bosons are produced with kinematic properties that are similar to those in Standard Model processes. Results are found to be compatible with the Standard Model expectations in the signal regions targeting large and intermediate mass-splittings, and chargino--neutralino masses up to 600 GeV are excluded at 95\% confidence level for a massless lightest supersymmetric particle. Excesses of data above the expected background are found in the signal regions targeting low mass-splittings, and the largest local excess amounts to 3.0 standard deviations.}

\DeclareUnicodeCharacter{2212}{-}

\begin{document}

\maketitle


\clearpage
\section{Introduction}
\label{sec:intro}
Supersymmetry (SUSY)~\cite{Golfand:1971iw,Volkov:1973ix,Wess:1974tw,Wess:1974jb,Ferrara:1974pu,Salam:1974ig} is a generalization of space--time symmetries which predicts new bosonic (fermionic) partners for the fermions (bosons) of the Standard Model (SM). If $R$-parity~\cite{Farrar:1978xj} is conserved, SUSY particles (called sparticles) are produced in pairs and the lightest supersymmetric particle (LSP) is stable and represents a possible dark-matter candidate~\cite{Goldberg:1983nd,Ellis:1983ew}. Superpartners of the charged and neutral electroweak (EW) and Higgs bosons mix, producing charginos ($\chinopm_{\!\!\!\!l},\ l=1,\ 2$) and neutralinos ($\nino_{\!\!\!m},\ m=1,\ 2,\ 3,\ 4$), collectively known as electroweakinos. The indices of these particles are ordered by mass in ascending order.

The production cross-sections of sparticles at the Large Hadron Collider (LHC) depend both on the type of interaction involved and on the sparticle masses. The colored sparticles (squarks and gluinos) are produced in strong interactions with significantly larger production cross-sections than non-colored sparticles of equal mass, such as the charginos and neutralinos. However, should the masses of gluinos and squarks prove to be out of reach at the LHC, the direct production of charginos and neutralinos could be the dominant sparticle production mode. With searches performed by the ATLAS and CMS Collaborations during LHC Run~2, the exclusion limits on colored-sparticle masses extend up to approximately 2~\TeV~\cite{Aaboud:2017vwy,Sirunyan:2017cwe,Sirunyan:2017kqq}, making electroweak production an increasingly promising probe for SUSY signals at the LHC.

This paper presents a search for pair-produced electroweakinos ($\chinoonepm\ninotwo$), with each of $\chinoonepm$ and $\ninotwo$ decaying to a $\ninoone$ (assumed to be the LSP) and a $W$ or $Z$ gauge boson, respectively, leading to final states with two or three isolated leptons (here taken to be electrons or muons only) which may be accompanied by jets and missing transverse momentum. The analysis uses an integrated luminosity of 36.1~fb$^{-1}$ of proton--proton ($pp$) collision data delivered by the LHC at a center-of-mass energy of $\sqrt{s}=13$~\TeV. The search employs the recursive jigsaw reconstruction (RJR) technique~\cite{Jackson:2016mfb,Jackson:2017gcy} in the construction of a suite of complementary discriminating variables. Signal regions are defined to probe a wide range of 
$\chinoonepm/\ninotwo$ (assumed to be mass degenerate) and $\ninoone$ masses, with mass differences $\Delta m = m_{\chinoonepm/\ninotwo}-m_{\ninoone}$ ranging from $\approx$~100~\GeV~to $\approx$~600~\GeV. This search has improved sensitivity to supersymmetric models previously studied by the ATLAS~\cite{SUSY-2013-11,SUSY-2013-12,Aad:2015eda,Aaboud:2018jiw} and CMS~\cite{Sirunyan:2017lae,Sirunyan:2017qaj,Sirunyan:2018ubx} Collaborations with the same integrated luminosity, which had expected exclusion sensitivities at 95\% confidence level (CL) of $\chinoonepm/\ninotwo$ masses up to 530~\GeV\ and 570~\GeV, respectively, for a massless LSP. 

In a separate search by ATLAS detailed in Ref.~\cite{Aaboud:2018jiw}, where the same SUSY scenarios are considered and the same dataset is used, an approach based on conventional variables complements the use of recursive jigsaw variables herein. In both cases, regions are enriched with events containing two or three leptons sensitive to the production of sparticles. In the approach described in Ref.~\cite{Aaboud:2018jiw}, selection criteria are imposed on object momenta, missing transverse momentum and angular parameters to reduce the background and define regions sensitive to signal events. On the other hand, the RJR approach provides a way to reconstruct the event from the detected particles in the presence of kinematic and combinatoric ambiguities by factorizing missing information according to decays and rest frames of intermediate particles. This yields a basis of largely uncorrelated variables that are subsequently used to design the search presented herein. The two different approaches yield event samples that are largely unique and non-overlapping in the signal regions targeted, with improved sensitivity in the simplified model used to optimize the search. The main SM backgrounds to the search arise from diboson and $Z+$jet processes.

\section{The ATLAS detector}
\label{sec:detector}
\interfootnotelinepenalty=10000
The ATLAS detector~\cite{Aad:2008zzm} is a multipurpose particle detector with a forward--backward symmetric cylindrical geometry and nearly 4$\pi$ coverage in solid angle.\footnote{ATLAS uses a right-handed coordinate system with its origin at the nominal interaction point in the center of the detector. The positive $x$-axis is defined by the direction from the interaction point to the center of the LHC ring, with the positive $y$-axis pointing upwards, while the beam direction defines the $z$-axis. Cylindrical coordinates $(r,\phi)$ are used in the transverse plane, $\phi$ being the azimuthal angle around the $z$-axis. The pseudorapidity $\eta$ is defined in terms of the polar angle $\theta$ by $\eta=-\ln\tan(\theta/2)$ and the rapidity is defined as $y = (1/2)\ln[(E+p_z)/(E-p_z)]$ where $E$ is the energy and $p_{\textrm z}$ the longitudinal momentum of the object of interest. The transverse momentum $\pt$, the transverse energy $\ET$ and the missing transverse momentum $\met$ are defined in the $x$--$y$ plane unless stated otherwise.} The inner detector (ID) tracking system consists of silicon pixel and microstrip detectors covering the pseudorapidity region $|\eta|<2.5$, surrounded by a transition radiation tracker, which improves electron identification over the region $|\eta|<2.0$. The innermost pixel layer, the insertable B-layer~\cite{B-layerRef}, was added between Run~1 and Run~2 of the LHC, at an average radius of 33 mm around a new, narrower and thinner beam pipe. The ID is surrounded by a thin superconducting solenoid providing an axial 2~T magnetic field and by a fine-granularity lead/liquid-argon (LAr) electromagnetic calorimeter covering $|\eta|<3.2$. A steel/scintillator-tile hadronic calorimeter provides coverage in the central pseudorapidity range ($|\eta|<1.7$). The end-cap and forward regions are instrumented with LAr calorimeters for both EM and hadronic energy measurements up to $|\eta| = 4.9$. The muon spectrometer with an air-core toroid magnet system surrounds the calorimeters. Three layers of high-precision tracking chambers provide coverage in the range $|\eta|<2.7$, while dedicated chambers allow triggering in the region $|\eta|<2.4$.

The trigger system~\cite{ATLASTrigger2016Paper} consists of two levels. The first level is a hardware-based system and uses a subset of the detector information. The second is a software-based system called the high-level trigger which runs offline reconstruction and calibration software, reducing the event rate to about 1 kHz.

\section{Data and Monte Carlo samples}
\label{sec:montecarlo}
The data were collected by the ATLAS detector during 2015 with a peak instantaneous luminosity of $L = 5.2 \times 10^{33} ~\textrm{cm}^{-2} \textrm{s}^{-1}$, and during 2016 with a maximum of $L = 1.37\times10^{34}$~cm$^{-2}\textrm{s}^{-1}$. The mean number of $pp$ interactions per bunch crossing (pileup) in the dataset was $\langle\mu\rangle$ = 14 in 2015 and $\langle\mu\rangle$ = 24 in 2016. 
Application of beam, detector and data-quality criteria resulted in a total integrated luminosity of 36.1~\ifb. The uncertainty in the integrated luminosity is $\pm$2.1\%. It is derived, following a methodology similar to that detailed in Ref.~\cite{Aaboud:2016hhf}, from a calibration of the luminosity scale using $x$--$y$ beam-separation scans performed in August 2015 and May 2016. 

A set of Monte Carlo (MC) background and signal samples of simulated events is used to optimize the selection criteria and assess the sensitivity to specific SUSY signal models. Where applicable, the MC samples are used in the background estimation as well. 

The production of $Z$ bosons in association with jets~\cite{ATL-PHYS-PUB-2016-003} was performed with the \sherpa~2.2.1 generator~\cite{Gleisberg:2008ta}. The NNPDF3.0NNLO~\cite{Ball:2014uwa} parton distribution function (PDF) was used in conjunction with dedicated parton shower tuning developed by the \sherpa\ authors. The matrix elements (ME) were calculated for up to two partons at next-to-leading order (NLO) and with up to two additional partons at leading order (LO) using the \textsc{Comix}~\cite{comix} and \textsc{Open Loops}~\cite{openloops} matrix-element generators, and merged with the \sherpa\ parton shower (PS)~\cite{sherpashower} using the ME+PS@NLO prescription~\cite{mepsnlo}. For MC closure studies of the data-driven $Z$+jets background estimate (described in Section~\ref{sec:background2L}), $\gamma$+jets events were generated at LO with up to four additional partons using the \sherpa\ 2.1.1 generator with CT10~\cite{CT10pdf} PDF set.

The \textsc{Powheg-Box} v2~\cite{powheg-box} generator was used for the generation of $t\bar{t}$ and single-top-quark processes in the $Wt$- and $s$-channels~\cite{ATL-PHYS-PUB-2016-004}, while $t$-channel single-top production was modeled using \textsc{Powheg-Box} v1~\cite{Frixione:2007vw}. For the latter process, the decay of the top quark was simulated using {\textsc MadSpin}~\cite{10a} preserving all spin correlations. For all processes the CT10~\cite{CT10pdf} PDF set was used for the matrix element, while the parton shower, fragmentation, and the underlying event were generated using \pythia~6.428~\cite{pythia6} with the CTEQ6L1~\cite{Pumplin:2002vw} PDF set and a set of tuned parameters called the {\textsc Perugia 2012} tune~\cite{perugia}. The top-quark mass in all samples was set to 172.5~\GeV. The $t\bar{t}$ and the $Wt$-channel single-top events were normalized to cross-sections calculated at next-to-next-to-leading-order plus next-to-next-to-leading-logarithm (NNLO+NNLL)~\cite{Czakon:2013goa,Czakon:2011xx,Kidonakis:2010ux,Kidonakis:2011wy} accuracy, while $s$- and $t$-channel single-top-quark events were normalized to the NLO cross-sections~\cite{Aliev:2010zk,Kant:2014oha}. The production of $Zt$ events was generated with the MG5\_aMC@NLO~2.2.1~\cite{Alwall:2014hca} generator at LO with the CTEQ6L1 PDF set.

The MG5\_aMC@NLO~2.2.2 (2.2.3 for $t\bar{t} + Z/\gamma^{*}$) generator at LO, interfaced to the \pythia~8.186~\cite{Sjostrand:2014zea} parton-shower model, was used for the generation of $t\bar{t}$ + EW processes ($t\bar{t} + W/Z/WW$)~\cite{ATL-PHYS-PUB-2016-005}, with up to two ($t\bar{t}+W$, $t\bar{t}+Z(\to \nu\nu/qq)$), one ($t\bar{t}+Z(\to \ell\ell)$~\footnote{The letter $\ell$ stands for the charged leptons (electrons, muons and taus). While the contributions from tau leptons are included in all the Monte Carlo samples, in the next sections the symbol $\ell$ refers to electrons and muons only.}) or no ($t\bar{t}+WW$) extra partons included in the matrix element. The events were normalized to their respective NLO cross-sections~\cite{Lazopoulos:2008de,Campbell:2012dh}.

Diboson processes ($WW$, $WZ$, $ZZ$)~\cite{ATL-PHYS-PUB-2016-002} were simulated using the  \sherpa~2.2.1 generator and contain off-shell contributions. For processes with four charged leptons (4$\ell$), three charged leptons and a neutrino (3$\ell$+1$\nu$) or two charged leptons and two neutrinos (2$\ell$+2$\nu$), the matrix elements contain all diagrams with four electroweak couplings, and were calculated for up to one (4$\ell$, 2$\ell$+2$\nu$) or no extra partons (3$\ell$+1$\nu$) at NLO. All diboson samples were also simulated with up to three additional partons at LO using the \textsc{Comix} and \textsc{OpenLoops} matrix-element generators, and were merged with the \sherpa\ parton shower using the ME+PS@NLO prescription. The diboson events were normalized to their NLO cross-sections~\cite{Campbell:1999ah,Campbell:2011bn}. Additional MC simulation samples of events with a leptonically decaying vector boson and photon, $V\gamma$, were generated at LO using Sherpa 2.1.1~\cite{Gleisberg:2008ta}.  Matrix elements including all diagrams with three electroweak couplings were calculated with up to three partons at LO and merged with the Sherpa parton shower~\cite{Schumann:2007mg} according to the ME+PS@LO prescription~\cite{Hoeche:2009rj}. The CT10 PDF set is used in conjunction with dedicated parton shower tuning developed by the Sherpa authors. 

Triboson processes ($WWW,\ WWZ,\ WZZ$ and $ZZZ$) were simulated with the \sherpa~2.2.1 generator with matrix elements calculated at LO with up to one additional parton. The triboson events were normalized to their LO cross-sections~\cite{ATL-PHYS-PUB-2017-005}. 

Higgs-boson production processes (including gluon--gluon fusion, associated vector-boson production, $VH$,\footnote{The letter $V$ represents the $W$ or $Z$ gauge boson.} and vector-boson fusion, VBF) were generated using \POWHEG v2~\cite{ATL-PHYS-PUB-2016-004} + \PYTHIA 8.186 and normalized to cross-sections calculated at NNLO with soft gluon emission effects added at NNLL accuracy, whilst $\ttbar H$ events were produced using \AMCatNLO 2.2.2 + \herwig~2.7.1~\cite{Corcella:2000bw} and normalized to the NLO cross-section~\cite{Dittmaier:2012vm}. All samples assume a Higgs boson mass of 125~\GeV.

\begin{table}[H]
\scriptsize
\begin{center}
\caption{The SUSY signals and the Standard Model background Monte Carlo samples used in this paper. The generators, the order in $\alpha_{\textrm s}$ of cross-section calculations used for yield normalization, PDF sets, parton showers and parameter tunes used for the underlying event are shown.}
\begin{tabular}{ l l l l l l }
\toprule
Physics process & Generator& Cross-section & PDF set & Parton shower & Tune \\
&& normalization & & & \\
\hline
\hline
SUSY processes & \madgraph~v2.2.3 & NLO+NLL & NNPDF2.3LO & \pythia~8.186 & A14 \\
$Z/\gamma^{*}(\rightarrow \ell \bar \ell)$ + jets & \sherpa~2.2.1         & NNLO  &  NNPDF3.0NNLO   & \sherpa\      & \sherpa~default\\
$\gamma $ + jets & \sherpa~2.1.1         & LO  &    CT10  & \sherpa\   & \sherpa~default\\

$H(\rightarrow \tau\tau),\ H(\rightarrow WW)$ & \textsc{Powheg-Box}~v2   &  NLO  & CTEQ6L1 & \pythia~8.186 & A14    \\
$HW,\ HZ$      &  MG5\_aMC@NLO~2.2.2  & NLO  & NNPDF2.3LO & \pythia~8.186 & A14    \\
$t\bar{t}+H$          &  MG5\_aMC@NLO~2.2.2  & NLO  & CTEQ6L1 & \herwig~2.7.1 & A14   \\
$t\bar{t}$              & \textsc{Powheg-Box}~v2   & NNLO+NNLL                   &  CT10 &  \pythia~6.428  & {\textsc Perugia2012} \\
Single top ($Wt$-channel) &  \textsc{Powheg-Box} v2  &  NNLO+NNLL  &  CT10 &  \pythia~6.428   & {\textsc Perugia2012} \\ 
Single top ($s$-channel)           &  \textsc{Powheg-Box} v2  & NLO  &  CT10 &  \pythia~6.428   &  {\textsc Perugia2012}\\
Single top ($t$-channel)           & \textsc{Powheg-Box} v1  & NLO  &  CT10f4 &  \pythia~6.428   &  {\textsc Perugia2012}\\
Single top ($Zt$-channel) & MG5\_aMC@NLO~2.2.1 & LO & CTEQ6L1 & \pythia~6.428 & {\textsc Perugia2012} \\
$t\bar{t}+W/WW$       &  MG5\_aMC@NLO~2.2.2  & NLO  & NNPDF2.3LO & \pythia~8.186 & A14    \\
$t\bar{t}+Z$          &  MG5\_aMC@NLO~2.2.3  & NLO  & NNPDF2.3LO & \pythia~8.186 & A14    \\
$WW$, $WZ$, $ZZ$      &  \sherpa~2.2.1       & NLO  &  NNPDF30NNLO & \sherpa\   & \sherpa~default \\
$V\gamma$             &  \sherpa~2.1.1       & LO  &  CT10 & \sherpa\   & \sherpa~default \\
Triboson              &  \sherpa~2.2.1       & NLO  &  NNPDF30NNLO & \sherpa\   & \sherpa~default \\

\toprule
\end{tabular}
\label{tab:montecarlo}
\end{center}
\end{table}

Simplified models~\cite{Alwall:2008ag} are defined by an effective Lagrangian describing the interactions of a small number of new particles, assuming one production process and one decay channel with a 100\% branching ratio. 
Specifically, the SUSY production modes considered in this paper are studied in the context of simplified models, assuming wino-like chargino--neutralino production with decays via Standard Model $W$ and $Z$ gauge bosons and a bino-like LSP, leading to two- and three-lepton final states. As illustrated in Figure~\ref{fig:TreeDiagramsPhysScenarios}, two scenarios are considered: one where the $W$ boson decays leptonically resulting in a three-lepton plus missing-transverse-momentum ($\met$) final state (Figure~\ref{fig:C1N2treeWZ}), and one where the $W$ boson decays hadronically, yielding two leptons with same flavor and opposite-sign charge plus two jets plus $\met$ in the final state, as in Figure~\ref{fig:C1N2treeWZj}. Figures~\ref{fig:C1N2treeWZ3L} and \ref{fig:C1N2treeWZ2Lj} show the diagrams where the $\chinoonepm\ninotwo$ system is produced in association with an initial state radiation (ISR) jet leading again to three-lepton and two-lepton final states. 

The MC signal samples were generated from leading-order matrix elements with up to two extra partons using \madgraph~v2.2.3~\cite{madgraph1} interfaced to $\pythia$ version 8.186, with the A14 parameter tune~\cite{ATL-PHYS-PUB-2014-021}, for the modeling of the SUSY decay chain, parton showering, hadronization and the description of the underlying event. Parton luminosities were provided by the NNPDF23LO PDF set~\cite{CT10pdf}. Jet--parton matching follows the CKKW--L prescription~\cite{Lonnblad:2011xx}, with a matching scale set to one quarter of the $\chinoonepm/\ninotwo$ mass. Signal cross-sections were calculated at NLO in the strong coupling constant, with soft gluon emission effects added at next-to-leading-logarithm (NLL) accuracy~\cite{Beenakker:1996ch,Kulesza:2008jb,Kulesza:2009kq,Beenakker:2009ha,Beenakker:2011fu}. The nominal cross-section and the uncertainty were taken from an envelope of cross-section predictions using different PDF sets and factorization and renormalization scales, as described in Ref.~\cite{Borschensky:2014cia}. For $\chinoonepm$ and $\ninotwo$ with a mass of 500~\GeV, the production cross-section is $46\pm4$~fb at $\sqrt{s}= 13$~\TeV.

A summary of the SUSY signals and the SM background processes together with the MC generators, cross-section calculation orders in $\alpha_{\textrm s}$, PDFs, parton shower and parameter tunes used is given in Table~\ref{tab:montecarlo}. 

 \begin{figure}[h]
   \centering   
   \subfigure[\label{fig:C1N2treeWZ}]{\includegraphics[width=0.45\textwidth]{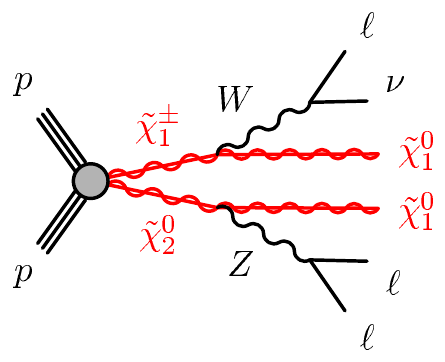}}
   \subfigure[\label{fig:C1N2treeWZj}]{\includegraphics[width=0.45\textwidth]{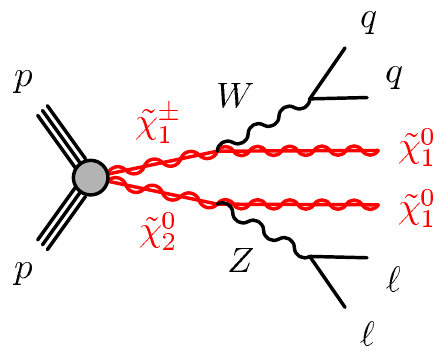}}
   \subfigure[\label{fig:C1N2treeWZ3L}]{\includegraphics[width=0.45\textwidth]{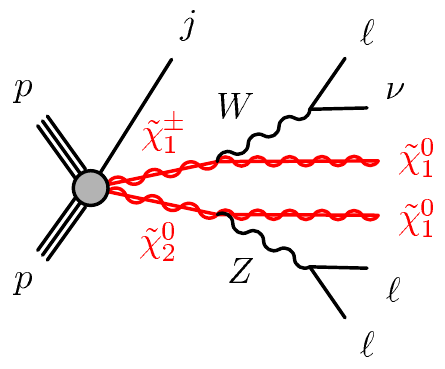}}
   \subfigure[\label{fig:C1N2treeWZ2Lj}]{\includegraphics[width=0.45\textwidth]{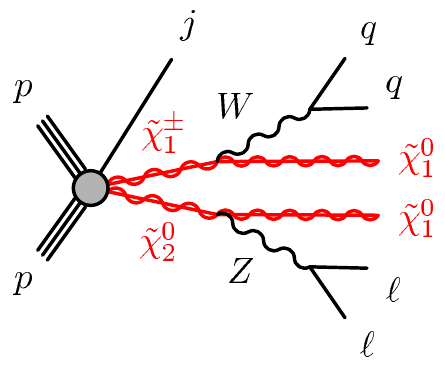}}

   \caption{Diagrams for the  physics scenarios studied in this paper:
   (a) $\chinoonepm\ninotwo$ with decays via leptonically decaying $W$ and $Z$ bosons, (b) $\chinoonepm\ninotwo$ with decays to two-lepton plus two-jet plus $\met$ final states through a hadronically decaying $W$ boson and a leptonically decaying $Z$ boson, (c) $\chinoonepm\ninotwo$ production in association with an initial state radiation jet (labeled `$j$' in the figure) with decays via leptonically decaying $W$ and $Z$ bosons and (d) $\chinoonepm\ninotwo$ production in association with an initial state radiation jet with decays to two-lepton plus two-jet plus $\met$ final states through a hadronically decaying $W$ boson and a leptonically decaying $Z$ boson.}
   \label{fig:TreeDiagramsPhysScenarios}
 \end{figure}

The {\textsc EvtGen}~v1.2.0 program~\cite{evtgen} was used to model the decays of $b$- and $c$-hadrons in the SM background samples except for those produced with \sherpa. All simulated events were overlaid with multiple $pp$ collisions simulated with the soft QCD processes of \pythia~8.186 using the A2 tune~\cite{A14tune} and the MSTW2008LO parton distribution functions~\cite{Martin:2009iq}. The MC samples were generated with a variable number of additional $pp$ interactions in the same and neighboring bunch crossings, and were reweighted to match the distribution of the mean number of interactions observed in data. 

For all SM background samples the response of the detector to particles was modeled with a full ATLAS detector simulation~\cite{:2010wqa} based on \textsc{Geant4}~\cite{Agostinelli:2002hh}. Signal samples were prepared using a fast simulation based on a parameterization of the performance of the ATLAS electromagnetic and hadronic calorimeters and on \textsc{Geant4} elsewhere.

\section{Object reconstruction and identification}
\label{sec:objects}

The reconstructed primary vertex of the event is required to be consistent with the luminous region and to have at least two associated tracks with $\pt > 400$~\MeV. When more than one such vertex is found, the vertex with the largest  $\sum \pt^2$ of the associated tracks is chosen.

Two different classes of reconstructed lepton candidates (electrons or muons) are used in the analysis, labeled baseline and high-purity in the following. When selecting samples for the search, events must contain a minimum of two baseline electrons or muons. 

Baseline muon candidates are formed by combining information from the muon spectrometer and ID as described in Ref.~\cite{MuonPerfRun2}, must pass the \textit{medium} identification requirements defined therein, and have $\ourpt > 10~\GeV$ and $|\eta| <2.7$.  High-purity muon candidates must additionally have $|\eta|<2.4$, the significance of the transverse impact parameter relative to the primary vertex $|d_0^{\mathrm{PV}}|/\sigma(d_0^{\mathrm{PV}}) <$ 3, and the longitudinal impact parameter relative to the primary vertex  $|z_0^{\mathrm{PV}} \mathrm{sin}\theta|<$ 0.5~mm. Furthermore, high-purity candidates must satisfy the 
\textit{GradientLoose} isolation requirements described in Ref.~\cite{MuonPerfRun2}, which rely on tracking-based and calorimeter-based variables and implement a set of $\eta$- and $\pt$-dependent criteria. The highest-\pT (leading) high-purity muon is also required to have $\pt > 25~\GeV$. 

Baseline electron candidates are reconstructed from an isolated electromagnetic calorimeter energy deposit matched to an ID track. They are required to have $\ourpt > 10~\GeV$, $|\eta| < 2.47$, and to satisfy a set of quality criteria similar to the \textit{Loose} likelihood-based identification criteria described in Ref.~\cite{ATLAS-CONF-2016-024}, but including a requirement of a B-layer hit. High-purity electron candidates additionally must satisfy \textit{MediumLH} selection criteria described in Ref.~\cite{ATLAS-CONF-2016-024}. They are also required to have $|d_0^{\mathrm{PV}}|/\sigma(d_0^{\mathrm{PV}}) <$~5, $|z_0^{\mathrm{PV}} \mathrm{sin}\theta|<$~0.5~mm, and to satisfy isolation requirements that are the same as those applied to high-purity muons~\cite{ATLAS-CONF-2016-024}. The leading high-purity electron is also required to have $\pt > 25~\GeV$. 

Jet candidates are reconstructed using the anti-$k_{t}$ jet clustering algorithm~\cite{Cacciari:2008gp,Cacciari:2005hq,Cacciari:2011ma} with a jet radius parameter of $0.4$ starting from clusters of calorimeter cells~\cite{Topocluster_ATLAS}. The jets are corrected for energy from pileup using the method described in Ref.~\cite{Cacciari:2007fd}: a contribution equal to the product of the jet area and the median energy density of the event is subtracted from the jet energy~\cite{ATLAS-CONF-2013-083}. Further corrections, referred to as the jet energy scale corrections, are derived from MC simulation and data and are used to calibrate the average energies of jets to the scale of their constituent particles~\cite{JetCalibRunTwo}. In order to reduce the number of jets originating from pileup, a significant fraction of the tracks associated with each jet must have an origin compatible with the primary vertex, as defined by the jet vertex tagger (JVT) output~\cite{JVT_CONF}. Only corrected jet candidates with $\ourpt > 20~\GeV$ and $|\eta|<4.5$  are retained. High-purity jets are defined with the tighter requirement $|\eta|<2.4$. The chosen requirement corresponds to the \textit{Medium} working point of the JVT and is only applied to jets with $\pt < 60~\GeV$ and $|\eta| < 2.4$. This requirement reduces jets from pileup to 1\% with an efficiency for pure hard-scatter jets of 92\%.

An algorithm based on boosted decision trees, MV2c10~\cite{btag_paper,ATL-PHYS-PUB-2016-012}, is used to identify jets containing a $b$-hadron ($b$-jets), with an operating point corresponding to an efficiency of 77\%, and rejection factors of $134$ for light-quark and gluon jets and $6$ for charm jets~\cite{ATL-PHYS-PUB-2016-012}, for reconstructed jets with $\ourpt > 20~\GeV$ and $|\eta|<2.5$ in simulated $t\bar{t}$ events. Candidate $b$-tagged jets are required to have $\ourpt > 20~\GeV$ and $|\eta|<2.4$.

After the selection requirements described above, ambiguities between candidate jets with $|\eta|<4.5$ and baseline leptons are resolved as follows:
\begin{enumerate}
\item Any electron sharing an ID track with a muon is removed.
\item If a $b$-tagged jet (identified using the 85\% efficiency working point of the MV2c10 algorithm) is within $\Delta R\equiv \sqrt{(\Delta y)^2 + (\Delta \phi)^2}=0.2$ of an electron candidate, the electron is rejected, as it is likely to originate from a semileptonic $b$-hadron decay; otherwise, if a non-$b$-tagged jet is within $\Delta R = 0.2$ of an electron candidate then the electron is kept and the jet is discarded as it is likely to be due to the electron-induced shower.
\item Electrons within $\Delta R = 0.4$ of a remaining jet candidate are discarded, to suppress electrons from semileptonic decays of $c$- and $b$-hadrons.
\item Jets with fewer than three associated tracks that have a nearby muon that carries a significant fraction of the transverse momentum of the jet ($\pt^\mu > 0.7 \sum \pt^{\textrm{jet tracks}}$, where $\pt^\mu$ and $\pt^{\textrm{jet tracks}}$ are the transverse momenta of the muon and the tracks associated with the jet, respectively) are discarded either if the candidate muon is within $\Delta R=0.2$ or if the muon is matched to a track associated with the jet.
\item Muons within $\Delta R=0.4$ of a remaining jet candidate are discarded to suppress muons from semileptonic decays of $c$- and $b$-hadrons.  
\end{enumerate}

The events used by the searches described in this paper are selected using high-purity leptons and jets with a trigger logic that accepts events with either two electrons, two muons or an electron plus a muon. The trigger-level requirements on the $\pt$ of the leptons involved in the trigger decision (the $\pt$ thresholds range between 8~\GeV\ and 22~\GeV) are looser than those applied offline to ensure that trigger efficiencies remain high and are constant in the relevant phase space.

Events containing a photon and jets are used to estimate the $Z/\gamma^{*}$+jets background in the $2\ell$+jets channel. These events are selected with a set of prescaled single-photon triggers with $\pt$ thresholds in the range 35--100~\GeV\ and an unprescaled single-photon trigger with threshold $\ourpt>140$~\GeV. High-purity photons must have $\ourpt >37$~\GeV\ to be on the efficiency plateau of the lowest-threshold single-photon trigger and satisfy a \textit{tight} identification requirement and $\ourpt$-dependent requirements on both track- and calorimeter-based isolation~\cite{Aaboud:2016yuq}. The $\gamma+$jets control sample, used for the data-driven $Z+$jets background estimate described in Section~\ref{sec:background2L}, makes use of high-purity photons. The ambiguities between candidate photons, jets and leptons are resolved by applying the following two requirements:
\begin{enumerate}
\item Photons are removed if they reside within $\Delta R$ = 0.4 of a baseline electron or muon.
\item Any jet within $\Delta R$ = 0.4 of any remaining photon is discarded.
\end{enumerate} 

The measurement of the missing transverse momentum vector $\anactualptmissvecwithapinit$ (and its magnitude $\ourmagptmiss$) is based on the calibrated transverse momenta of all electron, photon, muon and jet candidates and all tracks originating from the primary vertex and not associated with such objects~\cite{Aaboud:2018tkc}. The missing transverse momentum is the negative of the vector sum of the object momenta. 

\section{Analysis strategy and background prediction}
\label{sec:strategy}

To search for a possible signal, selection criteria are defined to enhance the expected signal yield relative to the SM background. Signal Regions (SRs) are designed using the MC simulation for both SUSY signals and the SM background processes, before looking at the data in the relevant phase space. They are optimized to maximize the expected sensitivity for the exclusion of each model considered. To estimate the SM backgrounds in an accurate and robust fashion, corresponding control regions (CRs) are defined for each of the signal regions. They are chosen to be orthogonal to the SR selections in order to provide independent data samples enriched in particular backgrounds, and are used to normalize the background MC simulation. The CR selections are optimized to have negligible SUSY signal contamination for the models near the LHC Run 1 excluded region's boundary~\cite{Aad:2015eda}, while minimizing the systematic uncertainties arising from the extrapolation of the CR event yields to estimate backgrounds in the SR. Cross-checks of the background estimates are performed with data in several validation regions (VRs) selected with requirements such that these regions do not overlap with the CR and SR selections, and also have a low expected signal contamination. 

To extract the final results, three different classes of likelihood fits are employed, denoted background-only, model-independent and model-dependent fits, using the HistFitter framework~\cite{HistFitter}. The fits are performed using the total number of events in each region. To obtain a set of background predictions that are independent of the observations in the SRs, the fit can be configured to use only the CRs to constrain the fit parameters: the SR bins are removed from the likelihood and any potential signal contribution is neglected everywhere. This fit configuration is referred to as the background-only fit. The scale factors representing the normalizations of background components relative to MC predictions are determined in the fit to all the CRs associated with an SR. This is most notably the case for diboson production since it is the dominant background in several SRs. The expected backgrounds in an SR are based on the yields predicted by simulation, corrected by the scale factors derived from the fit. A dedicated data-driven method is used to estimate the $Z$+jets background yield for the two lepton regions. The systematic and MC statistical uncertainties are included in the fit as nuisance parameters that are constrained by Gaussian distributions with widths corresponding to the sizes of the uncertainties considered and by Poisson distributions, respectively. The background-only fit results are also used to estimate the background event yields in the VRs. 

A model-independent fit is used to quantify the level of agreement between background predictions and observed yields and to quantify the number of possible beyond the Standard Model (BSM) signal events in each SR. This fit proceeds in the same way as the background-only fit, except that the number of observed events in the SR is added as an input to the fit, and an additional parameter for the BSM signal strength, constrained to be non-negative, is included. The observed and expected upper limits at 95\% confidence level (CL) on the number of events from BSM phenomena for each signal region ($S_{\textrm{obs}}^{95}$ and $S_{\textrm{exp}}^{95}$) are derived using the CL$_\text{s}$ prescription~\cite{Read:2002hq}, neglecting any possible signal contamination in the CRs. These limits, when normalized by the integrated luminosity of the data sample, may be interpreted as upper limits on the visible cross-section of BSM processes ($\langle\epsilon\sigma\rangle_{\textrm{obs}}^{95}$), where the visible cross-section is defined as the product of production cross-section, acceptance and efficiency. The model-independent fit is also used to compute the one-sided $p$-value of the background-only hypothesis ($p_0$), which quantifies the statistical significance of an excess; $p_0$ cannot exceed 0.5.

Finally, a model-dependent fit is used to set exclusion limits on the signal cross-sections for specific SUSY models. Such a fit proceeds in the same way as the model-independent fit, except that the yields in both the SRs and the CRs are taken into account. Signal-yield systematic uncertainties due to detector effects and the theoretical uncertainties in the signal acceptance are included in the fit. Correlations between signal and background systematic uncertainties are taken into account where appropriate. Limits on the signal cross-section are then mapped into limits on sparticle masses in the two-dimensional simplified-model planes.


\section{The recursive jigsaw reconstruction technique}
\label{sec:rjigsaw_intro}

The RJR technique~\cite{Jackson:2016mfb,Jackson:2017gcy} is a method for decomposing measured properties event by event to provide a basis of kinematic variables. This is achieved by approximating the rest frames of intermediate particle states in each event. This reconstructed view of the event gives rise to a natural basis of kinematic observables, calculated by evaluating the momentum and energy of different objects in these reference frames. Background processes are reduced by testing whether  each event exhibits the anticipated properties of the imposed decay tree under investigation while only applying minimal selection criteria on visible object momenta and missing momenta. The RJR technique is described in detail in Refs~\cite{Jackson:2016mfb,Jackson:2017gcy}, and has been used in previous ATLAS searches~\cite{Aaboud:2017ayj,Aaboud:2017vwy,Aaboud:2017aeu}.

\begin{figure}
\begin{center}
\subfigure[]{\includegraphics[width=0.4\textwidth]{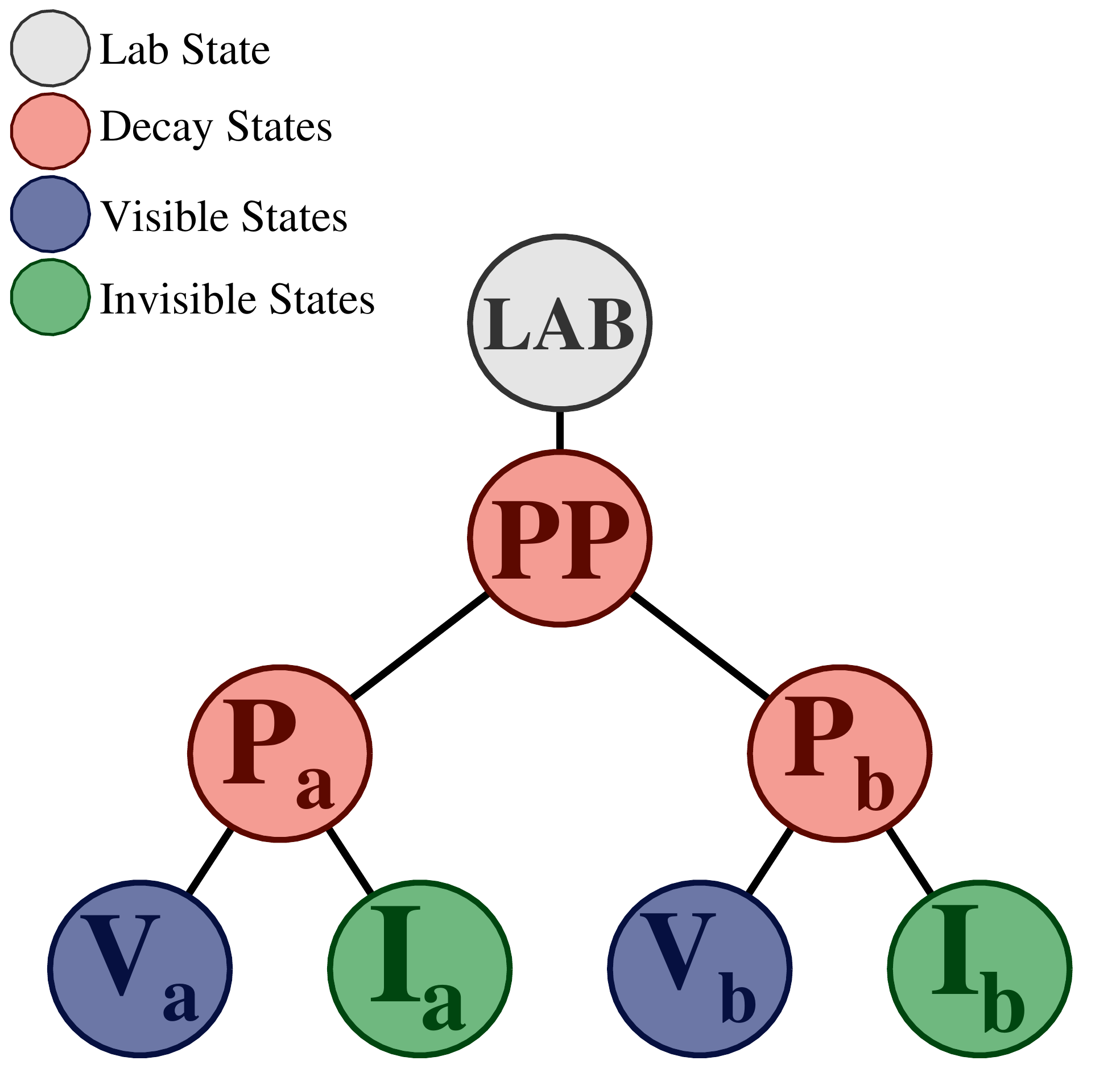}} \hspace*{0.05\textwidth}
\subfigure[]{\includegraphics[width=0.4\textwidth]{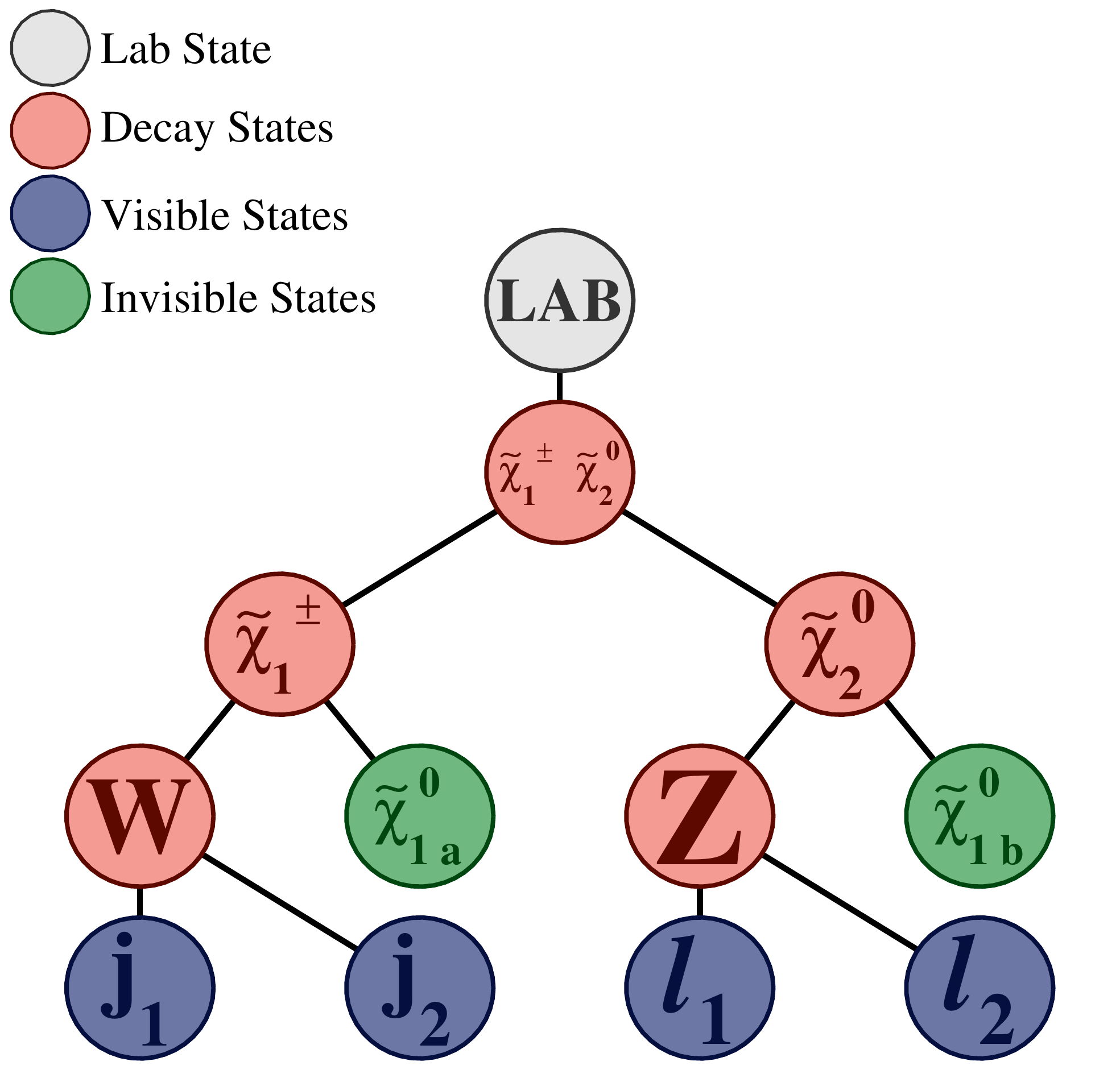}} 
\subfigure[]{\includegraphics[width=0.4\textwidth]{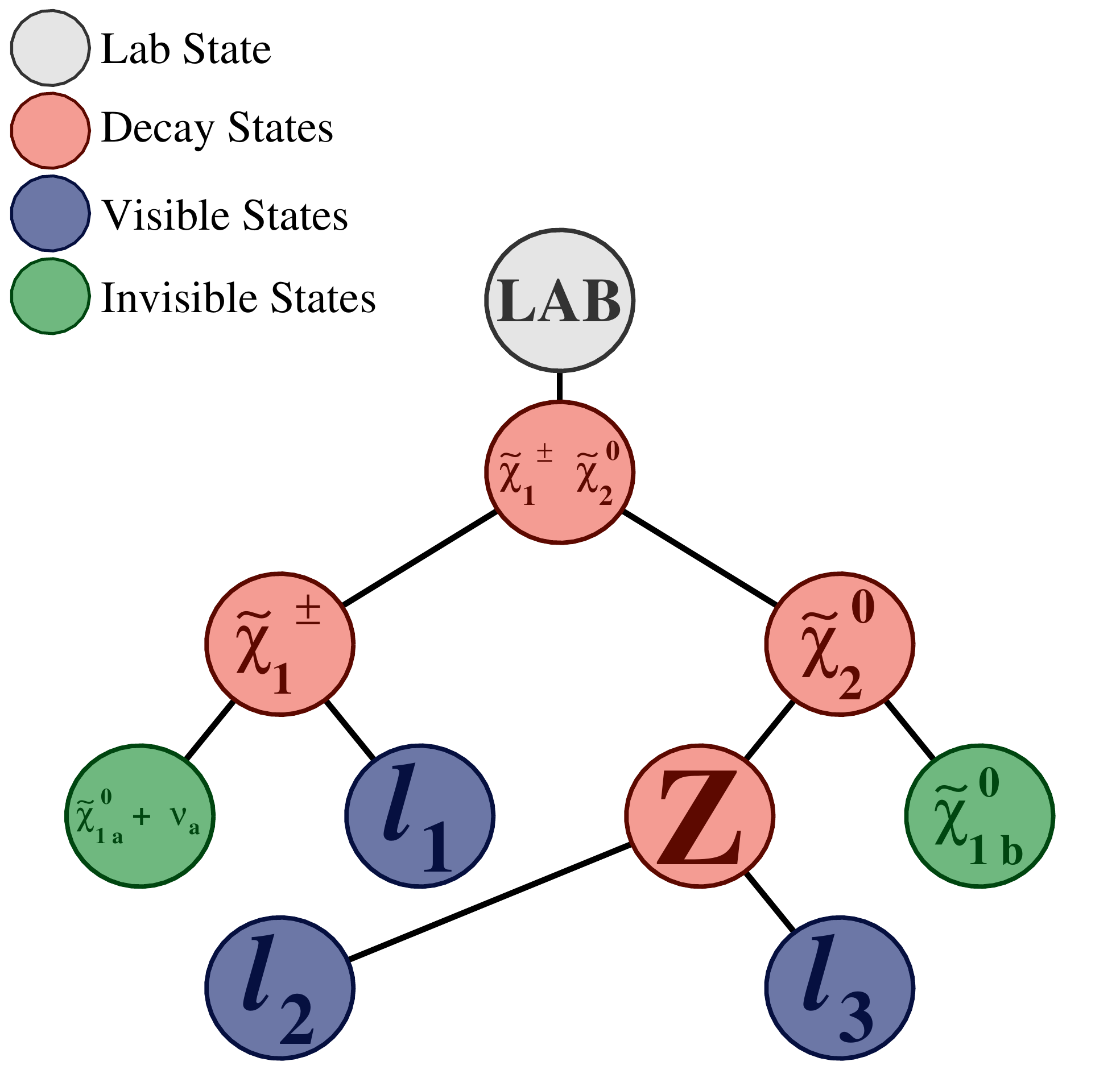}} \hspace*{0.05\textwidth}
\subfigure[]{\includegraphics[width=0.4\textwidth]{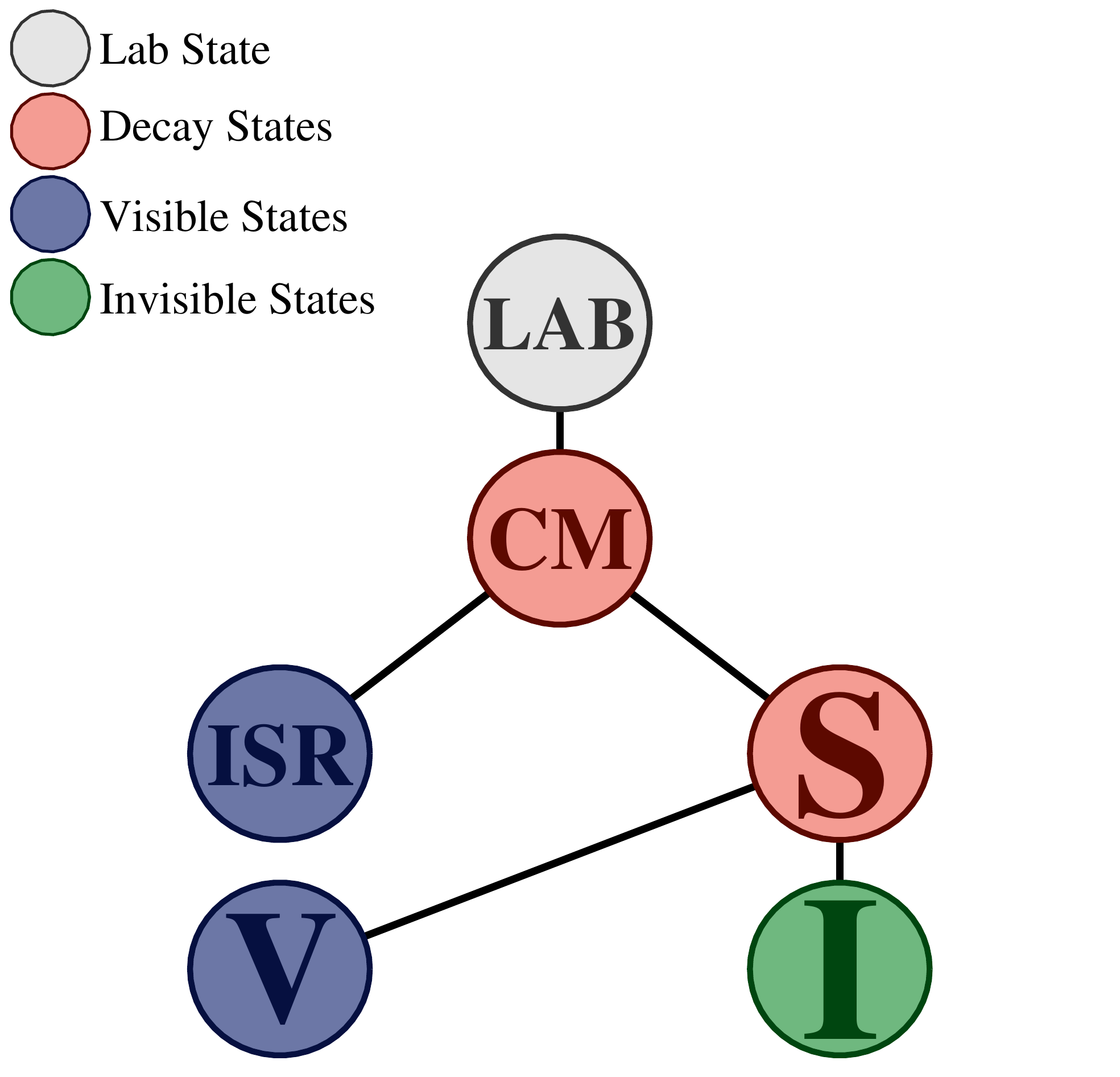}}
\end{center}
\caption{\label{fig:InclusiveTree} 
(a) The ``standard'' decay tree applied to pair-produced sparticles (``parent'' objects), P, decaying to visible states ``V'' and invisible states ``I''. (b) Decay trees for the $2\ell + 2$ jets final state and (c) $3\ell$ final state. (d) The ``compressed'' decay tree. CM denotes the center-of-mass frame. A signal sparticle system S decaying to a set of visible momenta V and invisible momentum I recoils from a jet-radiation system ISR.   
}
\end{figure}

Electrons, muons, hadronic jets and $\anactualptmissvecwithapinit$ (as defined in Section~\ref{sec:objects}) are used as input to the RJR algorithm. Motivated by searches for pair-production of sparticles in $R$-parity-conserving models, a decay tree is constructed following the canonical process in Figure~\ref{fig:InclusiveTree}(a), for the $2\ell$ (Figure~\ref{fig:InclusiveTree}(b)) and $3\ell$ (Figure~\ref{fig:InclusiveTree}(c)) search regions, used in the analysis of events. Each event is evaluated as if two sparticles (labeled PP) were produced, assigned to two hemispheres (P$_{\textrm{a}}$ and P$_{\textrm{b}}$) and then decayed to the particles observed in the detector with V denoting visible objects and I invisible objects. The benchmark signal models probed in this search give rise to signal events with at least two weakly interacting particles associated with two systems of invisible particles (shown in green), the respective children of the initially produced sparticles. For the $2\ell$ channel the lepton pair must be associated with the same visible collection, similarly for the jets, while for the $3\ell$ channel the opposite-charge, same-flavor pair most consistent with the $Z$-boson mass is selected as one visible collection, with the unpaired lepton being assigned to the opposite hemisphere (the $Z$ boson being associated with V$_{\textrm b}$, and the unpaired lepton with V$_{\textrm a}$).

After partitioning the visible objects, the remaining unknowns in the event are associated with the two collections of invisible particles: their masses, longitudinal momenta and information about how the two groups contribute to the $\anactualptmissvecwithapinit$. The RJR algorithm determines these unknowns by identifying the smallest Lorentz invariant function of the visible particles' four vectors that ensures the invisible particle mass estimators remain non-negative~\cite{Jackson:2017gcy}. In each of these newly constructed rest frames, all relevant momenta are defined and can be used to construct a set of variables such as multi-object invariant masses and angles between objects. The primary energy-scale-sensitive observables used in the search presented here are a suite of variables denoted by $H$. As shown in Eq.~(\ref{eq:Hdefinition}), the $H$ variables are constructed using different combinations of object momenta, including contributions from the invisible four-momenta, and are not necessarily evaluated in the lab frame, nor only in the transverse plane.

\begin{equation}
  \label{eq:Hdefinition}
  H_{n,m}^{\textrm{F}} = \sum_{i=1}^{n} |\vec{p}_{\textrm{vis},\ i}^{\textrm{~F}}| + \sum_{j=1}^{m} |\vec{p}_{\textrm{inv},\ j}^{\textrm{~F}}|
\end{equation}

The $H$ variables are labeled with a superscript F and two subscripts $n$ and $m$, $H_{n,m}^{\textrm{F}}$. The F represents the rest frame in which the momenta are evaluated. In this analysis, this may be the lab frame, the proxy for the sparticle--sparticle frame PP, or the proxy for the rest frame of an individual sparticle, P. The subscripts $n$ and $m$ represent the number of visible and invisible momentum vectors considered, respectively. For events with fewer than $n$ visible objects, the sum only runs over the available momenta. Only the leading $n-n_{\ell}$ jets are considered, where $n_{\ell}$ is the number of reconstructed leptons in the event. An additional subscript ``T'' denotes a transverse version of the variable, where the transverse plane is defined in a frame F as follows: the Lorentz transformation relating F to the lab frame is decomposed into a boost along the beam axis, followed by a subsequent transverse boost. The transverse plane is defined to be perpendicular to the longitudinal boost. In practice, this is the plane transverse to the beam-line. 

The following variables are used in the definition of the signal regions. The value of $n$ differs for the case of events with a leptonic $W$ decay where there are three visible objects and hence $n=3$, and for events with a hadronic $W$ decay where there are four visible objects, and thus $n=4$.

\begin{itemize}
\item $H_{n,1}^{~\textrm{PP}}$: scale variable as described above. Behaves similarly to the effective mass, $\meff$ (defined as the scalar sum of the transverse momenta of the visible objects and $\met$), used in previous ATLAS SUSY searches.
\item $H_{1,1}^{~\textrm{PP}}/H_{4,1}^{~\textrm{PP}}$: provides additional information in testing the balance of the two scale variables. This provides excellent discrimination against unbalanced events where the large scale is dominated by a particular object \pt\ or by large $\ourmagptmiss$. Behaves similarly to the $\met/\meff$. Utilized solely in the 2$\ell$ low mass signal region to mitigate the effects of $Z$+jets backgrounds, in cases where one high \pt\ jet dominates.
\item $p_{\textrm{T}~\textrm{PP}}^{\textrm{lab}}/(p^{\textrm{lab}}_{\textrm{T}~\textrm{PP}} + H_{\textrm T~n,1}^{~\textrm{PP}})$: compares the magnitude of the vector sum of the transverse momenta of all objects associated with the PP system in the lab frame ($p^{\textrm{lab}}_{\textrm{T}~\textrm{PP}}$) to the overall transverse scale variable considered. This quantity tests for significant boost in the transverse direction. For signal events this quantity peaks sharply towards zero while for background processes the distribution is broader. A test of how much a given process resembles the imposed PP system in the decay tree. 
\item $H_{\textrm{T}~3,1}^{~\textrm{PP}}/H_{3,1}^{~\textrm{PP}}$: a measure of the fraction of the momentum that lies in the transverse plane.
\item $\textrm{min}(H^{\textrm{P}_{\textrm{a}}}_{1,1},H^{\textrm{P}_{\textrm{b}}}_{1,1})/\textrm{min}(H^{\textrm{P}_{\textrm{a}}}_{2,1},H^{\textrm{P}_{\textrm{b}}}_{2,1})$: compares the scale due to one visible object and \met\ ($H^{\textrm{P}_{\textrm{a}}}_{1,1}$ and $H^{\textrm{P}_{\textrm{b}}}_{1,1}$ in their respective production frames) as opposed to two visible objects ($H^{\textrm{P}_{\textrm{a}}}_{2,1}$ and $H^{\textrm{P}_{\textrm{b}}}_{2,1}$). The numerator and denominator are each defined by finding the minimum value of these quantities. In the three-lepton case this corresponds to the hemisphere with the $Z$ boson as it is the only one with two visible objects, and the variable takes the form $H^{\textrm{P}_{\textrm{b}}}_{1,1}/H^{\textrm{P}_{\textrm{b}}}_{2,1}$. This variable tests against a single object taking a large portion of the hemisphere momentum. This is particularly useful in discriminating against $Z$+jets backgrounds.  
\item $\Delta\phi_{\textrm{V}}^{\textrm{P}}$: the azimuthal opening angle between the visible system V in frame P and the direction of the boost from the PP to P frame.
Standard Model backgrounds from diboson, top and $Z$+jets processes peak towards zero and $\pi$ due to their topologies not obeying the imposed decay tree while signals tend to have a flat distribution in this variable. 
\end{itemize}

In addition to trying to resolve the entirety of the signal event, it can be useful for sparticle spectra with smaller mass-splittings and lower intrinsic $\ourmagptmiss$ to instead select events with a partially resolved sparticle system recoiling from a high-\pt\ jet from ISR. To target such topologies, a separate decay tree for compressed spectra is shown in Figure~\ref{fig:InclusiveTree}(d). This tree is somewhat simpler and attempts to identify visible (V) and invisible (I) systems that are the result of an intermediate state corresponding to the system of sparticles and their decay products (S). As the \met is used to choose which jets are identified as ISR, a transverse view of the reconstructed event is used which ignores the longitudinal momentum of the jets and leptons, as described in Ref~\cite{Jackson:2016mfb}. The reference frames appearing in the decay tree shown in Figure~\ref{fig:InclusiveTree}(d), such as the center-of-mass (CM) frame of the whole reaction, are then approximations in this transverse projection. This tree yields a slightly different set of variables:

\begin{itemize}
\item $p_{\mathrm{T\ ISR}}^{~\textrm{CM}}$: the magnitude of the vector-summed transverse momenta of all jets assigned to the ISR system. 
\item $p_{\mathrm{T\ I}}^{~\textrm{CM}}$: the magnitude of the vector-summed transverse momenta of the invisible system. Behaves similarly to $\met$.
\item $p_{\mathrm{T}}^{~\textrm{CM}}$: the magnitude of the vector-summed transverse momenta of the CM system.
\item $R_{\textrm{ISR}} \equiv \vec{p}_{\textrm{I}}^{~\textrm{CM}}\cdot \hat{p}_{\mathrm{T}~\textrm{S}}^{~\textrm{CM}}/p_{\mathrm{T}~\textrm{S}}^{~\textrm{CM}}$: serves as an estimate of $m_{\tilde{\chi}^{0}_{1}}/m_{\tilde{\chi}^{0}_{2}/\tilde{\chi}^{\pm}_{1}}$. This corresponds to the fraction of the momentum of the S system that is carried by its invisible system I, with momentum $\vec{p}_{\textrm{I}}^{~\textrm{CM}}$ in the CM frame. As $p_{\mathrm{T}~\textrm{S}}^{~\textrm{CM}}$ grows it becomes increasingly hard for backgrounds to possess a large value in this ratio --- a feature exhibited by compressed signals~\cite{Jackson:2016mfb}.
\item $N_{\textrm{jet}}^{\textrm{S}}$: number of jets assigned to the signal system (S).
\item $N_{\textrm{jet}}^{~\textrm{ISR}}$: number of jets assigned to the ISR system.
\item $\Delta\phi_{~\textrm{ISR},\textrm{I}}^{~\textrm{CM}}$: the azimuthal opening angle between the ISR system and the invisible system in the CM frame.
\item $m_{Z}$: mass of the dilepton pair assigned to the signal system. In the 3-lepton final state, the $Z$ candidate is formed by finding the same-flavor opposite-charge pair closest to the $Z$ mass.
\item $m_{J}$: mass of the jet system assigned to the signal system.  
\end{itemize}

\section{Event selection: control, validation and signal region definitions}
\label{sec:selection}

Following the object reconstruction described in Section~\ref{sec:objects} and analysis strategy outlined in Section~\ref{sec:strategy}, the variables described in Section~\ref{sec:rjigsaw_intro} are used to define a set of SRs sensitive to the topologies of interest. 

Both the $2\ell$ and $3\ell$ SRs are designed to cover a wide range of $\chinoonepm/\ninotwo$ masses and different mass-splittings, $\Delta m$ $=$ $m_{\chinoonepm/\ninotwo}-m_{\ninoone}$. Specifically, the high-mass regions target high $\chinoonepm/\ninotwo$ masses and large mass-splittings ($\Delta m\gtrsim400$~\GeV) and the intermediate-mass regions probe mass-splittings of $\approx$200~\GeV. The low-mass and ISR SRs are constructed in order to probe similar regions of the two-dimensional SUSY parameter space and particularly the mass-splittings of $\approx$100~\GeV. In this region it is difficult to distinguish the signal from SM processes, due to the limited momentum that the LSPs carry. Improved sensitivity is achieved by designing the two low-mass and ISR SRs to be mutually exclusive, with each providing sensitivity to the parameter space under scrutiny. A statistical combination of these regions subsequently leads to further improved sensitivities. A schematic representation of the mass regions targeted by each SR can be seen in Figure~\ref{fig:SRsketch}.

For selections involving three charged leptons, the $W$-boson transverse mass, $m^{W}_{\textrm{T}}$, is used and is derived from $\anactualptmissvecwithapinit$ and the transverse momentum of the charged lepton ($p_{\textrm{T}}^{\ell}$) not associated with the $Z$ boson as follows: 

\begin{equation*}
  m^{W}_{\textrm{T}} = \sqrt{2 p_{\textrm{T}}^{\ell}\MET (1 - \cos\Delta\phi) },
\end{equation*}

where $\Delta\phi$ is the azimuthal opening angle between the charged lepton associated with the $W$ boson and the missing transverse momentum.

\begin{figure}[htbp]
  \begin{center} 
    \includegraphics[width=0.7\textwidth]{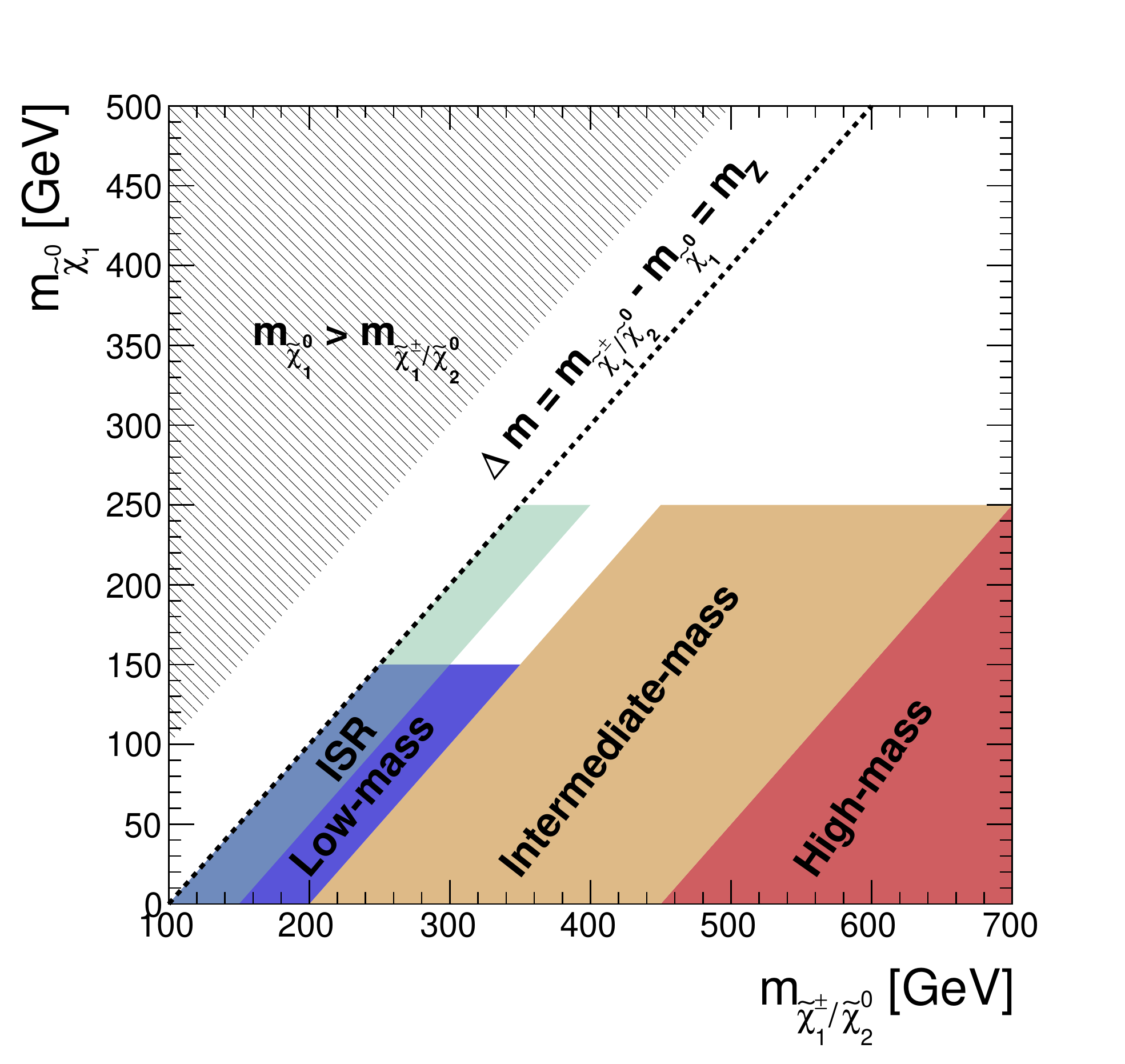}
    \vspace*{-0.03\textheight}
    \caption{\label{fig:SRsketch} Sketch of the regions that are probed by each signal region in the two-dimensional parameter space $m_{\chinoonepm/\ninotwo}$--$m_{\ninoone}$.}
  \end{center}
\end{figure}

\subsection{Event selection in the two-lepton channel}

The $2\ell$ search channel, using the standard decay tree, is designed with three SRs, two CRs to constrain the $VV$ background (where $V=W,\ Z$) and the processes with top quarks ($Wt+t\bar{t}$, where the sign symbolizes the sum of the two processes) and four VRs for validating the main background processes (including the $Z$+jets data-driven estimate described in Section~\ref{sec:background2L}). The preselection criteria used for the definition of the standard-decay-tree regions are listed in Table~\ref{tab:StandardRJPreselections} and include requirements on the lepton multiplicity ($n_{\textrm{leptons}}$), the jet multiplicity ($n_{\textrm{jets}}$), the $b$-tag jet multiplicity ($n_{b\textrm{-tag}}$), the transverse momenta of the leading ($p_{\textrm{T}}^{\ell_{1}},\ p_{\textrm{T}}^{j_{1}}$) and subleading ($p_{\textrm{T}}^{\ell_{2}},\ p_{\textrm{T}}^{j_{2}}$) leptons and jets and the invariant mass of the dilepton ($m_{\ell\ell}$) and dijet ($m_{jj}$) system. Most of  the regions are defined with exactly two opposite-charge, same-flavor leptons with transverse momentum greater than 25~\GeV\ and an invariant mass consistent with arising from a $Z$ boson. Exceptions to this are the diboson CR (CR$2\ell$-VV) and top VR (VR$2\ell$-Top). The CR$2\ell$-VV requires three or four leptons, which helps to select a sample enriched in diboson events as well as to ensure orthogonality with the SRs. The lepton pair is selected by choosing the opposite-charge, same-flavor pair closest to the $Z$ mass, while the remaining lepton(s) are treated as invisible objects contributing to $\anactualptmissvecwithapinit$. The additional requirement on $m_{\textrm{T}}^{W}$, which is applied only in the events containing exactly three charged leptons, ensures orthogonality with the $3\ell$ regions described in Section~\ref{sec:eventSel3L}. Both the top CR (CR$2\ell$-Top) and VR (VR$2\ell$-Top) are defined with a $b$-tag jet requirement while orthogonality with each other is ensured by inverting the dilepton invariant mass requirement. In all regions the dijet invariant mass is formed using the two leading jets in $\pt$. The SRs require the $m_{jj}$ to be consistent with a $W$ boson while the $Z+$jets (VR$2\ell$\_High-Zjets and VR$2\ell$\_Low-Zjets) and diboson (VR$2\ell$-VV) VRs select events outside of the $W$ mass window.

In addition to the preselection criteria, further selection requirements are applied in each region according to the parameter space probed. These selection requirements are shown in Table~\ref{tab:StandardRJSelections}. The min$\Delta\phi(j_1/j_2,\anactualptmissvecwithapinit)$ variable corresponds to the minimum azimuthal angle between the jets and $\anactualptmissvecwithapinit$ and is applied only in SR$2\ell$\_Low to further suppress the $Z+$jets contribution. The selection criteria applied in VR$2\ell$\_High-Zjets and VR$2\ell$\_Low-Zjets differ so as to be closer and orthogonal to their respective SRs. As such the $0.35 < H_{1,1}^{\textrm{PP}}/H_{4,1}^{\textrm{PP}} < 0.6$ requirement is retained only for VR$2\ell$\_Low-Zjets. VR$2\ell$-VV is the only region with an $H_{1,1}^{\textrm{PP}}$ requirement, but one that is necessary since it further suppresses the $Z+$jets background while keeping the VRs close to the SRs.

Similar to the $2\ell$ standard-decay-tree regions, another set of $2\ell$ regions is defined by taking advantage of the compressed decay tree. SR$2\ell$\_ISR has a requirement of at least three jets which makes it orthogonal to SR$2\ell$\_Low, where the jet multiplicity is defined with exactly two jets. The lepton and jet multiplicities as well as the requirements on the transverse momenta of these objects defining the preselection requirements in the ISR analysis are summarized in Table~\ref{tab:CompressedRJPreselections2L}. All the regions require at least one jet assigned to the ISR system ($N_{\textrm{jet}}^{~\textrm{ISR}}$) and at least two jets in the signal system ($N_{\textrm{jet}}^{\textrm{S}}$) in the construction of the compressed decay tree. The assignment of the jets in the two systems results from a mass minimization performed in the CM frame. Following the same strategy as for the CR$2\ell$\_VV, both CR$2\ell$\_ISR-VV and VR$2\ell$\_ISR-VV are defined with three or four leptons. To increase the number of events in VR$2\ell$\_ISR-VV, the transverse momentum requirement for jets is relaxed to 20~\GeV\ compared to 30~\GeV\ in the other regions.

\begin{table}[!htb]
  \centering
  \scriptsize
\caption{Preselection criteria for the three standard-decay-tree $2\ell$ SRs and the associated CRs and VRs. The variables are defined in the text.}
\begin{tabular}{l r r r r r r r r}
\toprule
Region                    &   $n_{\textrm{leptons}}$   &        $n_{\textrm{jets}}$  &  $n_{b\textrm{-tag}}$ & $\pt^{\ell_{1},\ell_{2}}$~[\GeV] & $\pt^{j_{1},j_{2}}$~[\GeV]  & $m_{\ell\ell}$~[\GeV] & $m_{jj}$~[\GeV] & $m_{\textrm{T}}^{W}$~[\GeV]\\
\hline
\hline
CR$2\ell$-VV              &  $\in[3,4]$  & $\geq 2$ & =0    & $>25$ & $>30$ & $\in(80,100)$& $>20$ & $\in(70,100)$\\
                          &    &  &    &  &  & &  & if $n_{\textrm{leptons}} =3$\\

CR$2\ell$-Top             &    $=2$     & $\geq 2$ & $=$1 & $>25$ & $>30$ & $\in(80,100)$& $\in(40,250)$ & $-$\\
\hline
VR$2\ell$-VV              &    $=2$     & $\geq 2$ & =0    & $>25$ & $>30$ & $\in(80,100)$&$\in(40,70)$& $-$\\
                          &          &          &     &       &       &              &or $\in(90,500)$& $-$\\
VR$2\ell$-Top             &    $=2$     & $\geq 2$ & $=$1    & $>25$ & $>30$ & $\in(20,80)$ &$\in(40,250)$& $-$\\
                          &          &          &      &       &       & or $>100$    &                & $-$\\ 
VR$2\ell$\_High-Zjets     &    $=2$     & $\geq 2$ & $=0$    & $>25$ & $>30$ & $\in(80,100)$&$\in(0,60)$      & $-$\\
                          &          &          &      &       &       &              &or $\in(100,180)$& $-$\\
VR$2\ell$\_Low-Zjets      &    $=2$     & $=2$     & $=0$    & $>25$ & $>30$ & $\in(80,100)$&$\in(0,60)$      & $-$\\   
                          &          &          &      &       &       &              &or $\in(100,180)$& $-$\\   
\midrule
SR$2\ell$\_High           &    $=2$     & $\geq 2$ & $=0$    & $>25$ & $>30$ & $\in(80,100)$& $\in(60,100)$   & $-$\\
SR$2\ell$\_Int            &    $=2$     & $\geq 2$ & $=0$    & $>25$ & $>30$ & $\in(80,100)$& $\in(60,100)$   & $-$\\
SR$2\ell$\_Low            &    $=2$     & $=2$     & $=0$    & $>25$ & $>30$ & $\in(80,100)$& $\in(70,90)$    & $-$\\
\toprule
\end{tabular}
\label{tab:StandardRJPreselections}
\end{table}

\begin{table}
  \centering
  \scriptsize
\caption{Selection criteria for the three standard-decay-tree $2\ell$ SRs and the associated CRs and VRs. The variables are defined in the text}
\begin{tabular}{l r r r r r r r}
\toprule
Region                    &   $H_{4,1}^{\textrm{PP}}$~[\GeV]  & $H_{1,1}^{\textrm{PP}}$~[\GeV] & $\frac{p^{\textrm{lab}}_{\textrm{T}~\textrm{PP}}}{p^{\textrm{lab}}_{\textrm{T}~\textrm{PP}}+H^{\textrm{PP}}_{\textrm{T}\ 4,1}}$ &
$\frac{\textrm{min}(H^{\textrm{P}_{\textrm{a}}}_{1,1},H^{\textrm{P}_{\textrm{b}}}_{1,1})}{\textrm{min}(H^{\textrm{P}_{\textrm{a}}}_{2,1},H^{\textrm{P}_{\textrm{b}}}_{2,1})}$
& $\frac{H_{1,1}^{\textrm{PP}}}{H_{4,1}^{\textrm{PP}}}$& $\Delta\phi_{\textrm{V}}^{\textrm{P}}$  & min$\Delta\phi(j_1/j_2,\anactualptmissvecwithapinit)$ \\
\hline
\hline
CR$2\ell$-VV              &$>200$&$-$   &$<0.05$& $>0.2$     &$-$             &$\in(0.3,2.8)$&$-$ \\
CR$2\ell$-Top             &$>400$&$-$   &$<0.05$& $>0.5$     &$-$             &$\in(0.3,2.8)$&$-$ \\
\hline
VR$2\ell$-VV              &$>400$&$>250$&$<0.05$& $\in(0.4,0.8)$&$-$             &$\in(0.3,2.8)$&$-$ \\
VR$2\ell$-Top             &$>400$&$-$   &$<0.05$& $>0.5$     &$-$             &$\in(0.3,2.8)$&$-$ \\
VR$2\ell$\_High-Zjets     &$>600$&$-$   &$<0.05$& $>0.4$     &$-$             &$\in(0.3,2.8)$&$-$ \\
VR$2\ell$\_Low-Zjets      &$>400$&$-$   &$<0.05$& $-$        &$\in(0.35,0.60)$&$-$           &$-$\\   
\hline
SR$2\ell$\_High           &$>800$&$-$   &$<0.05$& $>0.8$     &$-$             &$\in(0.3,2.8)$&$-$ \\
SR$2\ell$\_Int            &$>600$&$-$   &$<0.05$& $>0.8$     &$-$             &$\in(0.6,2.6)$&$-$ \\
SR$2\ell$\_Low            &$>400$&$-$   &$<0.05$ &  $-$      &$\in(0.35,0.60)$&$-$           &$>2.4$\\
\toprule
\end{tabular}
\label{tab:StandardRJSelections}
\end{table}

\begin{table}[h]
  \centering
  \scriptsize
\caption{Preselection criteria for the compressed-decay-tree $2\ell$ SR and the associated CRs and VRs. The variables are defined in the text.}
\begin{tabular}{l r r r r r r r}
\toprule
Region                &   $n_{\textrm{leptons}}$   & $N_{\textrm{jet}}^{~\textrm{ISR}}$ & $N_{\textrm{jet}}^{\textrm{S}}$ & $n_{\textrm{jets}}$  &  $n_{b\textrm{-tag}}$ & $\pt^{\ell_{1},\ell_{2}}$~[\GeV] & $\pt^{j_{1},j_{2}}$~[\GeV]\\
\hline
\hline
CR$2\ell$\_ISR-VV     &  $\in[3,4]$  & $\geq 1$  & $\geq 2$   & $> 2$        & $=0$    & $>25$ & $>30$\\
CR$2\ell$\_ISR-Top    &    $=2$     & $\geq 1$  & $=2$       & $\in[3,4]$   & $=1$ & $>25$ & $>30$\\
\hline
VR$2\ell$\_ISR-VV     &  $\in[3,4]$  & $\geq 1$  & $\geq 2$  & $\geq 3$      & $=0$    & $>25$ & $>20$ \\
VR$2\ell$\_ISR-Top    &    $=2$     & $\geq 1$  & $= 2$     & $\in[3,4]$    & $=1$ & $>25$ & $>30$ \\
VR$2\ell$\_ISR-Zjets  &    $=2$     & $\geq 1$  & $\geq 1$  & $\in[3,5]$    & $=0$    & $>25$ & $>30$ \\
\hline
SR$2\ell$\_ISR        &    $=2$     & $\geq 1$  & $= 2$     &$\in[3,4]$     & $=0$    & $>25$ & $>30$ \\
\toprule
\end{tabular}
\label{tab:CompressedRJPreselections2L}
\end{table}

\begin{table}[h]
  \centering
  \scriptsize
\caption{Selection criteria for the compressed-decay-tree $2\ell$ SR and the associated CRs and VRs. The variables are defined in the text.}
\begin{tabular}{l r r r r r r r}
\toprule
Region                 & $m_{Z}$~[\GeV] & $m_{J}$~[\GeV] & $\Delta\phi_{\textrm{ISR,I}}^{\textrm{CM}}$ & $R_{\textrm{ISR}}$        & $p^{\textrm{CM}}_{\textrm{T\ ISR}}$ [\GeV] & $p^{\textrm{CM}}_{\textrm{T\ I}}$ [\GeV] & $p^{\textrm{CM}}_{\textrm{T}}$ [\GeV] \\ 
\hline
\hline
CR$2\ell$\_ISR-VV      & $\in(80,100)$     &   $>20$           &  $>2.0$               &$\in(0.0,0.5)$    & $>50$               &  $>50$            &    $<30$         \\
CR$2\ell$\_ISR-Top     & $\in(50,200)$     &   $\in(50,200)$   &  $>2.8$               &$\in(0.4,0.75)$   & $>180$              &  $>100$           &    $<20$         \\   
\hline
VR$2\ell$\_ISR-VV      &  $\in(20,80)$     &   $>20$           & $>2.0$               & $\in(0.0,1.0)$   & $>70$               & $>70$             & $<30$            \\
                       &  or $>100$           &                   &                      &    &         &          &         \\                   
VR$2\ell$\_ISR-Top     &  $\in(50,200)$    &   $\in(50,200)$   &$>2.8$               & $\in(0.4,0.75)$  & $>180$              & $>100$             & $>20$           \\
VR$2\ell$\_ISR-Zjets   &  $\in(80,100)$    &    $<50$ or $>110$            &$-$                  & $-$              & $>180$              & $>100$            & $<20$            \\
\hline
SR$2\ell$\_ISR         &  $\in(80,100)$    &  $\in(50,110)$ & $>2.8$               & $\in(0.4,0.75)$  & $>180$              & $>100$            & $<20$                 \\
\toprule
\end{tabular}
\label{tab:2L2J_compressedCRs}
\end{table}

The ISR regions are further defined with a series of requirements based on the variables reconstructed from the compressed decay tree. These requirements are listed in Table~\ref{tab:2L2J_compressedCRs}. The ISR SR is defined by requiring a highly energetic ISR jet system which recoils against the entire signal system in the CM frame. In VR$2\ell$\_ISR-VV the $m_{Z}$ requirement is inverted in order to be orthogonal to the CR$2\ell$\_ISR-VV. The top CRs (CR$2\ell$\_ISR-Top) and VR (VR$2\ell$\_ISR-Top) are defined with a $b$-tag jet requirement and have broader $m_{Z}$ and $m_{J}$ windows. The broader mass windows help to increase the numbers of data events in these regions since in processes with top quarks the leptons and jets result from sources other than $Z$ and $W$ bosons, respectively. The orthogonality of the two regions is achieved by inverting the $p^{\textrm{CM}}_{\textrm{T}}$ requirement. A validation region for $Z+$jets (VR$2\ell$\_ISR-Zjets) is defined with exactly two leptons and between three and five jets, none of which are $b$-tagged; $m_{J}$ must be outside of the range expected from vector-boson decays ($<50$~\GeV\ or $>110$~\GeV).

Post-fit distributions of variables from the 2$\ell$ search for selected regions are shown in Figures~\ref{fig:crRJR2L} and~\ref{fig:vrRJR2L} for data and the different MC samples. In these figures, the background component labeled as ``Others'' includes the SM contributions from Higgs boson, $V\gamma,\ VVV,\ t\bar{t}V$ production and contributions from non-prompt and non-isolated leptons. The background estimate is described in Section~\ref{sec:background}.

\begin{figure}[!tbp]
\begin{center} 
\vspace*{-0.03\textheight}
\subfigure[]{\includegraphics[width=0.42\textwidth]{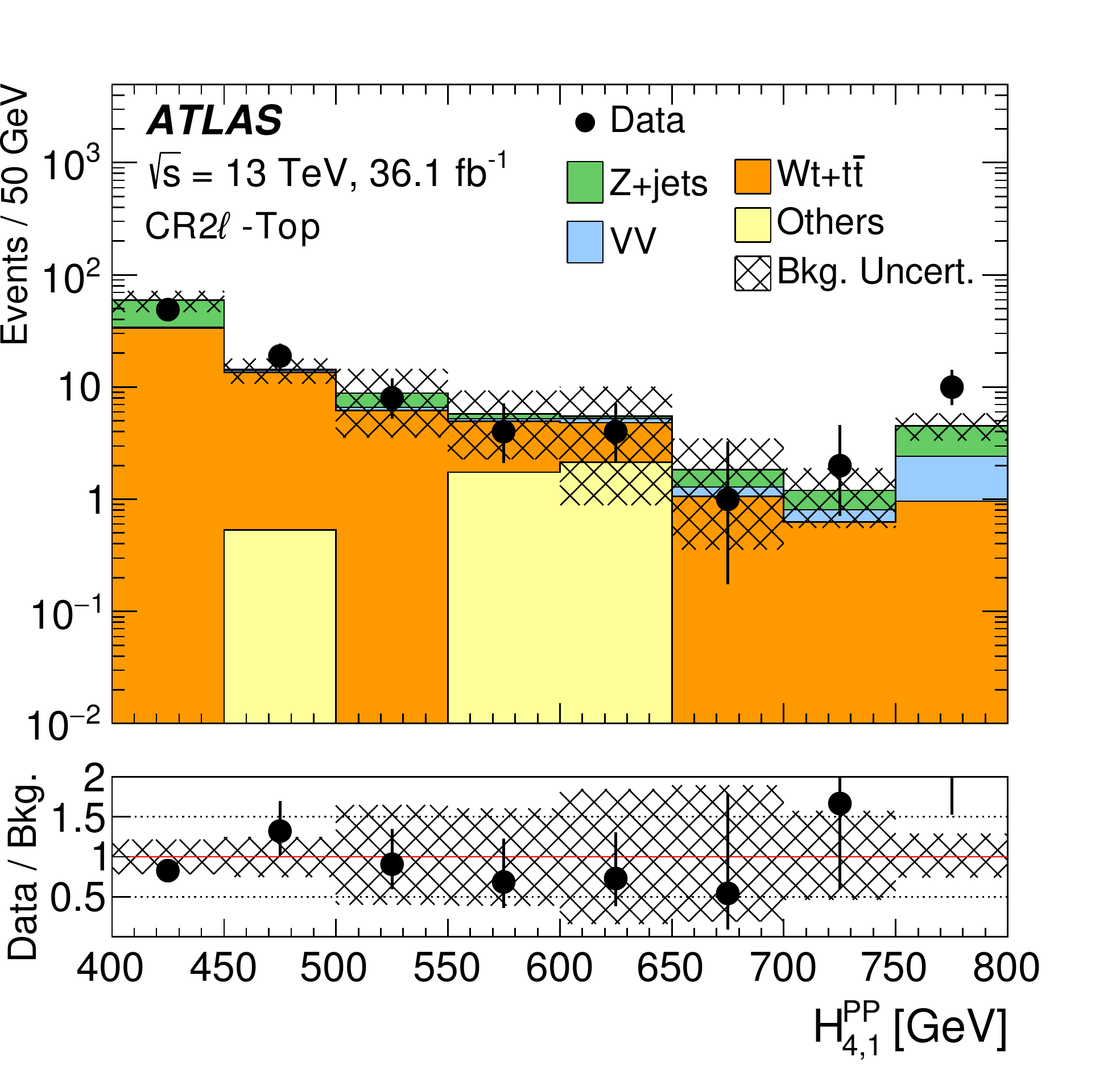}} 
\subfigure[]{\includegraphics[width=0.42\textwidth]{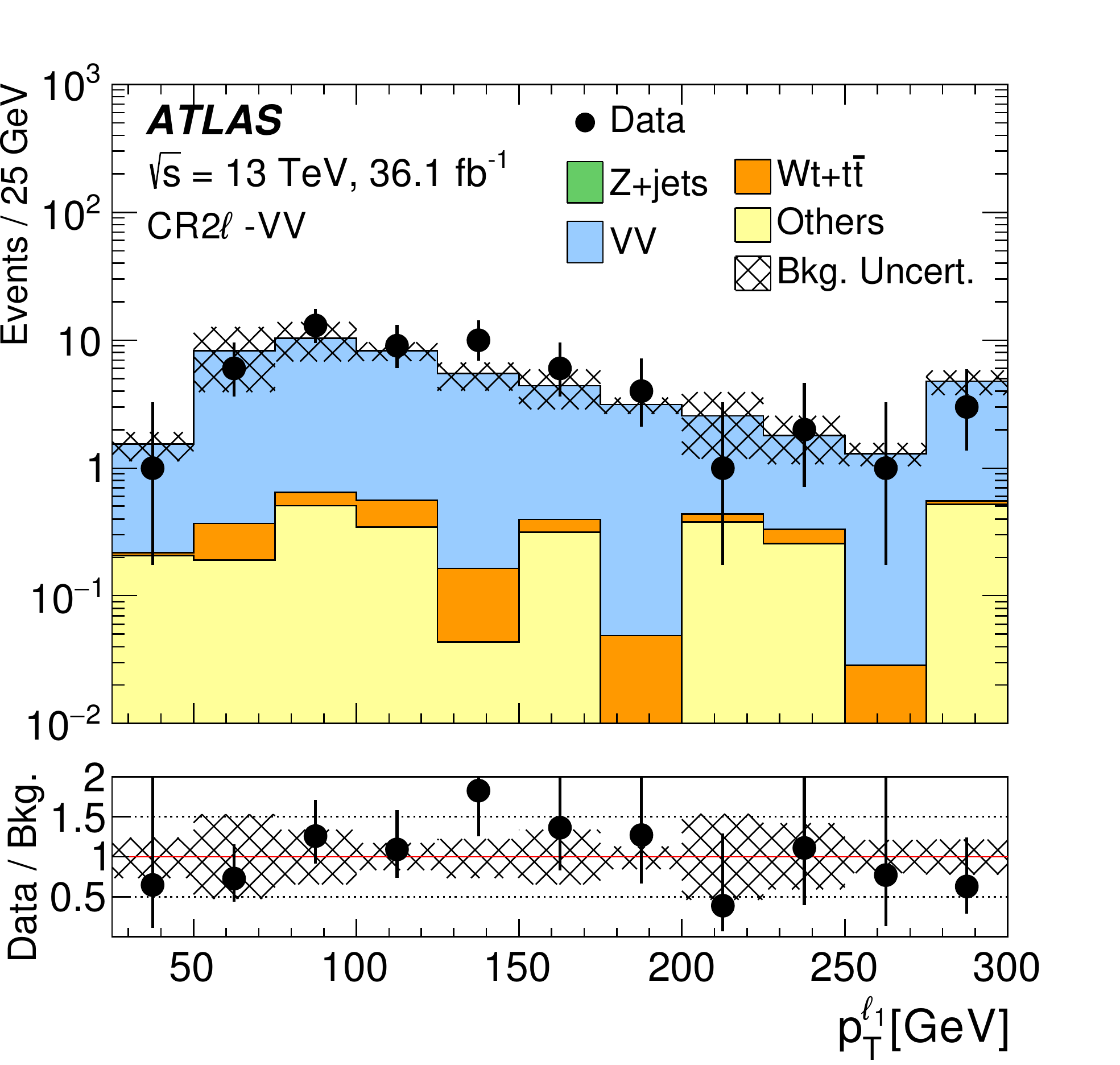}} \\
\vspace*{-0.005\textheight}\subfigure[]{\includegraphics[width=0.42\textwidth]{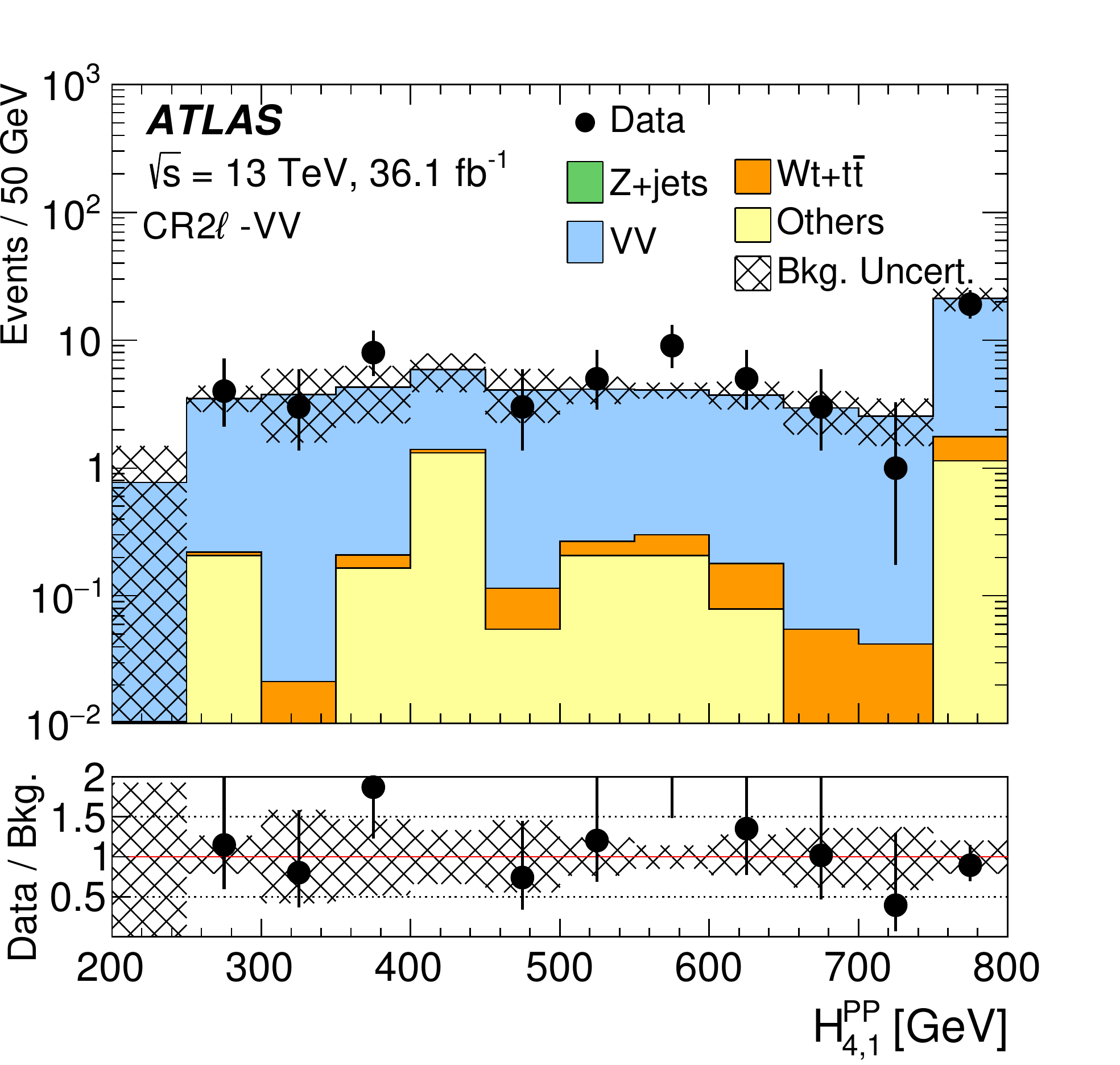}} 
\vspace*{-0.005\textheight}\subfigure[]{\includegraphics[width=0.42\textwidth]{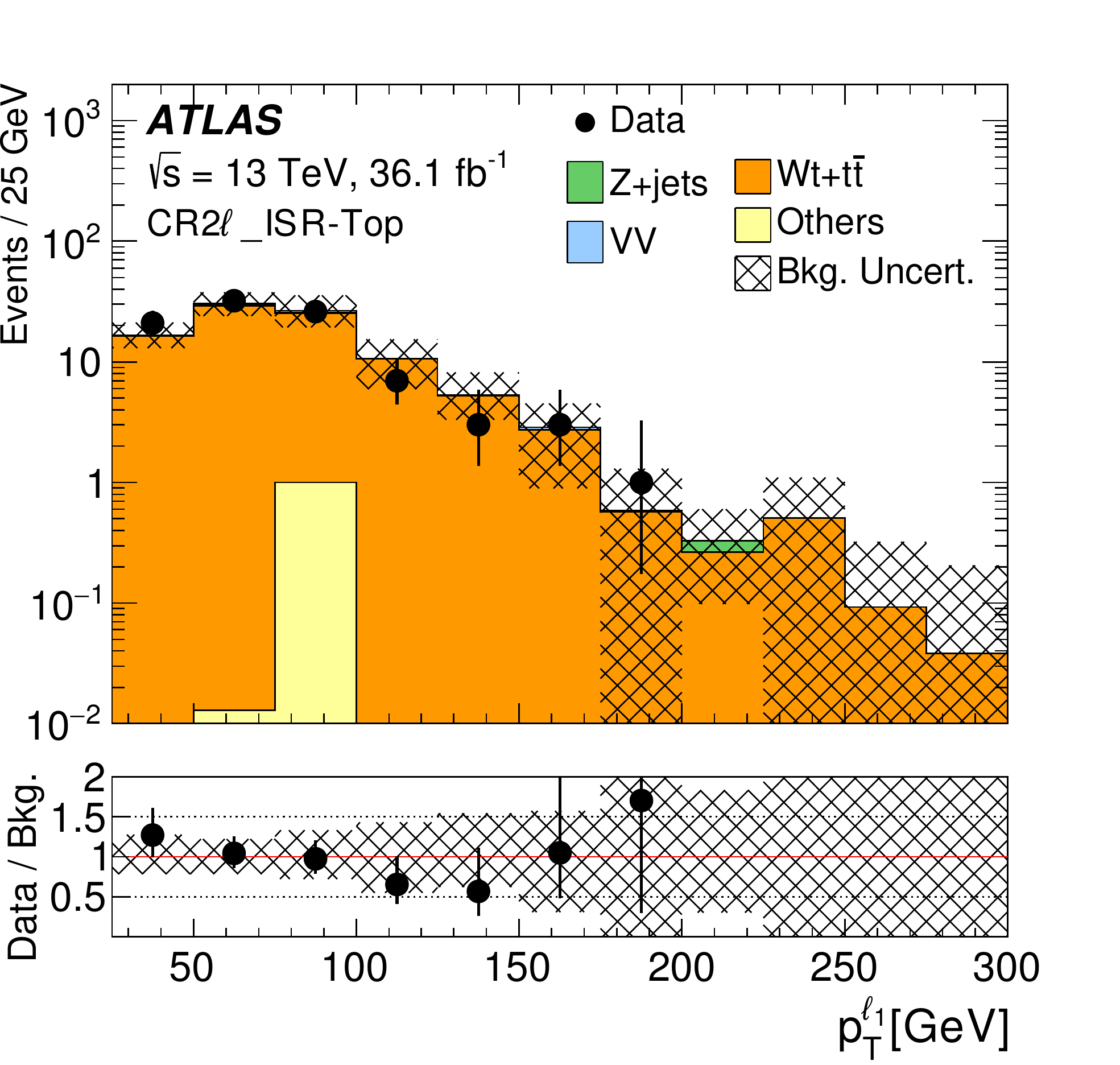}} \\
\vspace*{-0.005\textheight}\subfigure[]{\includegraphics[width=0.42\textwidth]{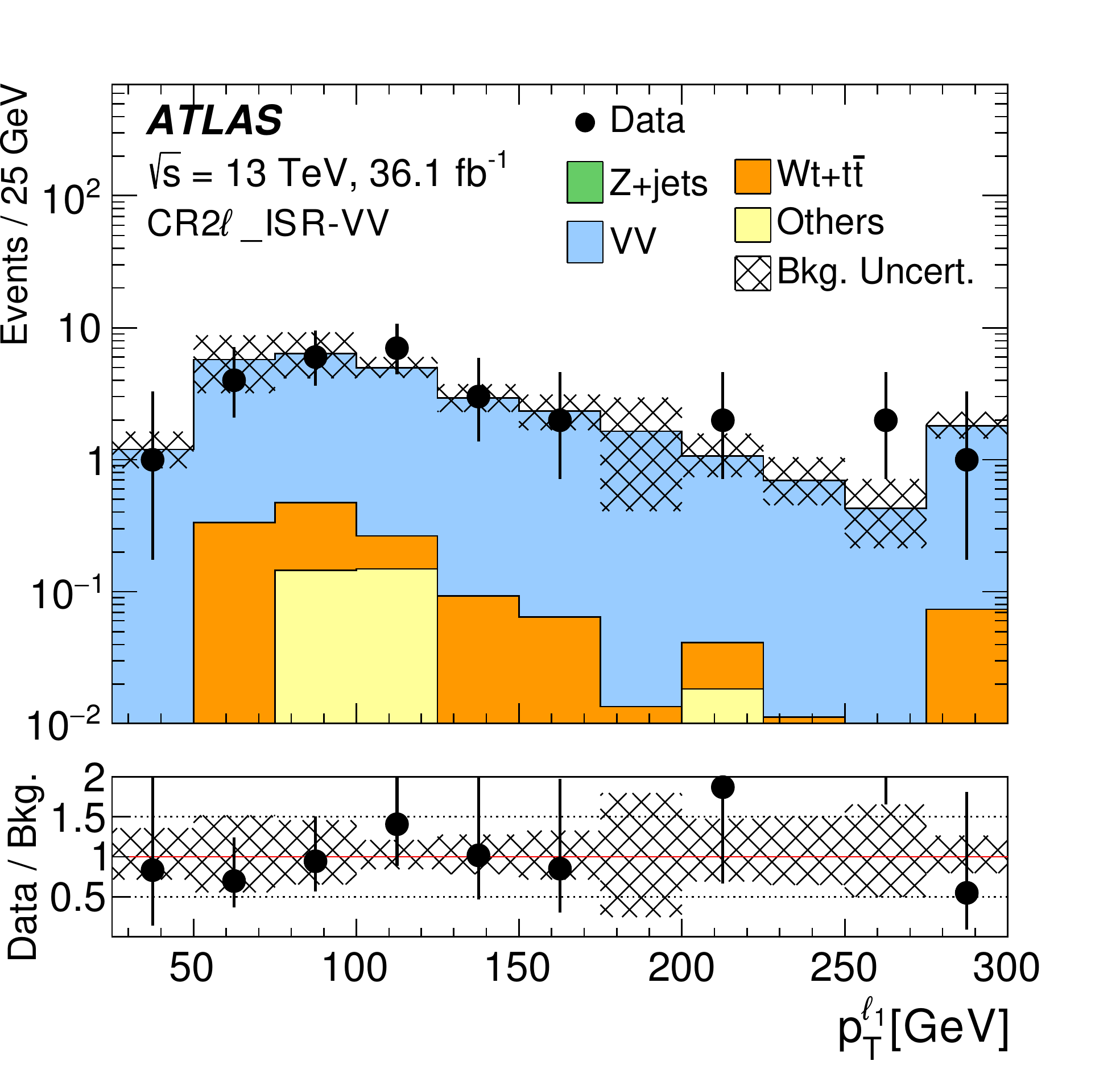}} 
\vspace*{-0.005\textheight}\subfigure[]{\includegraphics[width=0.42\textwidth]{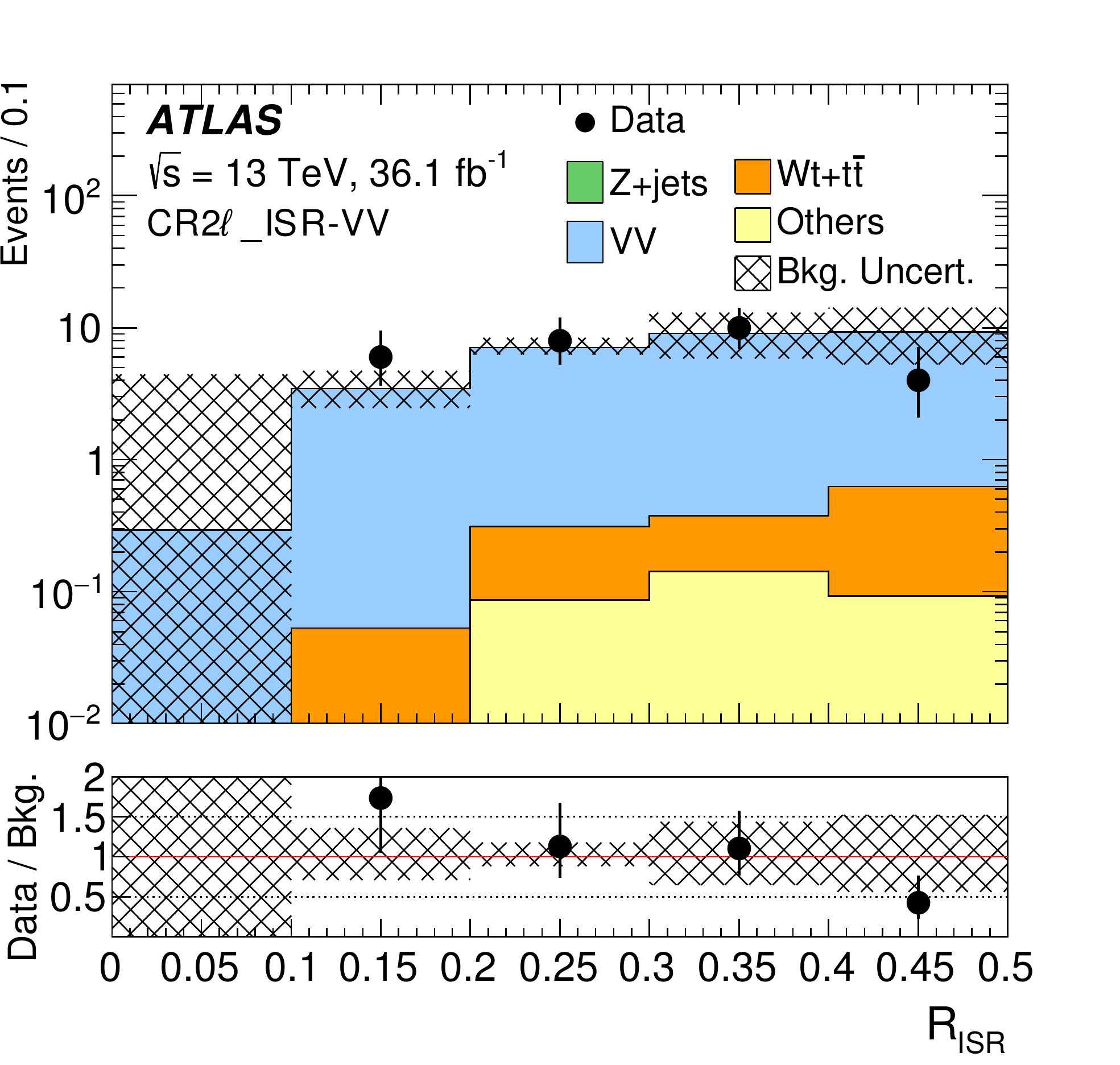}} 
\end{center}
\vspace*{-0.03\textheight}\caption{\label{fig:crRJR2L}
Distributions of kinematic variables in the control regions for the 2$\ell$ channel after applying all selection requirements in Tables~\ref{tab:StandardRJSelections} or~\ref{tab:2L2J_compressedCRs}. 
The histograms show the post-fit MC background predictions. The last bin includes the overflow. The FNP contribution is estimated from a data-driven technique and is included in the category ``Others''.
Distributions for the (a) $H_{4,1}^{\textrm{PP}}$ standard-decay-tree top CR, (b) $p_{\textrm{T}}^{\ell_{1}}$ and (c) $H_{4,1}^{\textrm{PP}}$ for the standard decay tree VV CR, (d) $p_{\textrm{T}}^{\ell_{1}}$ compressed-decay-tree top CR, and (e) $p_{\textrm{T}}^{\ell_{1}}$ compressed-decay-tree $VV$ CR and (f) $R_{\textrm{ISR}}$ compressed-decay-tree $VV$ CR are plotted. The hatched error bands indicate the combined theoretical, experimental and MC statistical uncertainties. 
}
\end{figure}

\begin{figure}[!tbp]
\begin{center} 
\vspace*{-0.03\textheight}
\subfigure[]{\includegraphics[width=0.42\textwidth]{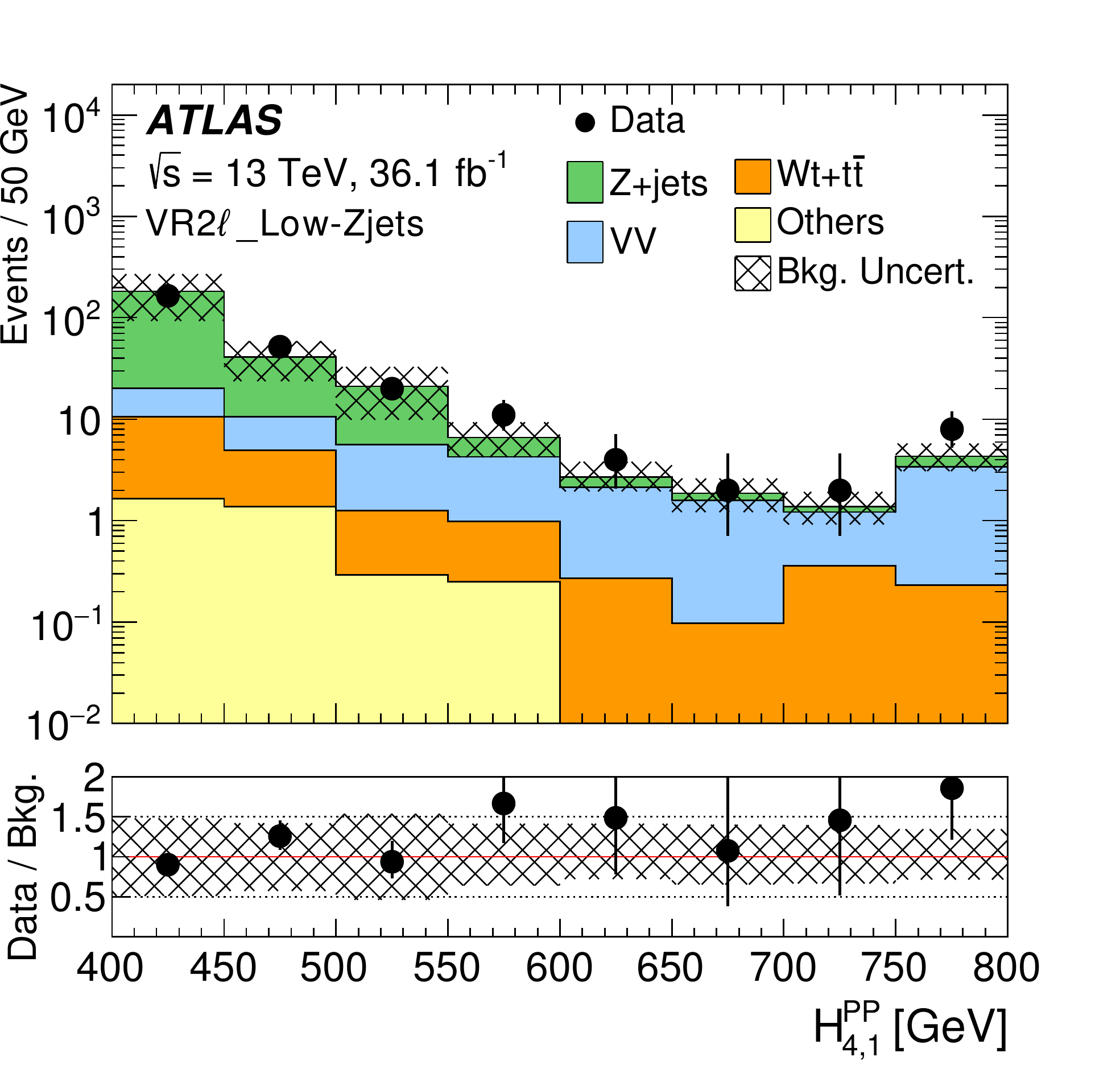}} 
\subfigure[]{\includegraphics[width=0.42\textwidth]{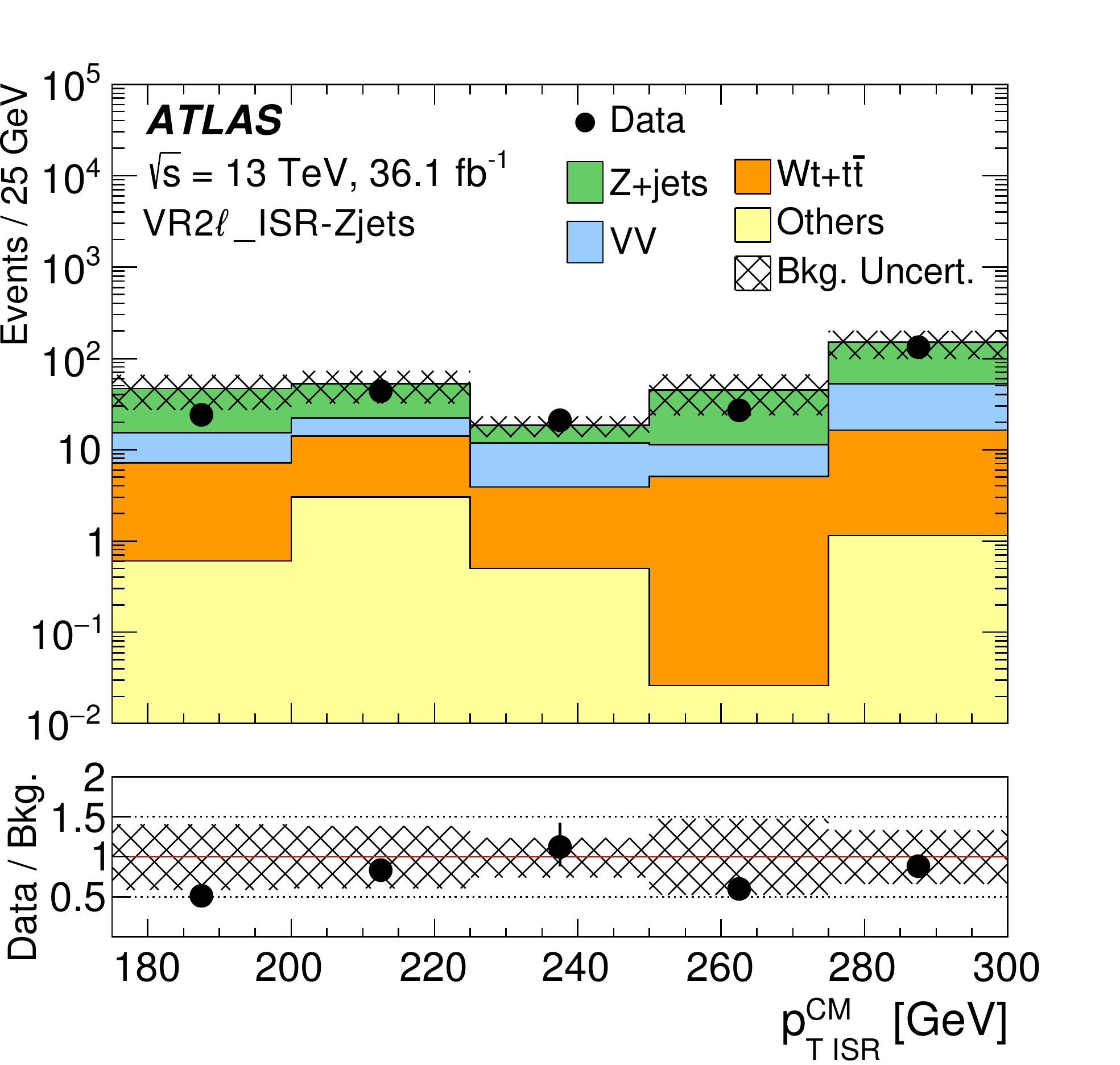}}\\
\subfigure[]{\includegraphics[width=0.42\textwidth]{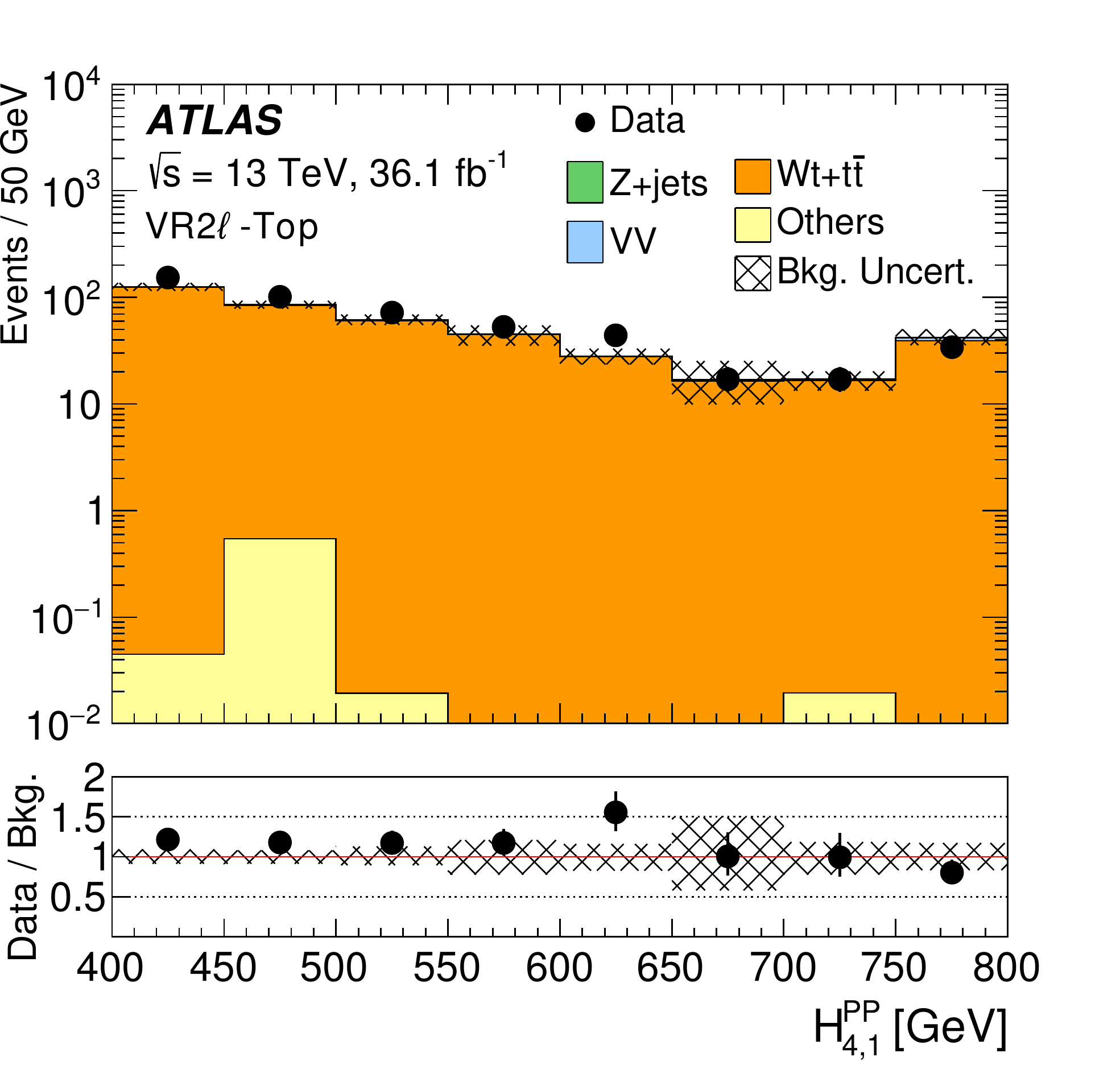}}  
\vspace*{-0.005\textheight}\subfigure[]{\includegraphics[width=0.42\textwidth]{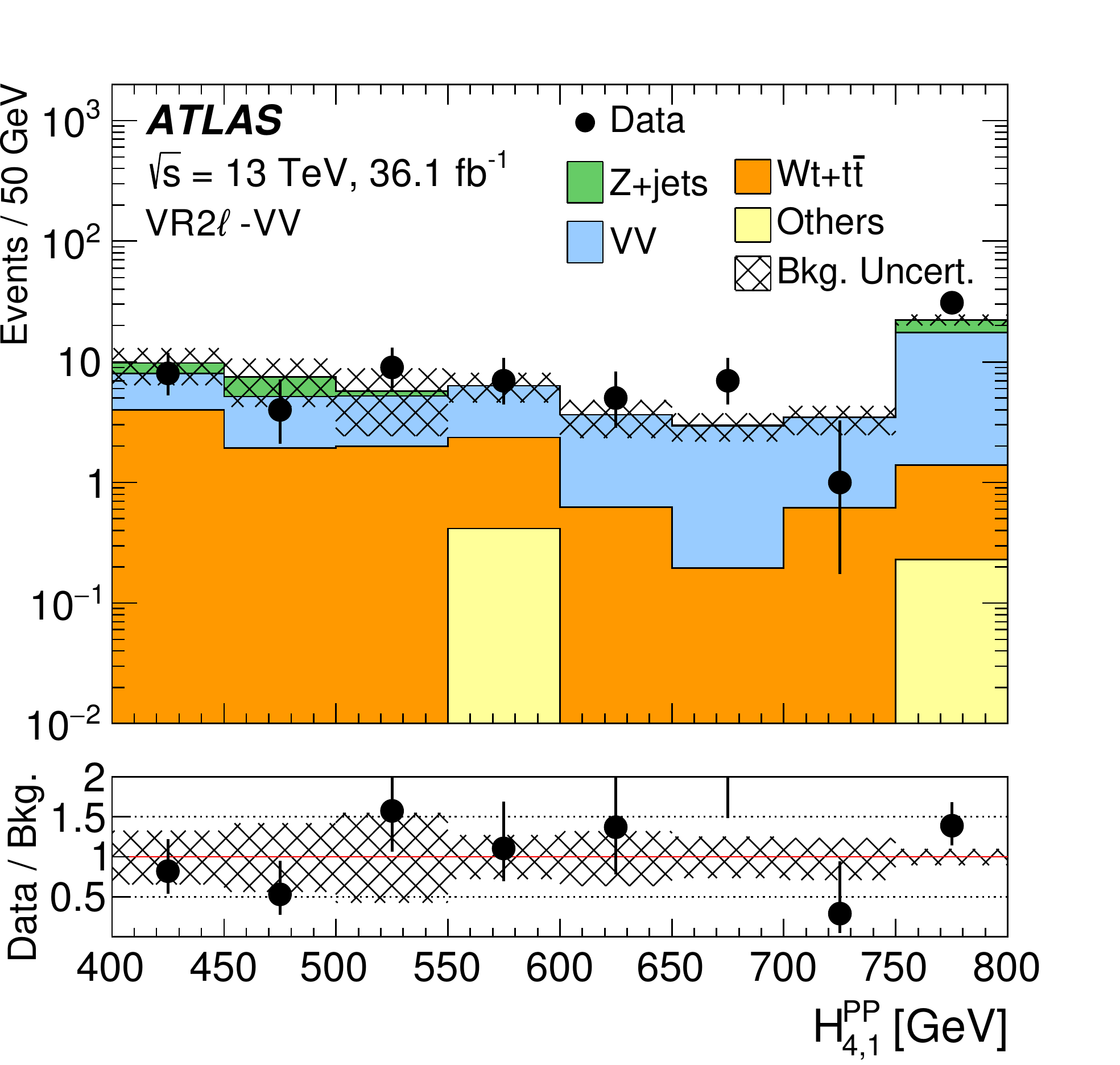}} \\  
\vspace*{-0.005\textheight}\subfigure[]{\includegraphics[width=0.42\textwidth]{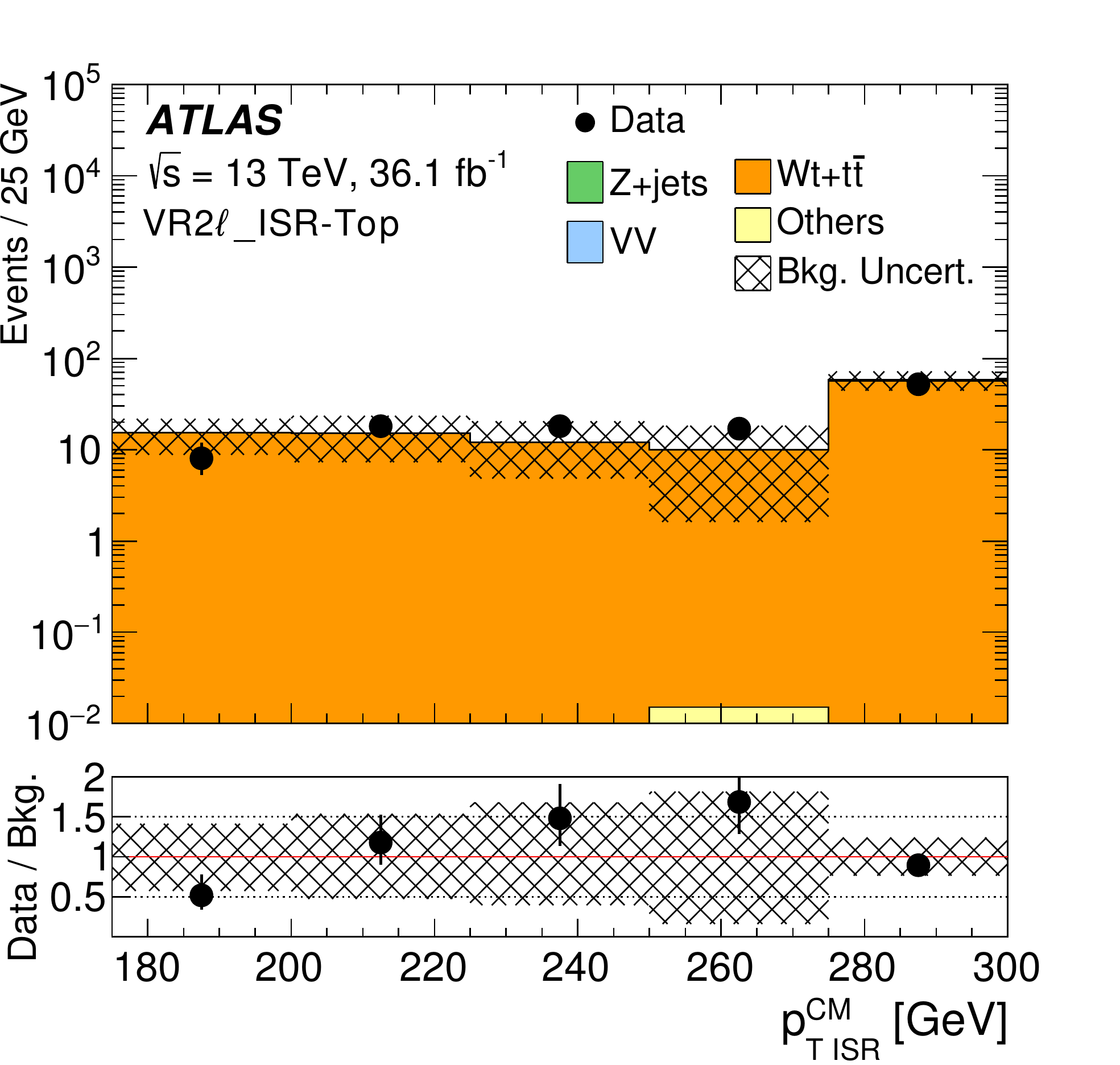}}
\vspace*{-0.005\textheight}\subfigure[]{\includegraphics[width=0.42\textwidth]{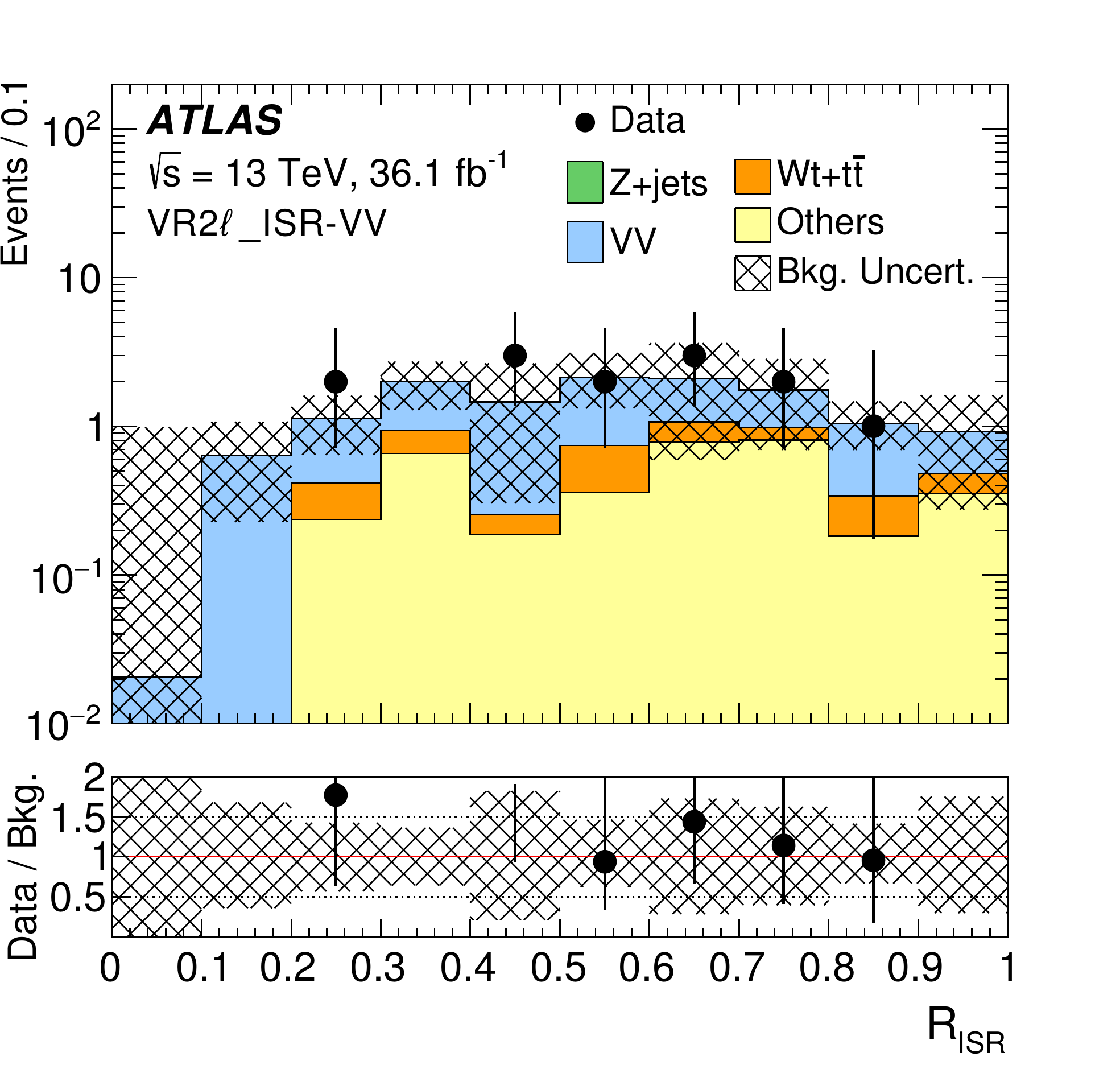}} 
\end{center}
\vspace*{-0.03\textheight}\caption{\label{fig:vrRJR2L}
Distributions of kinematic variables in the validation regions for the 2$\ell$ channel after applying all selection requirements in Tables~\ref{tab:StandardRJSelections} or ~\ref{tab:2L2J_compressedCRs}. 
The histograms show the post-fit MC background predictions. The last bin includes the overflow. The FNP contribution is estimated from a data-driven technique and is included in the category ``Others''.
Plots show (a) $H_{4,1}^{\textrm{PP}}$ and (b) $p_{\textrm{T\ ISR}}^{\textrm{CM}}$ in the $Z$+jets VRs for the standard and compressed decay trees respectively; (c) $H_{4,1}^{\textrm{PP}}$ in the top VR and (d) $H_{4,1}^{\textrm{PP}}$ in the diboson VR for the standard 
decay tree; (e) $p_{\textrm{T\ ISR}}^{\textrm{CM}}$ in the top VR and (f) $R_{\textrm{ISR}}$ in the diboson VR for the compressed decay tree. 
The hatched error bands indicate the combined theoretical, experimental and MC statistical uncertainties.
}
\end{figure}

\subsection{Event selection in the three-lepton channel}
\label{sec:eventSel3L}
The strategy followed for the design of the $3\ell$ search channel has many similarities with the $2\ell$ channel. Three SRs are defined with the standard decay tree (SR$3\ell$\_High, SR$3\ell$\_Int, SR$3\ell$\_Low) and the diboson background contribution is controlled and validated in a dedicated CR (CR$3\ell$-VV) and VR (VR$3\ell$-VV), which contain mutually exclusive events with respect to the SRs. The initial selection of events proceeds with preselection requirements summarized in Table~\ref{tab:3LPreselectionStandard}. All regions require exactly three energetic leptons with the transverse momentum of the third leading lepton in $\pt$, $p_{\textrm{T}}^{\ell_{3}}$, required to be at least 30~$\GeV$. The regions are additionally required to have low jet activity. A same-flavor opposite-charge lepton pair is required, formed by finding the pair with invariant mass closest to the $Z$-boson mass, while the remaining (unpaired) lepton is used to construct $m_{\textrm{T}}^{W}$. SR$3\ell$\_Low has a jet veto which makes it orthogonal to the ISR SR (SR$3\ell$\_ISR) that is described below.

The selection requirements defining the SRs, CR and VR can be seen in Table~\ref{tab:3L_standardCRs}. For signals targeting larger masses, and hence mass-splittings between the parent and LSP (`high' and `intermediate' regions), the selection criteria imposed on scale quantities are tighter, with looser requirements applied to ratio values. The opposite is true as the mass-splitting becomes smaller, where the selection criteria imposed on scale quantities are less stringent, since the produced objects are not expected to be too energetic; better sensitivity is obtained by applying selection criteria to ratios of quantities. Orthogonality between the CR, VR and SRs is achieved by inverting the requirement on $m_{\textrm{T}}^{W}$ and using different transverse-mass windows. 

SR$3\ell$\_Low requires no jet activity, so an orthogonal 3$\ell$ ISR region is defined when there are jets in the event. As with all uses of the compressed decay tree, at least one jet must be identified in the event, to populate the ISR system. For the SR$3\ell$\_ISR region all jets are associated with the ISR system. The highly energetic ISR system that accompanies the leptons reduces the contributions from fake or non-prompt (FNP) leptons and allows the relaxation of lepton $\pt$ thresholds. The exact preselection requirements applied in the ISR regions are shown in Table~\ref{tab:3LPreselectionCompressed}.

The lepton pair formation follows the same prescription used for the regions constructed with the standard decay tree. The selection criteria applied to the events after preselection are given in Table~\ref{tab:3L_compressedCRs}. The diboson CR (CR$3\ell$\_ISR-VV) is defined with an inverted $m_{\textrm{T}}^{W}$ requirement while the corresponding VR (VR$3\ell$\_ISR-VV) is defined with a relaxed requirement on $m_{\textrm{T}}^{W}$ and has the $p^{\textrm{CM}}_{\textrm{T}}$ requirement inverted.  

Post-fit distributions of variables from the 3$\ell$ search for selected regions, are shown in Figures~\ref{fig:crRJR3L} and~\ref{fig:vrRJR3L} for data and the different MC samples. The background component labeled ``Others'' refers to the processes with a Higgs boson, $t\bar{t}V$ and the non-prompt and non-isolated leptons. 

\begin{table}[!tp]
  \centering
  \scriptsize
\caption{Preselection criteria for the $3\ell$ CR, VR and SR with the standard decay tree. The variables are defined in the text.}
\begin{tabular}{l r r r r r r}
\toprule
Region & $n_{\textrm{leptons}}$ & $n_{\textrm{jets}}$ & $n_{b\textrm{-tag}}$ & $\pt^{\ell_{1}}$ [\GeV] & $\pt^{\ell_{2}}$ [\GeV] & $\pt^{\ell_{3}}$ [\GeV] \\
\hline
\hline
CR$3\ell$-VV    & $=3$ & $<3$ & $=0$ &$>60$ & $>40$ & $>30$\\
VR$3\ell$-VV    & $=3$ & $<3$ & $=0$ & $>60$ & $>40$ & $>30$\\
\hline
SR$3\ell$\_High & $=3$ & $<3$ & $=0$ &$>60$ & $>60$ & $>40$ \\
SR$3\ell$\_Int  & $=3$ & $<3$ & $=0$ &$>60$ & $>50$ & $>30$ \\
SR$3\ell$\_Low  & $=3$ & $=0$ & $=0$ &$>60$ & $>40$ & $>30$  \\
\toprule
\end{tabular}
\label{tab:3LPreselectionStandard}
\end{table}

\begin{table}[!tbp]
  \centering
    \scriptsize
\caption{Selection criteria for the $3\ell$ CR, VR and SR with the standard decay tree. The variables are defined in the text.}
\begin{tabular}{l r r r r r r}
\toprule
Region & $m_{\ell\ell}$ [\GeV] &  $m^{W}_{\textrm{T}}$ [\GeV] & $H^{\textrm{PP}}_{3,1}$ [\GeV]  & $\frac{p^{\textrm{lab}}_{\textrm{T}~\textrm{PP}}}{p^{\textrm{lab}}_{\textrm{T}~\textrm{PP}}+H^{\textrm{PP}}_{\textrm{T}\ 3,1}}$  & $\frac{H^{\textrm{PP}}_{\textrm{T}\ 3,1}}{H^{\textrm{PP}}_{3,1}}$ & $\frac{H^{\textrm{P}_{\textrm{b}}}_{1,1}}{H^{\textrm{P}_{\textrm{b}}}_{2,1}}$ \\
\hline
\hline
CR$3\ell$-VV    &$\in(75,105)$ &  $\in(0,70)$   & $>250$  & $<0.2$ & $>0.75$  & -- \\
VR$3\ell$-VV    &$\in(75,105)$ &  $\in(70,100)$ & $>250$  & $<0.2$ & $>0.75$  & -- \\
\hline
SR$3\ell$\_High &$\in(75,105)$ & $>150$         & $>550$  & $<0.2$  & $>0.75$ & $>0.8$ \\
SR$3\ell$\_Int  &$\in(75,105)$ & $>130$         & $>450$  & $<0.15$ & $>0.8$  & $>0.75$ \\
SR$3\ell$\_Low  &$\in(75,105)$ & $>100$         & $>250$  & $<0.05$ & $>0.9$  & -- \\
\toprule
\end{tabular}
\label{tab:3L_standardCRs}
\end{table}

\begin{table}[!tbp]
  \centering
    \scriptsize
\caption{Preselection criteria for the $3\ell$ CR, VR and SR with the compressed decay tree. The variables are defined in the text.}
\begin{tabular}{l r r r r r r}
\toprule
Region & $n_{\textrm{leptons}}$ & $n_{\textrm{jets}}$ & $n_{b\textrm{-tag}}$ & $\pt^{\ell_{1}}$ [\GeV] & $\pt^{\ell_{2}}$ [\GeV] & $\pt^{\ell_{3}}$ [\GeV] \\
\hline
\hline
CR$3\ell$\_ISR-VV    & $=3$ & $\geq 1$ & $=0$ &$>25$ & $>25$ & $>20$\\
VR$3\ell$\_ISR-VV    & $=3$ & $\geq 1$ & $=0$ &$>25$ & $>25$ & $>20$\\
\hline
    SR$3\ell$\_ISR  & $=3$ & $\in[1,3]$ & $=0$ &$>25$ & $>25$ & $>20$  \\
\toprule
\end{tabular}
\label{tab:3LPreselectionCompressed}
\end{table}

\begin{table}[!tbp]
  \centering
    \scriptsize
\caption{Selection criteria for the $3\ell$ CR, VR and SR with the compressed decay tree. The variables are defined in the text.}
\begin{tabular}{l r r r r r r r}
\toprule
Region     & $m_{\ell\ell}$ [\GeV] & $m^{W}_{\textrm{T}}$ [\GeV] & $\Delta\phi_{\textrm{ISR,I}}^{\textrm{CM}}$ & $R_{\textrm{ISR}}$    & $p^{\textrm{CM}}_{\textrm{T\ ISR}}$ [\GeV] & $p^{\textrm{CM}}_{\textrm{T\ I}}$ [\GeV] & $p^{\textrm{CM}}_{\textrm{T}}$ [\GeV]   \\ 
\hline
\hline
CR$3\ell$\_ISR-VV & $\in(75,105)$ & $<100$ & $>2.0$ & $\in(0.55,1.0)$ &  $>80$  & $>60$  & $<25$ \\
VR$3\ell$\_ISR-VV & $\in(75,105)$ & $>60$  & $>2.0$ & $\in(0.55,1.0)$ & $>80$   & $>60$  & $>25$ \\
\hline
SR$3\ell$\_ISR    & $\in(75,105)$ & $>100$ &$>2.0$  & $\in(0.55,1.0)$ &  $>100$ & $>80$  & $<25$ \\
\toprule
\end{tabular}
\label{tab:3L_compressedCRs}
\end{table}

\begin{figure}[htbp]
\begin{center} 
\vspace*{-0.00\textheight}
\subfigure[]{\includegraphics[width=0.42\textwidth]{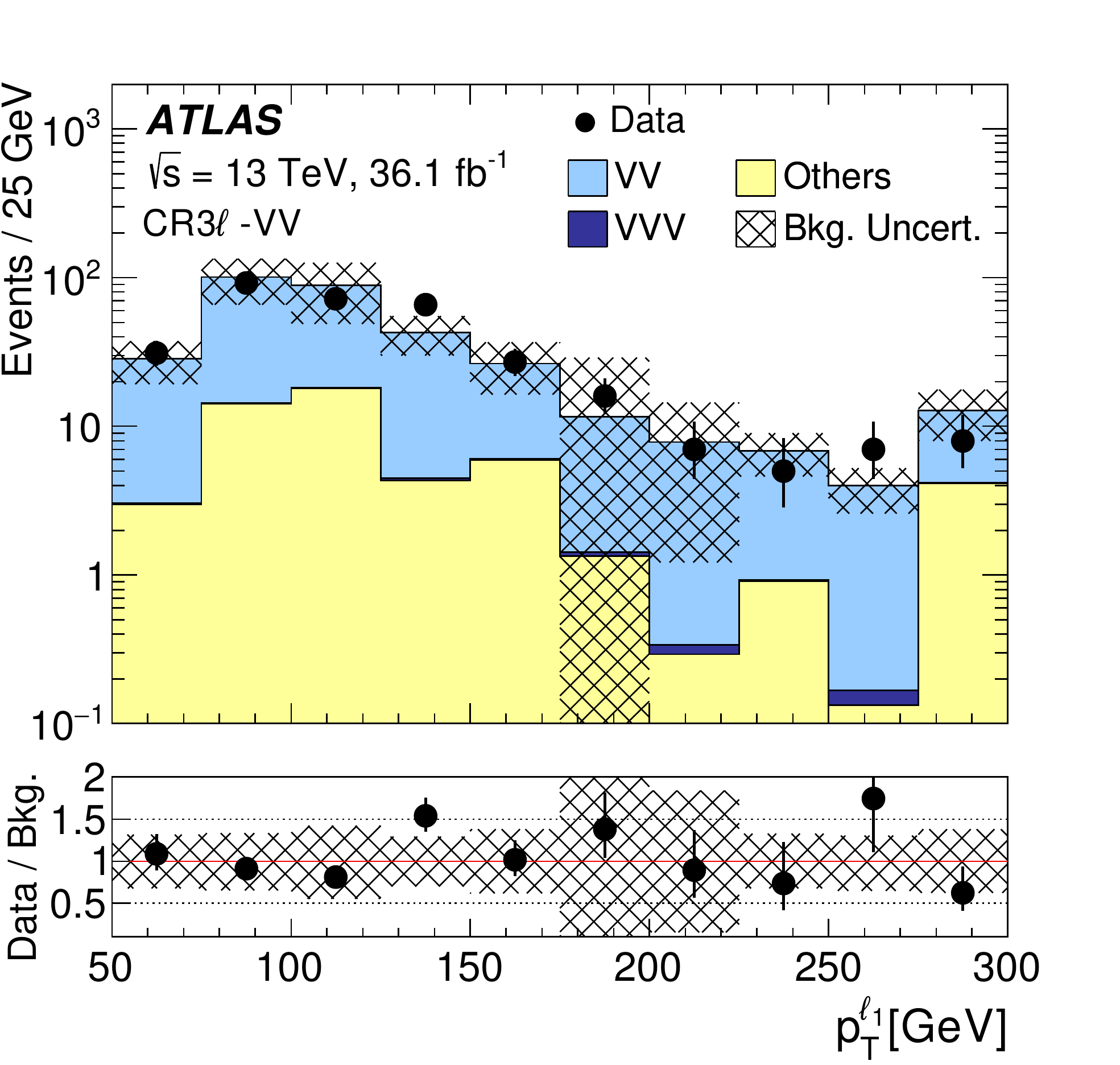}} 
\subfigure[]{\includegraphics[width=0.42\textwidth]{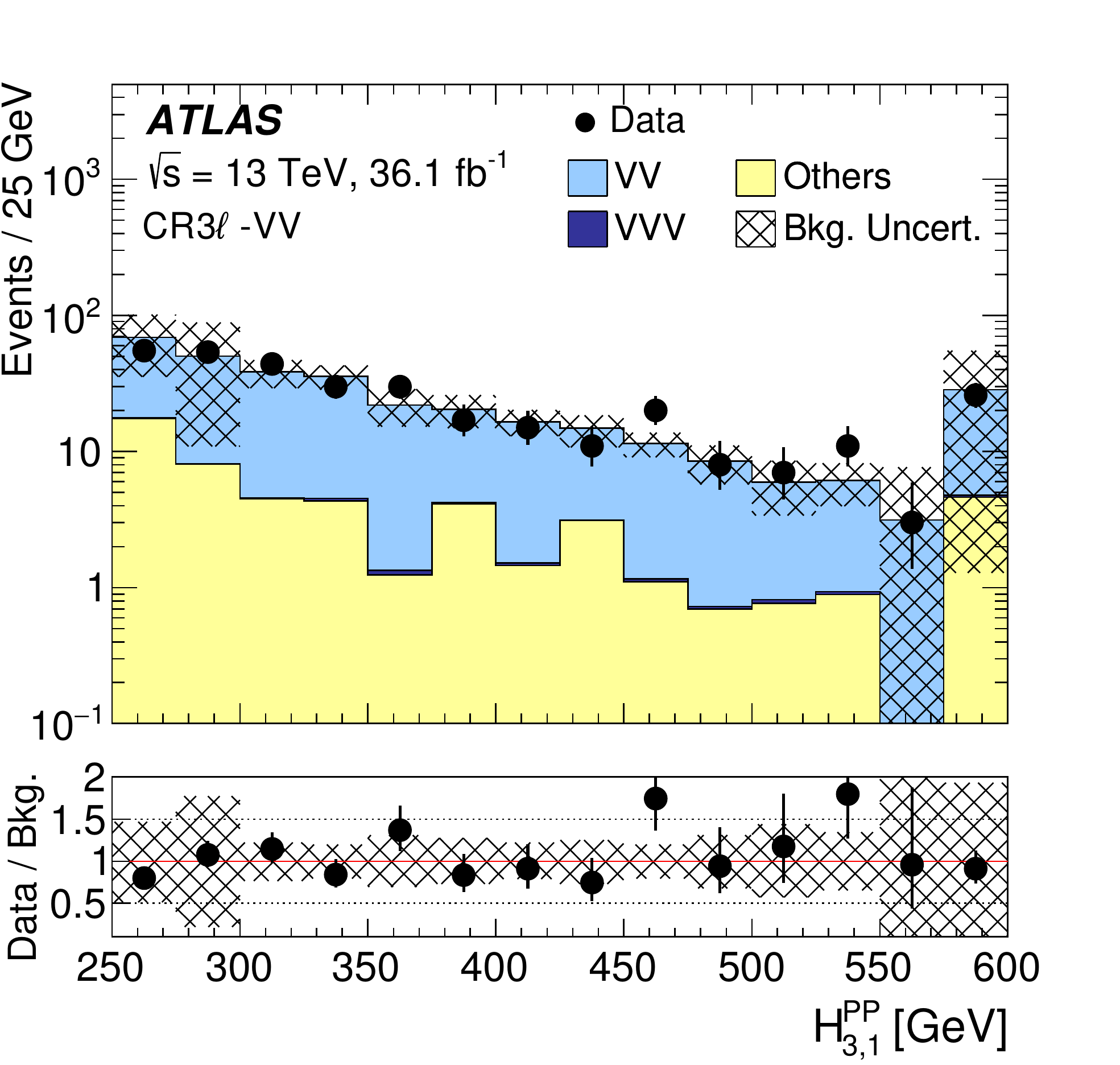}} \\ 
\vspace*{-0.005\textheight}\subfigure[]{\includegraphics[width=0.42\textwidth]{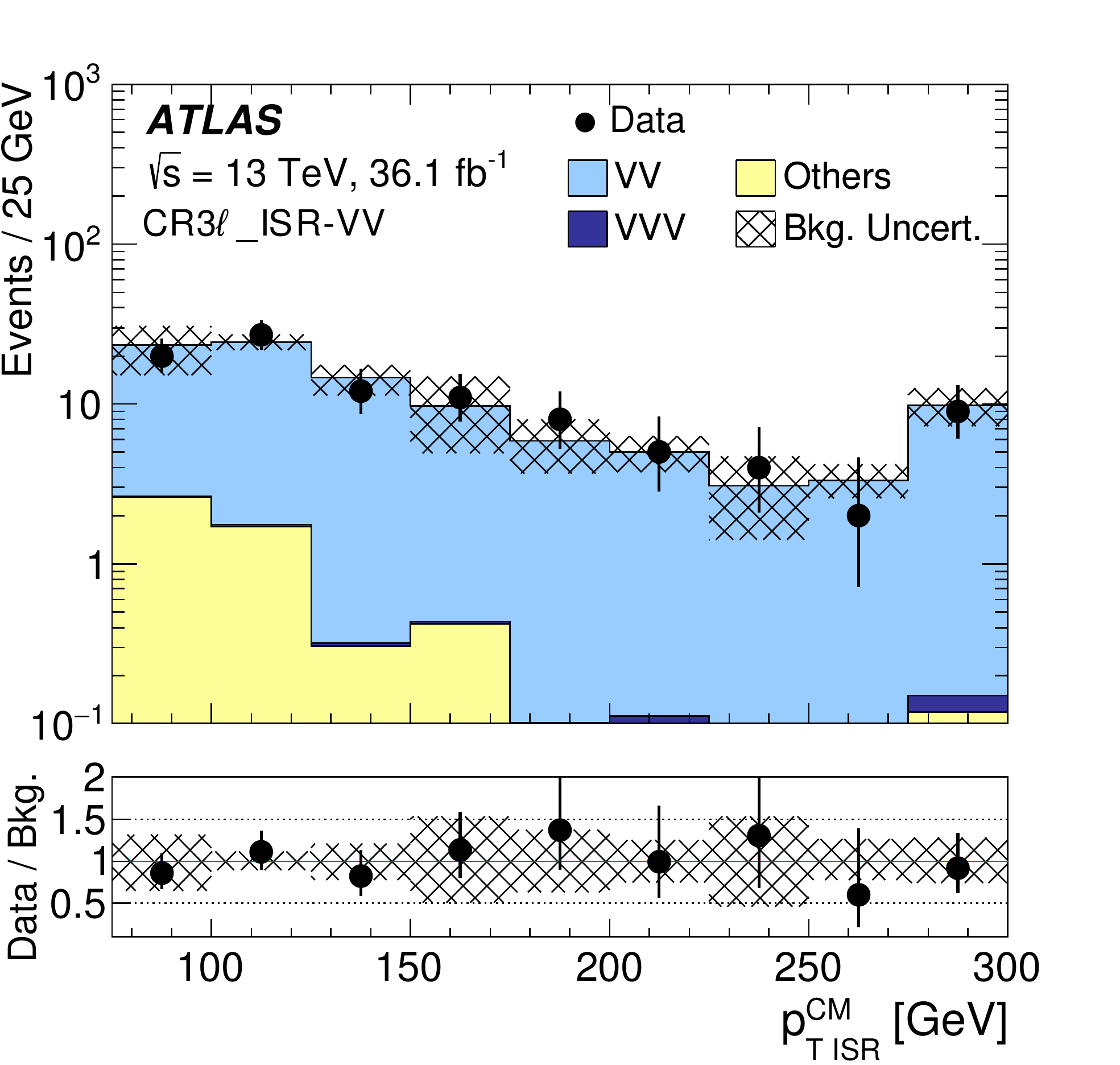}} 
\vspace*{-0.005\textheight}\subfigure[]{\includegraphics[width=0.42\textwidth]{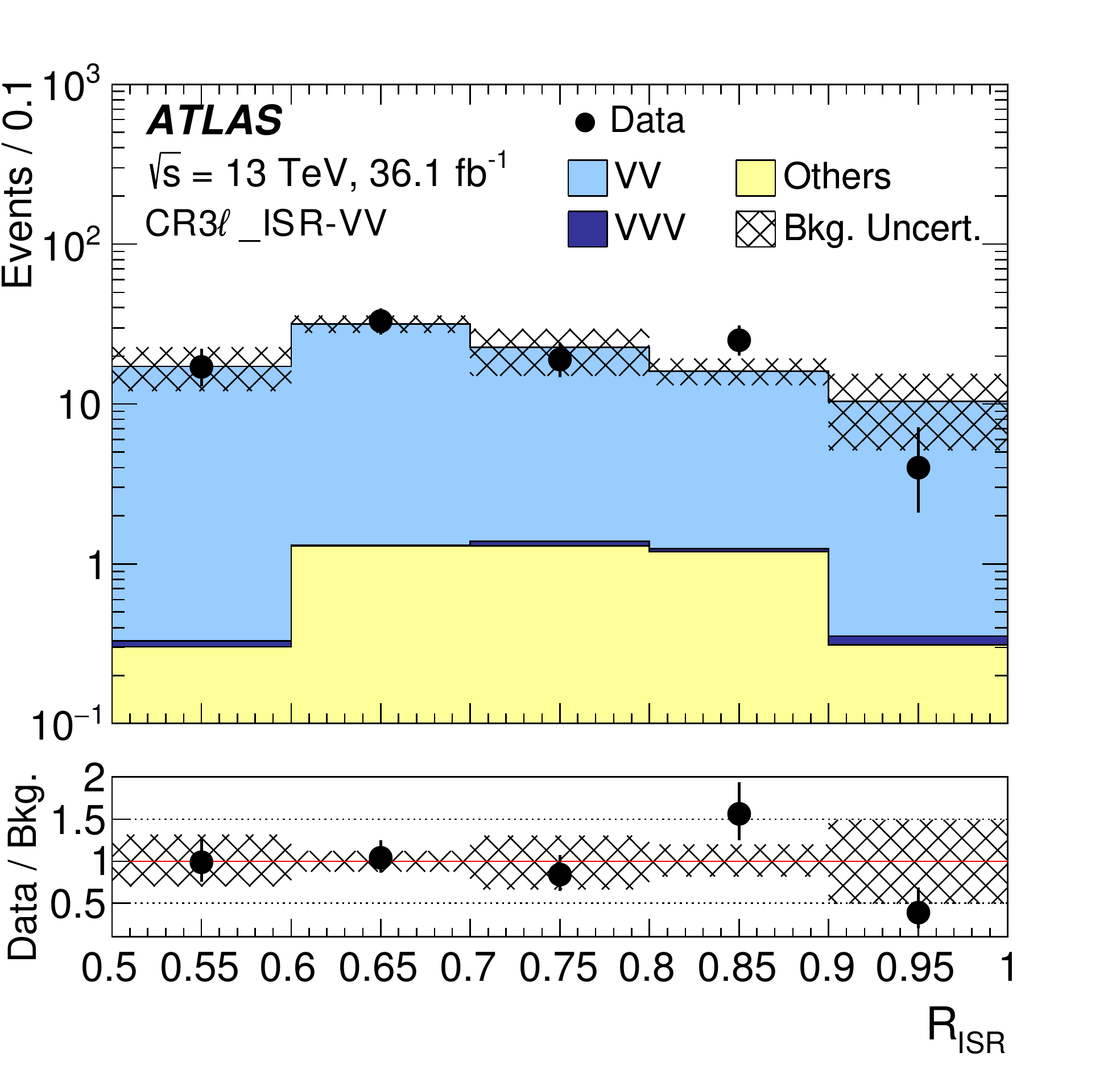}}  
\end{center}
\vspace*{-0.03\textheight}\caption{\label{fig:crRJR3L}
  Distributions of kinematic variables in the control regions for the 3$\ell$ channel after applying all selection criteria described in Tables~\ref{tab:3L_standardCRs} or~\ref{tab:3L_compressedCRs}. The histograms show the post-fit MC background predictions. The FNP contribution is estimated from a data-driven technique and is included in the category ``Others''. The last bin includes the overflow. Plots show (a) $p_{\textrm{T}}^{\ell_{1}}$ and (b) $H_{3,1}^{\textrm{PP}}$ for the diboson CR in the standard decay tree, (c) $p_{\textrm{T\ ISR}}^{\textrm{CM}}$  and (d) $R_{\textrm{ISR}}$ for the diboson CR in the compressed decay tree.  
The hatched error bands indicate the combined theoretical, experimental and MC statistical uncertainties.
}
\end{figure}

\begin{figure}[htbp]
\begin{center} 
\vspace*{-0.03\textheight}
\subfigure[]{\includegraphics[width=0.42\textwidth]{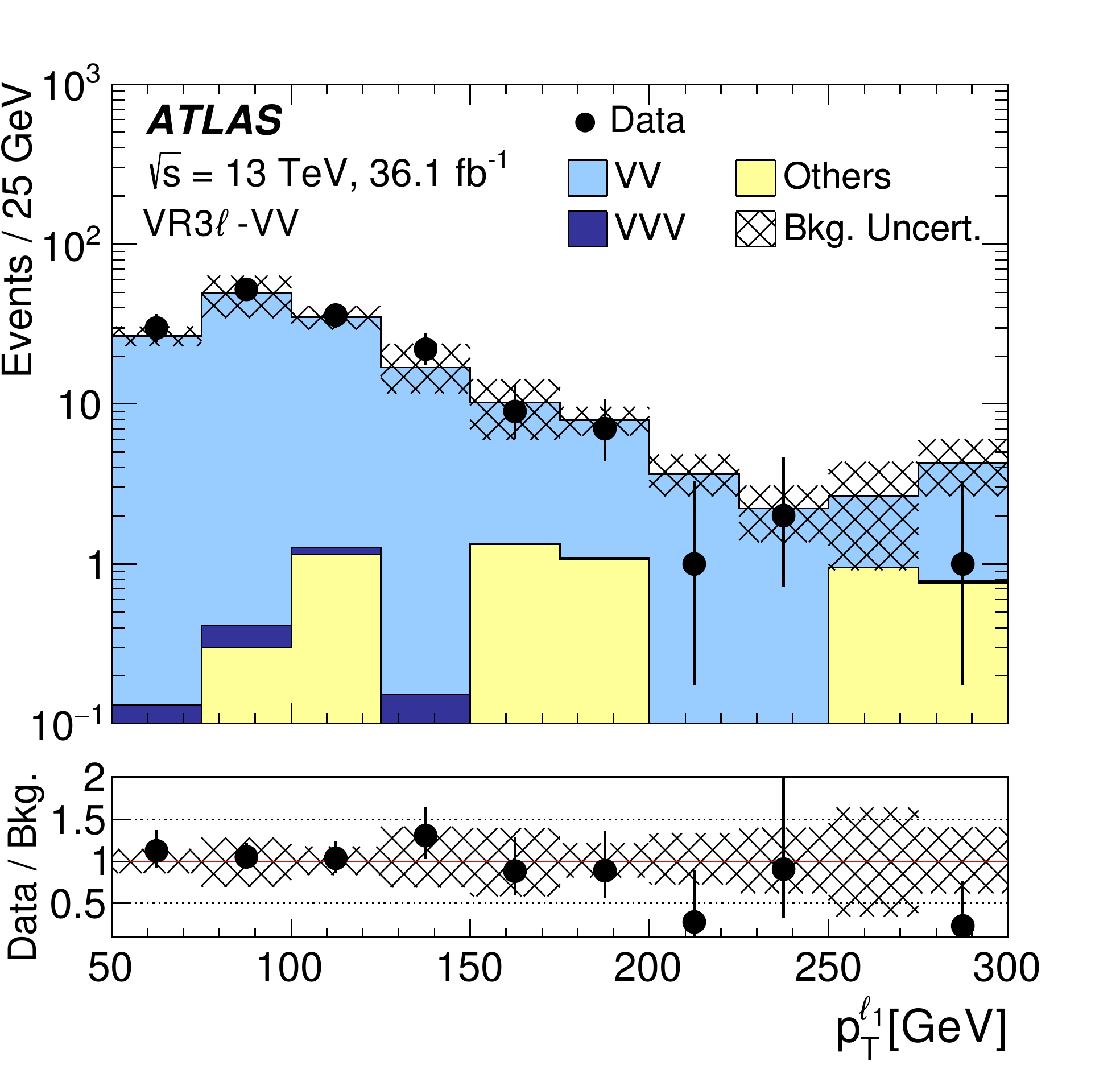}} 
\subfigure[]{\includegraphics[width=0.42\textwidth]{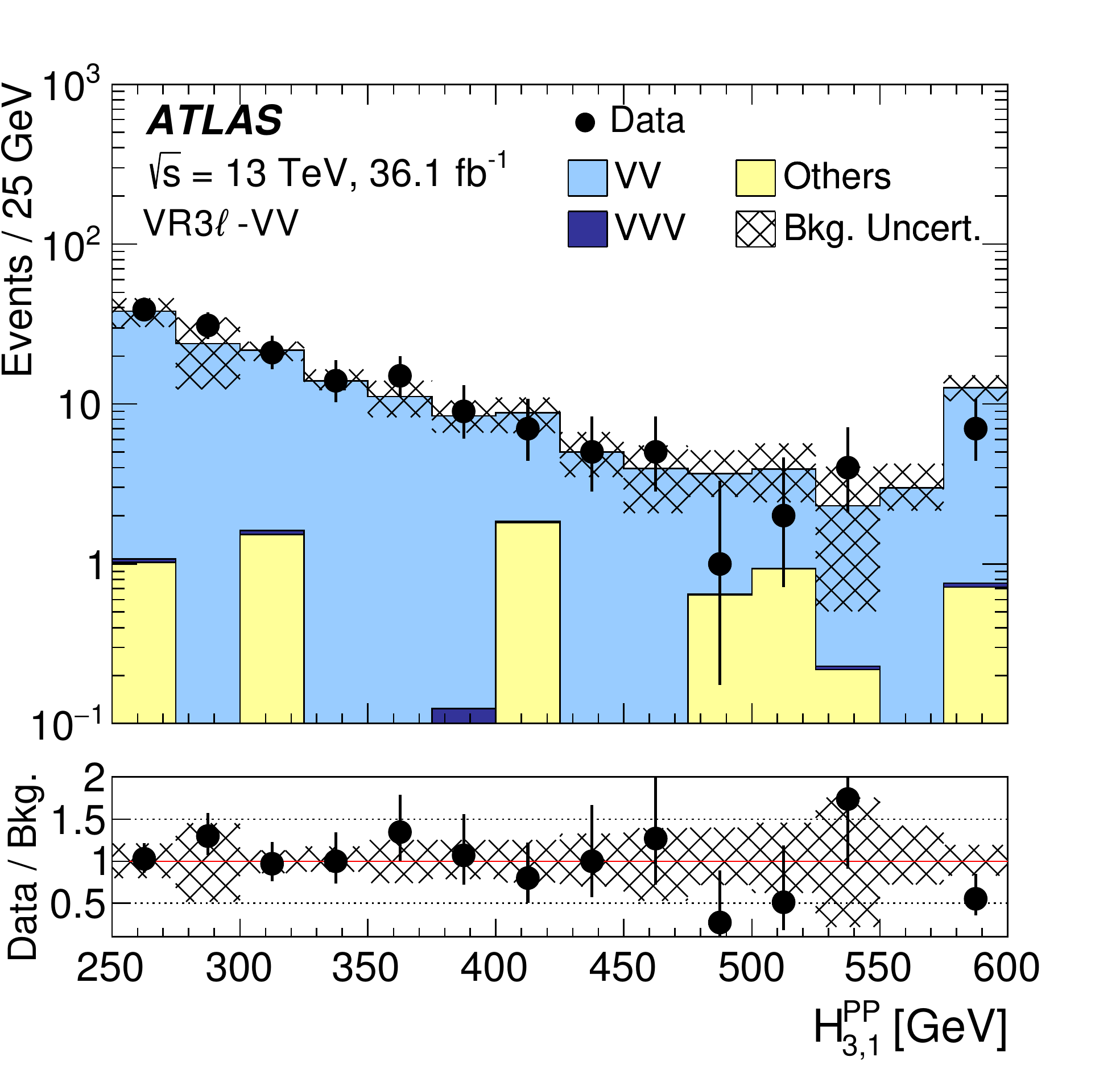}} \\ 
\vspace*{-0.005\textheight}\subfigure[]{\includegraphics[width=0.42\textwidth]{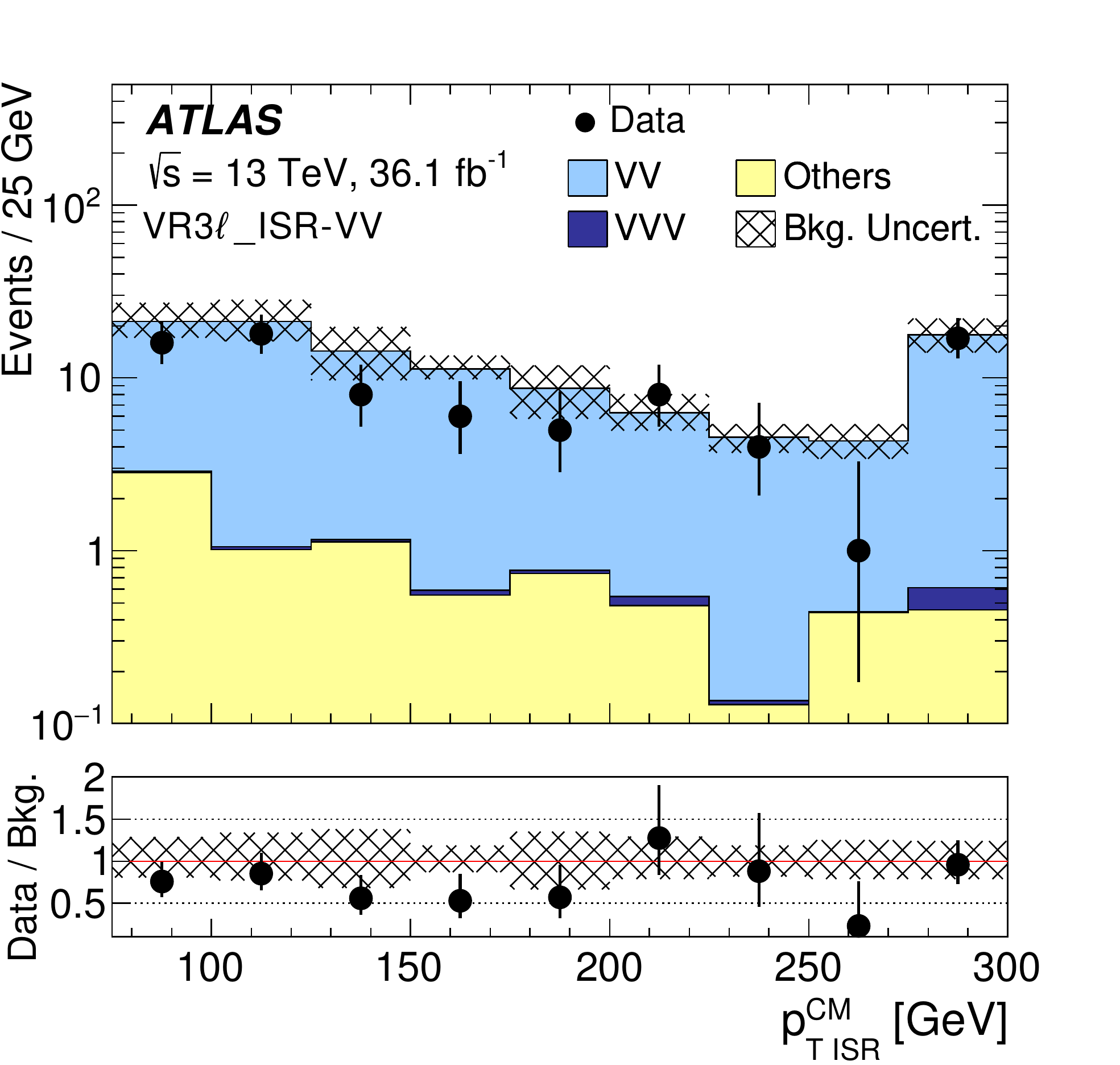}} 
\vspace*{-0.005\textheight}\subfigure[]{\includegraphics[width=0.42\textwidth]{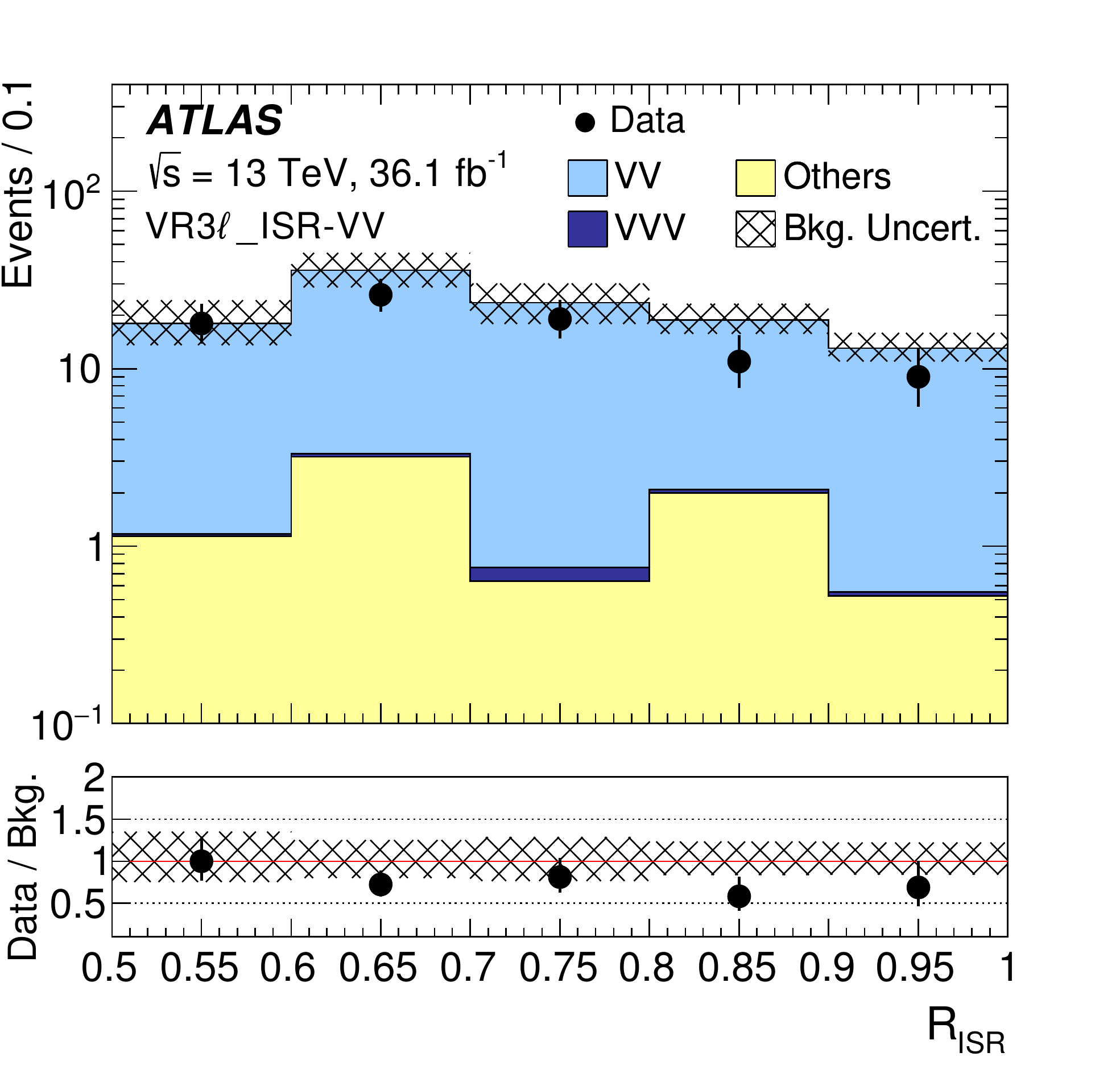}} 
\end{center}
\vspace*{-0.03\textheight}\caption{\label{fig:vrRJR3L}
Distributions of kinematic variables in the validation regions for the 3$\ell$ channel after applying all selection criteria in Tables~\ref{tab:3L_standardCRs} or~\ref{tab:3L_compressedCRs}. The histograms show the post-fit MC background predictions. The FNP contribution is estimated from a data-driven technique and is included in the category ``Others''. The last bin includes the overflow. Plots show (a) $p_{\textrm{T}}^{\ell_{1}}$ and (b) $H_{3,1}^{\textrm{PP}}$ for the standard decay tree, (c) $p_{\textrm{T\ ISR}}^{\textrm{CM}}$ and (d) $R_{\textrm{ISR}}$ for the compressed decay tree. 
The hatched error bands indicate the combined theoretical, experimental and MC statistical uncertainties.
}
\end{figure}

\section{Background estimation}
\label{sec:background}

Several SM background processes contribute to the event counts in the signal regions. The largest backgrounds arise from dibosons and $Z+$jets, with lesser contributions from top-quark pairs, single top quarks, tribosons and Higgs bosons. In general, these backgrounds can be classified into two categories, the irreducible backgrounds with prompt and isolated leptons (also referred to as real leptons) and genuine $\MET$ from neutrinos, and reducible backgrounds that contain one or more FNP lepton(s) or where experimental effects (e.g.\ detector mismeasurements of jets or leptons or imperfect removal of object double-counting) lead to significant ``fake'' $\MET$.

An FNP lepton can originate from a semileptonic decay of a $b$- or $c$- hadron, decays in flight of light hadrons, misidentification of a light-flavor jet, or photon conversions. In the $2\ell$ analysis such backgrounds originate from multijet, $W+$jets, single-top-quark and $t\bar{t}$ production events, while in the $3\ell$ analysis there are additional contributions from $Z+$jets and $WW$ and from any other physics process leading to less than three prompt and isolated leptons. In both analyses, this background is estimated using a data-driven technique, the matrix method~\cite{TOPQ-2010-01}. 

This method uses two types of lepton identification criteria: ``signal'', corresponding to high-purity leptons and ``baseline'', corresponding to the definition of Section~\ref{sec:objects}. The method makes use of the numbers of observed events containing baseline--baseline, baseline--signal, signal--baseline and signal--signal lepton pairs (ordered in $\pt$) in a given SR. In the $3\ell$ search channel the highest-$\pt$ electron or muon is taken to be real. Simulation studies show that this is a valid assumption in $> 95$\% of three-signal-lepton events. Knowing the probabilities for real and FNP leptons satisfying the baseline selection criteria to also satisfy the signal selection, the observed event counts with the different lepton selection criteria can be used to extract a data-driven estimate of the FNP background. The probabilities are calculated similarly to Ref.~\cite{Aaboud:2018jiw}.

\subsection{Background estimate in the two-lepton channel}
\label{sec:background2L}

The $Z$+jets process can provide a large background, particularly in the low-mass and compressed SRs, due to fake \MET\ from jet or lepton mismeasurements or from neutrinos in semileptonic decays of $b$- or $c$-hadrons. These effects are difficult to model in simulation, so instead $\gamma$+jets events in data are used to extract the \MET\ shape in $Z$+jets events. Similar methods were employed in searches for SUSY in events with two leptons, jets, and large \MET\ in ATLAS~\cite{SUSY-2016-05} and CMS~\cite{CMS-SUS-11-021,CMS-SUS-14-014}. The \MET shape is extracted from a data control sample of $\gamma$+jets events, which have a topology similar to $Z$+jets events, recorded using a set of single-photon triggers. The events selected with prescaled triggers correspond to photon~$\pt<140$~\GeV~and these events are weighted with the corresponding trigger prescale factor. Corrections for the different $\gamma$ versus $Z$-boson $\pt$ distributions and different momentum resolutions for electrons, muons, and photons are applied. Backgrounds from $W\gamma$ and $Z\gamma$ production, which contain a photon and genuine \MET\ from neutrinos, are subtracted using MC simulation that is normalized to data in a $V\gamma$ control region containing a selected lepton and photon. The $V\gamma$ normalization factor is found to be equal to 0.79$\pm$0.79.

To model quantities that depend on the individual lepton momenta, a \mll\ value is assigned to each $\gamma$+jets event by sampling from \mll\ distributions (parameterized as a function of boson $\pt$ and the component of \MET\ that is parallel to the boson $\pt$) extracted from $Z$+jets simulation. Each $\gamma$+jets event is boosted to the rest frame of the emulated $Z$ boson and the photon is split into two pseudo-leptons, assuming isotropic decays in the rest frame. In all the two-lepton SRs (except for SR2$\ell\_$Low) the $Z+$jets background is directly estimated by weighting appropriately the $\gamma+$jets events surviving the SR selections. In SR2$\ell\_$Low, the direct $Z+$jets background estimation lacks statistical precision due to the high prescale factors of the triggers used to select $\gamma$+jets events with low momentum ($\pt^{\gamma}<100$~\GeV), as opposed to the other SRs whose definitions, including an ISR-jet requirement, are such that 
events with a large dilepton system $p_{\textrm{T}}$ ($p_{\textrm{T}}^{\ell\ell}$) are selected. Due to this, an alternative approach is used for the $Z+$jets estimate in the low-mass SR, which relies on the robust $\gamma+$jets estimate of high-$p_{\textrm{T}}^{\ell\ell}$ ($p_{\textrm{T}}^{\ell\ell}>100$~\GeV) events. The $\gamma+$jets template is used to directly estimate the high-$p_{\textrm{T}}^{\ell\ell}$ $Z+$jets component of SR$2\ell$\_Low while the low-$p_{\textrm{T}}^{\ell\ell}$ ($p_{\textrm{T}}^{\ell\ell}<100$~\GeV) $Z+$jets contribution is estimated by using a transfer factor defined as the ratio of low-$p_{\textrm{T}}^{\ell\ell}$ to high-$p_{\textrm{T}}^{\ell\ell}$ events and is calculated from an orthogonal sample with an inverted $H_{4,1}^{\textrm{PP}}$ requirement. The ratio is found to be 3.9$\pm$2.1, while the high-$p_{\textrm{T}}^{\ell\ell}$ $Z+$jets estimate is 1.29$\pm$0.5. The uncertainties quoted are statistically only.

To validate the method, as well as to check the modeling of other SM backgrounds, validation regions are defined for each SR. The definitions of these regions (VR2$\ell$-VV, VR2$\ell$-Top, VR2$\ell$\_High-Zjets and VR2$\ell$\_Low-Zjets) are given for the standard decay tree in Table~\ref{tab:StandardRJSelections} and (VR2$\ell$\_ISR-VV, VR2$\ell$\_ISR-Top and VR2$\ell$\_ISR-Zjets) for the compressed decay tree in Table~\ref{tab:2L2J_compressedCRs}. The VRs targeting the validation of the $Z+$jets background estimation have an inverted dijet mass requirement with respect to the corresponding SR definitions as well as having some other selection criteria relaxed. In this way a potential signal contribution is rejected while the regions remain close but orthogonal to the SR selections. 

As described in Section~\ref{sec:selection}, the background contributions from $Wt+t\bar{t}$ and $VV$ are normalized to data in dedicated CRs and the extracted normalization factors from the fit are validated in orthogonal regions. The $VV$ process in the SRs has contributions from all diboson processes producing at least two leptons in the final state. The dominant diboson process in SR$2\ell$\_High and SR$2\ell$\_Int is $ZZ\to\ell\ell\nu\nu$ with a smaller contribution from $WZ\to\ell\nu\ell\ell$. The picture changes with lower $\chinoonepm/\ninotwo$ masses and smaller mass-splitting; in SR$2\ell$\_Low the dominant component is $WW\to\ell\nu\ell\nu$ followed by $WZ\to\ell\nu\ell\ell$ while in SR$2\ell$\_ISR the dominant contribution is from $WZ\to\ell\nu\ell\ell$ and to a lesser extent from $ZZ\to\ell\ell\nu\nu$. The semihadronic decays of dibosons, for example $ZV\to \ell\ell qq$, are accounted for by the $\gamma+$jets template since they do not lead to genuine \MET in the event. The CRs are designed to have compositions, in terms of diboson processes, similar to their respective SRs.  

The two-lepton diboson and top CRs defined with the standard decay tree do not contain an explicit selection to make them orthogonal to their respective compressed CRs. However, the two decay trees of the RJR method, by construction, probe different event topologies, hence they select events where the overlap is designed to be insignificant. For the top CR the overlap is less than 1$\%$ while for the diboson CR it is smaller than 3$\%$. Since the impact of this effect is negligible in comparison with the background uncertainties, it is not considered in the remainder of the analysis.

The normalization factors obtained from the background-only fit for $Wt+t\bar{t}$ and $VV$ for the selections applied to the standard (compressed) decay tree are 0.91$\pm$0.23 and 0.91$\pm$0.13 (0.99$\pm$0.12 and 0.94$\pm$0.18), respectively, where the uncertainties are dominated by the statistical uncertainty. The background fit results are summarized in Tables~\ref{table:twoLeptonCRBkgOnlyFit} and~\ref{table:twoLeptonVRBkgOnlyFit} for the CRs and VRs, respectively. The data are consistent with the expected background in all validation regions.

  \begin{table}
    \caption{Background fit results for the 2$\ell$ CRs. The normalization factors for $Wt+t\bar{t}$ and $VV$ for the standard and compressed decay trees are different and are extracted from separate fits. The nominal predictions from MC simulation are given for comparison for the $Wt+t\bar{t}$ and $VV$ backgrounds. The ``Other'' category contains the contributions from Higgs boson processes, $V\gamma,\ VVV,\ t\bar{t}V$ and non-prompt and non-isolated lepton production. The dashes indicate that these backgrounds are negligible and are included in the category ``Other''. Combined statistical and systematic uncertainties are given. The individual uncertainties can be correlated and do not necessarily add in quadrature to the total systematic uncertainty.}
\begin{center}
\setlength{\tabcolsep}{0.0pc}
{\scriptsize
\begin{tabular*}{\textwidth}{@{\extracolsep{\fill}}lP{-1}P{-1}P{-1}P{-1}}
\noalign{\smallskip}\toprule\noalign{\smallskip}
\multicolumn{1}{l}{Region}       & \multicolumn{1}{c}{CR2$\ell$-VV}            & \multicolumn{1}{c}{CR2$\ell$-Top}            & \multicolumn{1}{c}{CR2$\ell$\_ISR-VV}            & \multicolumn{1}{c}{CR2$\ell$\_ISR-Top}              \\[-0.05cm]
\noalign{\smallskip}\hline\hline\noalign{\smallskip}
Observed events          & 60              & 97              & 28              & 93                    \\
\noalign{\smallskip}\hline\noalign{\smallskip}
Total (post-fit) SM events         & 60 , 8          & 97 , 10          & 28 , 5          & 93 , 10              \\
\noalign{\smallskip}\hline\noalign{\smallskip}
        Other         & 3.5 , 0.3          & 1.4 , 0.3          & 0.72 , 0.31          & 0.50 , 0.15              \\
        Fit output, $Wt+t\bar{t}$         & -          & 60 , 11          & -          & 90 , 10              \\
        Fit output, $VV$         & 57 , 8          & 4.0 , 1.0          & 27 , 5          & 0.99 , 0.31              \\
        $Z+$jets          & -          & 31 , 15          & -          & 2.1 , 1.0              \\     
\noalign{\smallskip}\hline\noalign{\smallskip}
        Fit input, $Wt+t\bar{t}$         & -         & 66          & -          & 91              \\
        Fit input, $VV$                  & 62          & 4.4          & 29          & 1.1              \\
\noalign{\smallskip}\toprule\noalign{\smallskip}
\end{tabular*}
}
\end{center}
\label{table:twoLeptonCRBkgOnlyFit}
\end{table}

  \begin{table}
    \caption{Expected and observed yields from the background fit for the 2$\ell$ VRs. The nominal predictions from MC simulation are given for comparison for the $Wt+t\bar{t}$ and $VV$ backgrounds. The ``Other'' category contains the contributions from Higgs boson processes, $V\gamma,\ VVV,\ t\bar{t}V$ and non-prompt and non-isolated lepton production. The dashes indicate that these backgrounds are negligible and are included in the category ``Other''. Combined statistical and systematic uncertainties are given. The individual uncertainties can be correlated and do not necessarily add in quadrature to the total systematic uncertainty.}
\begin{center}
\setlength{\tabcolsep}{0.0pc}
{\tiny
\begin{tabular*}{\textwidth}{@{\extracolsep{\fill}}lP{-1}P{-1}P{-1}P{-1}P{-1}P{-1}P{-1}}
\noalign{\smallskip}\toprule\noalign{\smallskip}
\multicolumn{1}{l}{Region}         & \multicolumn{1}{c}{VR2$\ell$\_Low-Zjets}            & \multicolumn{1}{c}{VR2$\ell$\_High-Zjets}            & \multicolumn{1}{c}{VR2$\ell$-VV}       & \multicolumn{1}{c}{VR2$\ell$-Top}            & \multicolumn{1}{c}{VR2$\ell$\_ISR-VV}            & \multicolumn{1}{c}{VR2$\ell$\_ISR-Top}            & \multicolumn{1}{c}{VR2$\ell$\_ISR-Zjets}              \\[-0.05cm]
\noalign{\smallskip}\hline\hline\noalign{\smallskip}
Observed events          & 263              & 77              & 72              & 491              & 13              & 113              & 248                    \\
\noalign{\smallskip}\hline\noalign{\smallskip}
Total (post-fit) SM events         & 261, 130          & 69 , 26          & 61 , 13          & 423 , 105          & 12 , 4          & 110 , 18          & 310 , 100              \\
\noalign{\smallskip}\hline\noalign{\smallskip}
        Other         & 3.5 , 1.5          & 0.25\rlap{$_{-0.25}^{+0.62}$}          & 0.80 , 0.09          & 2.3 , 0.4          & 4.2 , 0.5          & 0.68 , 0.22          & 3.0 , 0.6              \\
        Fit output, $Wt+t\bar{t}$         & 15 , 5          & 1.7 , 0.7          & 12 , 4          & 415 , 105          & -          & 107 , 18          & 40 , 8              \\
        Fit output, $VV$          & 30 , 7          & 16 , 3          & 40 , 13          & 3.7 , 0.9          & 7.9 , 3.6          & 0.97 , 0.25          & 67 , 15              \\
        $Z+$jets         & 210 , 130          & 51 , 25          & 8.4 , 4.1          & 2.4 , 1.2          & -          & 1.6 , 0.8          & 200 , 100              \\
\noalign{\smallskip}\hline\noalign{\smallskip}
        Fit input, $Wt+t\bar{t}$         & 16          & 1.9          & 13          & 455          & -          & 108          & 41              \\
        Fit input, $VV$          & 33          & 17          & 43          & 4.1          & 8.4          & 1.1          & 71              \\
\noalign{\smallskip}\toprule\noalign{\smallskip}
\end{tabular*}
}
\end{center}
\label{table:twoLeptonVRBkgOnlyFit}
\end{table}

\subsection{Background estimate in the three-lepton channel}
  
The irreducible background in the 3$\ell$ channel is dominated by SM $WZ$ diboson production. The shape of the diboson background is taken from simulation but normalized to data in dedicated CRs. The normalization factors extracted from the background-only fit are found to be 1.09$\pm$0.10 and 1.13$\pm$0.13 for the standard- and compressed decay tree selections, respectively. The results of the background estimates are validated in a set of dedicated VRs. Other background sources such as $VVV$, $t\bar{t}V$ and processes with a Higgs boson contributing to the irreducible background are taken from simulation. A summary of the background fit results for the $3\ell$ CRs and VRs is given in Table~\ref{table:threeLeptonCRVRBkgOnlyFit}. 

Similar to the two-lepton CR design, the three-lepton diboson CR defined with the standard decay tree does not contain an explicit selection to make it orthogonal to its respective compressed CR. The overlap is less than 0.5$\%$. Since the impact of this effect is negligible in comparison with the background uncertainties, it is not considered in the remainder of the analysis.

\begin{table}
  \caption{Expected and observed yields from the background fit for the 3$\ell$ CRs and VRs. The normalization factors for $VV$ for the standard and compressed decay trees are different and are extracted from separate fits. The nominal predictions from MC simulation are given for comparison for the $VV$ background. The ``Other'' category contains the contributions from Higgs boson processes, $t\bar{t}V$ and non-prompt and non-isolated lepton production. Combined statistical and systematic uncertainties are given. The individual uncertainties can be correlated and do not necessarily add in quadrature to the total systematic uncertainty.}
  \begin{center}
\setlength{\tabcolsep}{0.0pc}
{\small
\begin{tabular*}{\textwidth}{@{\extracolsep{\fill}}lP{-1}P{-1}P{-1}P{-1}}
\noalign{\smallskip}\toprule\noalign{\smallskip}
\multicolumn{1}{l}{Region}           & \multicolumn{1}{c}{CR3$\ell$-VV}            & \multicolumn{1}{c}{VR3$\ell$-VV}            & \multicolumn{1}{c}{CR3$\ell$\_ISR-VV}            & \multicolumn{1}{c}{VR3$\ell$\_ISR-VV}              \\[-0.05cm]
\noalign{\smallskip}\hline\hline\noalign{\smallskip}
Observed events          & $331$              & $160$              & $98$              & $83$                    \\
\noalign{\smallskip}\hline\noalign{\smallskip}
Total (post-fit) SM events         & 331, 18          & 159, 38          & 98, 10          & 109, 24              \\
\noalign{\smallskip}\hline\noalign{\smallskip}
        Other         & 52, 13          & 5.6, 1.2          & 4.4, 1.2          & 7.1, 1.6              \\
        Tribosons         & 1.1, 0.1          & 0.44, 0.03          & 0.22, 0.14          & 0.42, 0.04              \\
        Fit output, $VV$          & 278, 18          & 153, 38          & 93, 10          & 102, 24              \\
 \noalign{\smallskip}\hline\noalign{\smallskip}
        Fit input, $VV$         & 255          & 140          & 83          & 90              \\
\noalign{\smallskip}\toprule\noalign{\smallskip}
\end{tabular*}
}
\end{center}
\label{table:threeLeptonCRVRBkgOnlyFit}
\end{table}

\section{Systematic uncertainties}
\label{sec:systematics}

Several sources of experimental and theoretical systematic uncertainties are considered in the SM background estimates and signal expectations and are included in the profile likelihood fits described in Section~\ref{sec:strategy}. The systematic uncertainties that are considered are related to the jet energy scale and resolution, the modeling of $\MET$ in the simulation, the lepton reconstruction and identification, the $VV$ theoretical modeling uncertainties, the non-prompt lepton background estimation and the data-driven $Z+$jets estimate. The effects of these uncertainties are evaluated for all signal event samples and background processes. The normalization of the $Wt+t\bar{t}$ and $VV$ background predictions is extracted in dedicated control regions and the systematic uncertainties thus only affect the extrapolation to the SRs. The statistical uncertainty due to the number of events in the MC samples is also included. The systematic uncertainty associated with the pileup reweighting of the simulated events is also considered and found to have a negligible impact on the final results.

The jet energy scale and resolution uncertainties are derived as a function of the \pT and $\eta$ of the jet, as well as of the pileup conditions and the jet flavor composition of the selected jet sample. They are determined using a combination of simulated events and data samples, through measurements of the jet response balance in multijet, $Z$+jets and $\gamma+$jets events~\cite{JetCalibRunTwo}.

The systematic uncertainties related to the modeling of \MET in the simulation are estimated by propagating the uncertainties in the energy and momentum scale of each of the physics objects, as well as the uncertainties in the soft-term resolution and scale~\cite{ATL-PHYS-PUB-2015-023}.

The remaining detector-related systematic uncertainties, such as those in the lepton reconstruction efficiency, $b$-tagging efficiency~\cite{ATLAS-CONF-2014-004,ATLAS-CONF-2014-046}, lepton energy scale, energy resolution and in the modeling of the trigger~\cite{ATLAS-CONF-2016-024, MuonPerfRun2}, are included but are found to be negligible in all channels.

The uncertainties arising from the modeling of diboson events in simulation are estimated by varying the renormalization, factorization and merging scales used to generate the samples, as well as the PDFs.

In the 2$\ell$ channel, uncertainties in the data-driven $Z$+jets estimate are calculated following the methodology used in Ref.~\cite{SUSY-2016-05}. An additional uncertainty is based on the difference between the expected background yield from the nominal method (which produces 6.3 events in SR$2\ell\_$Low and 0.1 events in SR$2\ell\_$ISR) and from a second method implemented as a cross-check, which extracts the dijet mass shape from data validation regions, normalizes the shape to the sideband regions of the SRs, and extrapolates the background into the $W$ mass region. The $Z+$jets background estimations obtained from the sideband method are 5.9 and 0.2 events for SR$2\ell\_$Low and SR$2\ell\_$ISR, respectively. Moreover, a 100\% uncertainty in the $V\gamma$ normalization factor is included. To cover any statistical limitations on the $Z+$jets estimate that may be present in SR2$\ell\_$ISR, an upper limit on the $Z+$jets estimate is considered as an additional systematic uncertainty. The upper limit is calculated by multiplying the sum of the nominal $Z+$jets background estimate, adding the statistical uncertainty, with the ratio of low-$p_{\mathrm{T}}^{\ell\ell}$ to high-$p_{\mathrm{T}}^{\ell\ell}$ events calculated with a looser requirement on $p_{\mathrm{T\ I}}^{\mathrm{CM}}$. This is the dominant uncertainty in the ISR region and accounts for 95\% of the total uncertainty in the $Z+$jets estimate.  

Systematic uncertainties are also assigned to the estimated background from FNP leptons in both the $2\ell$ and $3\ell$ channels to account for potentially different compositions (heavy flavor, light flavor or conversions) between the signal and control regions. An additional uncertainty is associated with the subtraction of prompt leptons from this CR using simulation.

A summary of the dominant uncertainties in the $2\ell$ SRs is shown in Table~\ref{table:twoLeptonSystematics}. The uncertainties with the largest impact in these SRs are those in the data-driven $Z+$jets estimate, followed by the $VV$ modeling uncertainties, the statistical uncertainties in the MC background samples and the uncertainty in the fitted normalization factor for $VV$ related to the number of events in the corresponding CRs.     

\begin{table}
\caption{Summary of the main systematic uncertainties and their impact (in \%) on the total SM background prediction in each of the $2\ell$ SRs. The total systematic uncertainty can be different from the sum in quadrature of individual sources due to the correlations between them resulting from the fit to the data.}
\begin{center}
\setlength{\tabcolsep}{0.0pc}
{\small
\begin{tabular*}{\textwidth}{@{\extracolsep{\fill}}lrrrr}
\noalign{\smallskip}\toprule\noalign{\smallskip}
Signal Region           & SR2$\ell$\_High            & SR2$\ell$\_Int            & SR2$\ell$\_Low            & SR2$\ell$\_ISR              \\[-0.05cm]
\noalign{\smallskip}\hline\hline\noalign{\smallskip}
Total uncertainty [$\%$]              & 42   & 38   & 70   & 103\\
\noalign{\smallskip}\hline\noalign{\smallskip}
$Z+$jets data-driven estimate   & 42   & 31   & 69   & 96\\
$VV$ theoretical uncertainties  & 28   & 27   & 6    & 34\\
MC statistical uncertainties    & 16   & 12   & 5    & 9\\
$VV$ fitted normalization       & 13   & 14   & 2    & 16\\
FNP leptons      & -    &  5   & 13   & 12\\
Jet energy resolution           & 5    & 10   & 4    & 3\\
Jet energy scale                & 1    & 2    & $<1$ & 3\\
$\MET$ modeling                    & 3    & 4    & $<1$ & $<1$\\
$t\bar{t}$ fitted normalization & $<1$ & $<1$ & 2    & 2\\
Lepton reconstruction / identification                  & $<1$ & $<1$ & $<1$ & $<1$\\
\noalign{\smallskip}\toprule\noalign{\smallskip}
\end{tabular*}
}
\end{center}
\label{table:twoLeptonSystematics}
\end{table}

A similar summary of the systematic uncertainties impacting the $3\ell$ SRs is given in Table~\ref{table:threeLeptonSystematics}. These are dominated by the statistical uncertainties in the MC background samples, the modeling uncertainties in the $VV$ processes and the uncertainties related to the fitted normalization factors for $VV$.

\begin{table}
\caption{Summary of the main systematic uncertainties and their impact (in \%) on the total SM background prediction in each of the $3\ell$ SRs. The total systematic uncertainty can be different from the sum in quadrature of individual sources due to the correlations between them resulting from the fit to the data.}
\begin{center}
\setlength{\tabcolsep}{0.0pc}
{\small
\begin{tabular*}{\textwidth}{@{\extracolsep{\fill}}lrrrr}
\noalign{\smallskip}\toprule\noalign{\smallskip}
 Signal Region           & SR3$\ell$\_High            & SR3$\ell$\_Int            & SR3$\ell$\_Low            & SR3$\ell$\_ISR              \\[-0.05cm]
\noalign{\smallskip}\hline\hline\noalign{\smallskip}
Total uncertainty [$\%$]              & 44   & 22      & 19   & 26 \\
\noalign{\smallskip}\hline\noalign{\smallskip}
$VV$ theoretical uncertainties  & 18   & 9       & 12   & 19 \\
MC statistical uncertainties    & 37   & 17      & 8    & 10\\
$VV$ fitted normalization       & 8   &  7       & 9    & 11\\
FNP leptons      & 7    & $<1$    & 3   & 5\\
Jet energy resolution           & 4    & $<1$    & 7    & 3\\
Jet energy scale                & 7    & $<1$    & 2 & 3\\
$\MET$ modeling                   & 2    & $<1$    & 1 & 4\\
Lepton reconstruction / identification                  & 3    & 4       & 2 & 2\\
\noalign{\smallskip}\toprule\noalign{\smallskip}
\end{tabular*}
}
\end{center}
\label{table:threeLeptonSystematics}
\end{table}

\FloatBarrier

\section{Results and interpretation}
\label{sec:results}

The observed numbers of events in the 2$\ell$ channel are compared with the expected background contributions in Table~\ref{table:twoLeptonSRBkgOnlyFit} and Figure~\ref{fig:2LPullPlotAllregions}; those in the 3$\ell$ channel are shown in Table~\ref{table:threeLeptonSRBkgOnlyFit} and Figure~\ref{fig:3LPullPlotAllRegions}. No significant excesses above the SM expectation are observed in the SRs targeting intermediate- and high-mass signal models. An excess of events above the background estimate is observed in each of the four low-mass and ISR signal regions. To quantify the level of agreement of the observed data with the SM expectations, a model-independent fit is performed separately for each SR. The results of this fit for the 2$\ell$ and 3$\ell$ searches are given in Table~\ref{table.upperlimit}.

Selected kinematic distributions in the low-mass and ISR regions for the $2\ell$ and $3\ell$ selections after applying all the selection requirements defining these SRs are shown in Figures~\ref{fig:srRJR2L} and \ref{fig:srRJR3L}, respectively. In all figures a SUSY signal benchmark model is shown for illustration. This simplified model assumes $m_{\chinoonepm/\ninotwo}=200$~\GeV\ and $m_{\ninoone}=100$~\GeV, and was used to optimize the event selection criteria for the low-mass and ISR SRs.

With the complementarity of the $3\ell$ low-mass and ISR regions, a study of events that fall in either one or the other is possible. Many of the discriminating variables are specific to the decay trees, hence events in the ISR and low-mass SRs cannot be displayed together in these observables. Figure~\ref{fig:EitherOR3L} shows the transverse mass distribution, calculated using the unpaired lepton prior to the selection imposed on this variable, for events passing the 3$\ell$ low-mass (\ref{fig:EitherOR3L_low}) and the 3$\ell$ ISR SR requirements (\ref{fig:EitherOR3L_isr}). These distributions show events with no additional jet activity, along with those including a jet identified as emanating from an ISR system. In both figures there is an excess of events with transverse mass above the minimum value of 100~\GeV\ required in both SR$3\ell$\_Low and SR$3\ell$\_ISR.

Exclusion limits for simplified models, in which pairs of $\chinoonepm\ninotwo$ decay with 100\% branching ratio into $W/Z$ vector bosons, are shown in Figure~\ref{fig:Money}. Figures \ref{fig:Money2L} and \ref{fig:Money3L} show the exclusion limits obtained from the $2\ell$ and 3$\ell$ channels respectively and after selecting the SR with the highest expected sensitivity for each signal-model assumption. The low-mass and ISR regions are statistically combined. Figure~\ref{fig:MoneyStatCombo} corresponds to the statistical combination of the $2\ell$ and $3\ell$ search channels. The combination proceeds by statistically combining the SRs of the two channels which target the same region in the two-dimensional parameter space (e.g.\ SR2$\ell$\_High with SR3$\ell$\_High) since they contain mutually exclusive events. Once the statistical combination is performed then the combined SR producing the best expected CL$_\text{s}$ value for each model assumption is chosen. The last step is needed since the high-, intermediate- and low-mass SRs have event overlap while the low-mass and ISR SRs are mutually exclusive and can be statistically combined. Finally, Figure \ref{fig:MoneyComparison} compares the expected and observed exclusion limits obtained from the recursive jigsaw approach with those described in Ref.~\cite{Aaboud:2018jiw}.

\begin{table}[!tbp]
  \begin{center}
    \caption{Expected and observed yields from the background-only fit for the 2$\ell$ SRs. The errors shown are the statistical plus systematic uncertainties. Uncertainties in the predicted background event yields are quoted as symmetric, except where the negative error reaches down to zero predicted events, in which case the negative error is truncated.
      \label{table:twoLeptonSRBkgOnlyFit}}
    \small{
      \setlength{\tabcolsep}{0.0pc}
      \begin{tabular*}{\textwidth}{@{\extracolsep{\fill}}lP{-1}P{-1}P{-1}P{-1}}    
        \noalign{\smallskip}\toprule\noalign{\smallskip}
        \multicolumn{1}{l}{Signal region}           & \multicolumn{1}{c}{SR2$\ell$\_High}            & \multicolumn{1}{c}{SR2$\ell$\_Int}            & \multicolumn{1}{c}{SR2$\ell$\_Low}            & \multicolumn{1}{c}{SR2$\ell$\_ISR}\\[-0.05cm]
        \noalign{\smallskip}\hline\hline\noalign{\smallskip}
        Total observed events          & 0              & 1              & 19              & 11                    \\
        \noalign{\smallskip}\hline\noalign{\smallskip}
        Total background events         & 1.9, 0.8          & 2.4, 0.9          & 8.4, 5.8          & 2.7\rlap{$_{-2.7}^{+2.8}$}              \\
        \noalign{\smallskip}\hline\noalign{\smallskip}
        Other        & 0.02, 0.01          & 0.05\rlap{$_{-0.05}^{+0.12}$}          & 0.02\rlap{$_{-0.02}^{+1.07}$}          & 0.06\rlap{$_{-0.06}^{+0.33}$}              \\
        Fit output, $Wt+t\bar{t}$         & 0.00, 0.00         & 0.00, 0.00          & 0.57, 0.20          & 0.28\rlap{$_{-0.28}^{+0.34}$}              \\
        Fit output, $VV$         & 1.8, 0.7          & 2.4, 0.8          & 1.5, 0.9          & 2.3, 1.1              \\
        $Z+$jets         & 0.07\rlap{$_{-0.07}^{+0.78}$}          & 0.00\rlap{$_{-0.00}^{+0.74}$}            & 6.3, 5.8          & 0.10\rlap{$_{-0.10}^{+2.58}$}              \\
        \noalign{\smallskip}\hline\noalign{\smallskip}
        Fit input, $Wt+t\bar{t}$ & 0.00 & 0.00 & 0.63 & 0.28\\
        Fit input, $VV$ & 1.9 & 2.6 & 1.6 & 2.4\\
        \noalign{\smallskip}\toprule\noalign{\smallskip}
      \end{tabular*}
    } 
  \end{center}  
\end{table}

\begin{figure}
\centering
\includegraphics[scale=0.7]{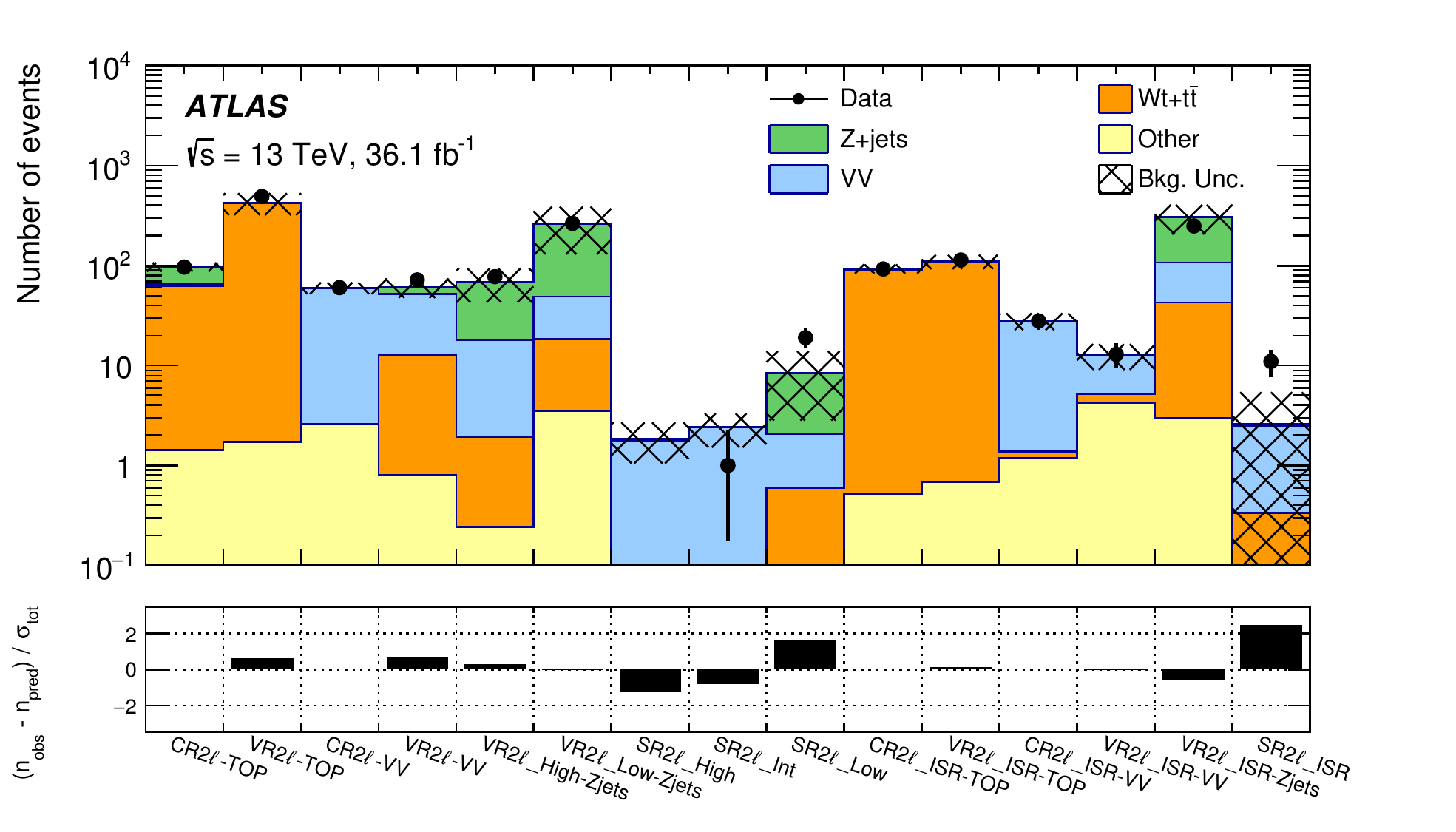}
\caption{The observed and expected SM background yields in the CRs, VRs and SRs considered in the 2$\ell$ channel. The statistical uncertainties in the background prediction are included in the uncertainty band, as well as the experimental and theoretical uncertainties. The bottom panel shows the difference in standard deviations between the observed and expected yields.}
\label{fig:2LPullPlotAllregions}		
\end{figure}

\begin{table}
  \caption{Expected and observed yields from the background-only fit for the $3\ell$ SRs. The errors shown are the statistical plus systematic uncertainties. Uncertainties in the predicted background event yields are quoted as symmetric, except where the negative error reaches down to zero predicted events, in which case the negative error is truncated.}
  \begin{center}
\setlength{\tabcolsep}{0.0pc}
{\small
\begin{tabular*}{\textwidth}{@{\extracolsep{\fill}}lP{-1}P{-1}P{-1}P{-1}}
\noalign{\smallskip}\toprule\noalign{\smallskip}
\multicolumn{1}{l}{Signal region}           & \multicolumn{1}{c}{SR$3\ell$\_High}            & \multicolumn{1}{c}{SR$3\ell$\_Int}            & \multicolumn{1}{c}{SR$3\ell$\_Low}            & \multicolumn{1}{c}{SR3$\ell$\_ISR}              \\[-0.05cm]
\noalign{\smallskip}\hline\hline\noalign{\smallskip}
Total observed events          & 2              & 1              & 20              & 12                    \\
\noalign{\smallskip}\hline\noalign{\smallskip}
Total background events         & 1.1, 0.5          & 2.3, 0.5          & 10, 2          & 3.9, 1.0              \\
\noalign{\smallskip}\hline\noalign{\smallskip}
Other          & 0.03\rlap{$_{-0.03}^{+0.07}$}          & 0.04, 0.02          & 0.02\rlap{$_{-0.02}^{+0.34}$}          & 0.06\rlap{$_{-0.06}^{+0.19}$}              \\
Triboson          & 0.19, 0.07          & 0.32, 0.06          & 0.25, 0.03          & 0.08, 0.04              \\
Fit output, $VV$         & 0.83, 0.39          & 1.9, 0.5          & 10, 2          & 3.8, 1.0              \\
\noalign{\smallskip}\hline\noalign{\smallskip}
Fit input, $VV$ & 0.76 & 1.8 & 9.2 & 3.4\\
\noalign{\smallskip}\toprule\noalign{\smallskip}
\end{tabular*}
}
\end{center}
\label{table:threeLeptonSRBkgOnlyFit}
\end{table}

\begin{figure}
\centering
		\includegraphics[scale=0.7]{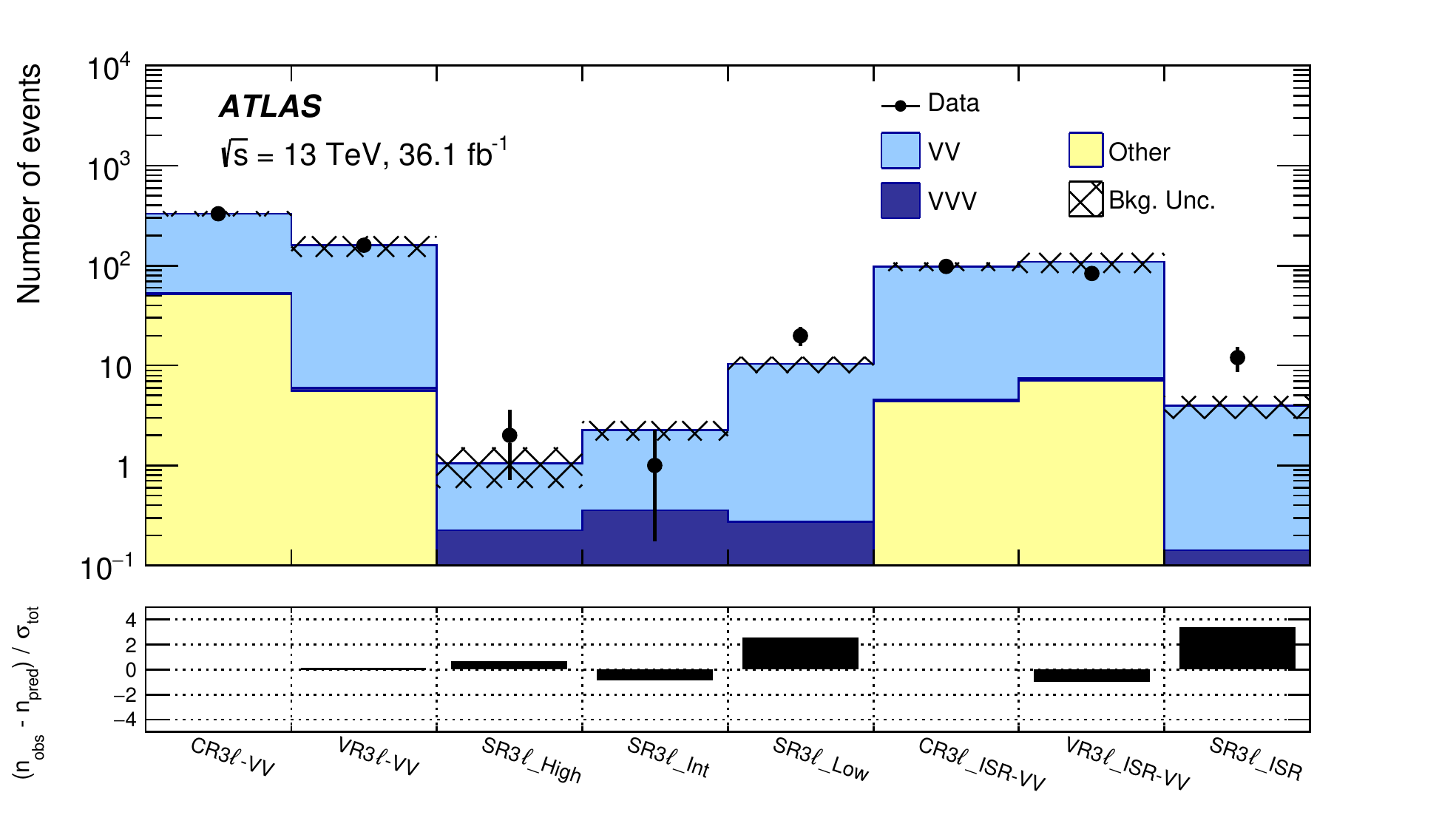}
                \caption{The observed and expected SM background yields in the CRs, VRs and SRs considered in the 3$\ell$ channel. The statistical uncertainties in the background prediction are included in the uncertainty band, as well as the experimental and theoretical uncertainties. The bottom panel shows the difference in standard deviations between the observed and expected yields.}
\label{fig:3LPullPlotAllRegions}		
\end{figure}

\begin{table}
  \caption{Model-independent fit results for all SRs. The first column shows the SRs, the second and third columns show the 95\% CL upper limits on the visible cross-section ($\langle\epsilon\sigma\rangle_{\textrm{ obs}}^{95}$) and on the number of
signal events ($S_{\textrm{ obs}}^{95}$ ).  The fourth column
($S_{\textrm{ exp}}^{95}$) shows the 95\% CL upper limit on the number of
signal events, given the expected number (and $\pm 1\sigma$
excursions of the expectation) of background events. The last column
indicates the discovery $p_{0}$-value and its associated significance ($Z$).}
\centering
\setlength{\tabcolsep}{0.0pc}
\begin{tabular*}{\textwidth}{@{\extracolsep{\fill}}lcrrrr}
\noalign{\smallskip}\toprule\noalign{\smallskip}
Signal region                        & $\langle\epsilon{\sigma}\rangle_{\textrm{ obs}}^{95}$[fb]  &  $S_{\textrm{ obs}}^{95}$  & $S_{\textrm{ exp}}^{95}$ & $p_{0}$ ($Z$)  \\
\noalign{\smallskip}\hline\hline\noalign{\smallskip}
SR3$\ell$\_ISR    & $0.42$ &  $15.3$ & $ { 6.9 }^{ +3.1 }_{ -2.2 }$ & $ 0.001$~$(3.02)$ \\
SR2$\ell$\_ISR    & $0.43$ &  $15.4$ & $ { 9.7 }^{ +3.6 }_{ -2.5 }$ & $ 0.02$~$(1.99)$ \\
SR3$\ell$\_Low    & $0.53$ &  $19.1$ & $ { 9.5 }^{ +4.2 }_{ -1.8 }$ & $ 0.016$~$(2.13)$ \\
SR2$\ell$\_Low    & $0.66$ &  $23.7$ & $ { 16.1 }^{ +6.3 }_{ -4.3 }$ & $ 0.08$~$(1.39)$ \\
SR3$\ell$\_Int    & $0.09$ &  $3.3$ & $ { 4.4 }^{ +2.5 }_{ -1.5 }$ & $ 0.50$~$(0.00)$ \\
SR2$\ell$\_Int    & $0.09$ &  $3.3$ & $ { 4.6 }^{ +2.6 }_{ -1.5 }$ & $ 0.50$~$(0.00)$ \\ 
SR3$\ell$\_High    & $0.14$ &  $5.0$ & $ { 3.9 }^{ +2.2 }_{ -1.3 }$ & $ 0.23$~$(0.73)$ \\
SR2$\ell$\_High   & $0.09$ &  $3.2$ & $ { 4.0 }^{ +2.3 }_{ -1.2 }$ & $ 0.50$~$(0.00)$ \\
\noalign{\smallskip}\toprule\noalign{\smallskip}
\end{tabular*}
\label{table.upperlimit}
\end{table}

\begin{figure}[htbp]
\begin{center}  
\vspace*{-0.005\textheight}\subfigure[]{\includegraphics[width=0.42\textwidth]{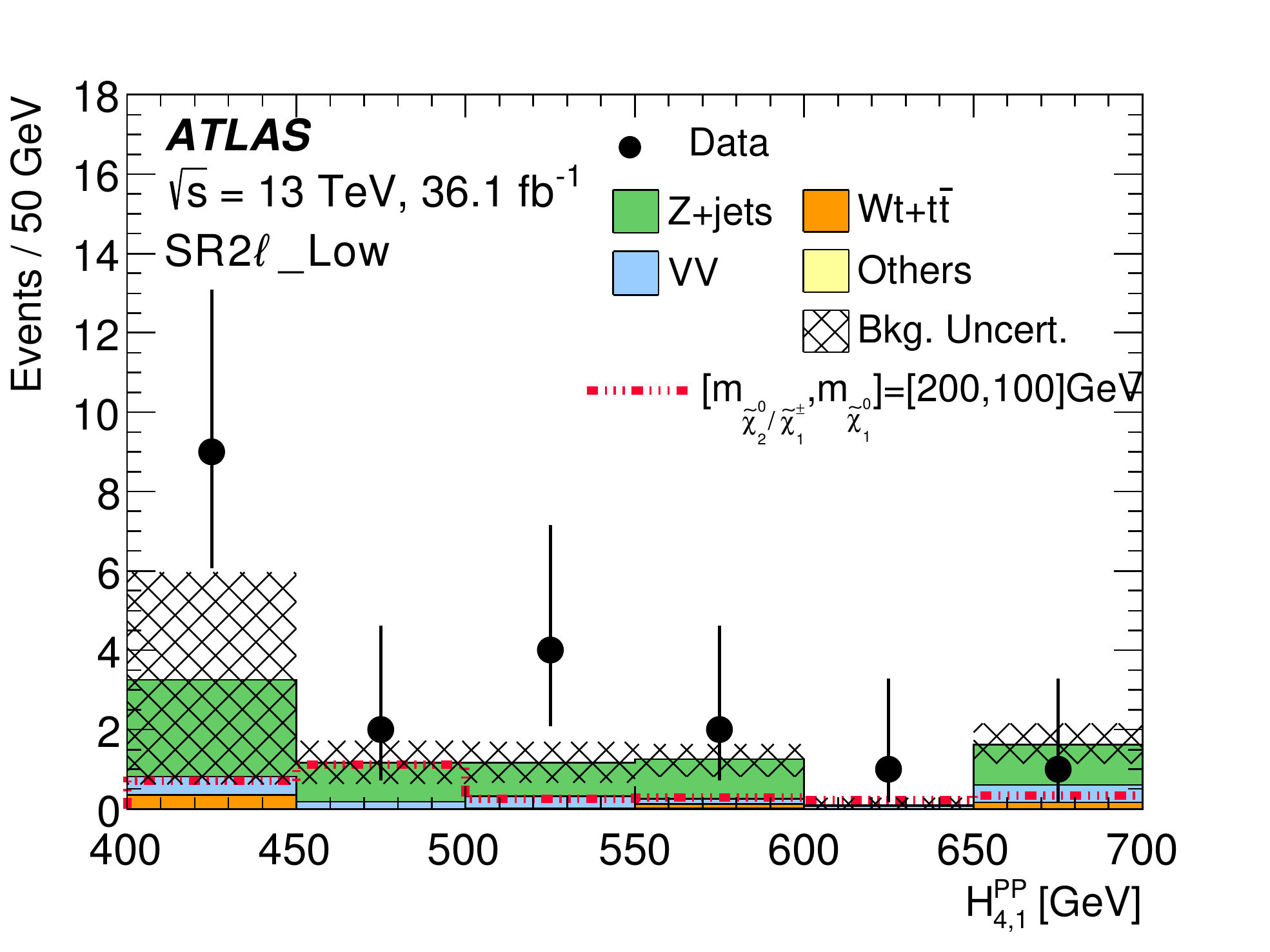}} 
\vspace*{-0.005\textheight}\subfigure[]{\includegraphics[width=0.42\textwidth]{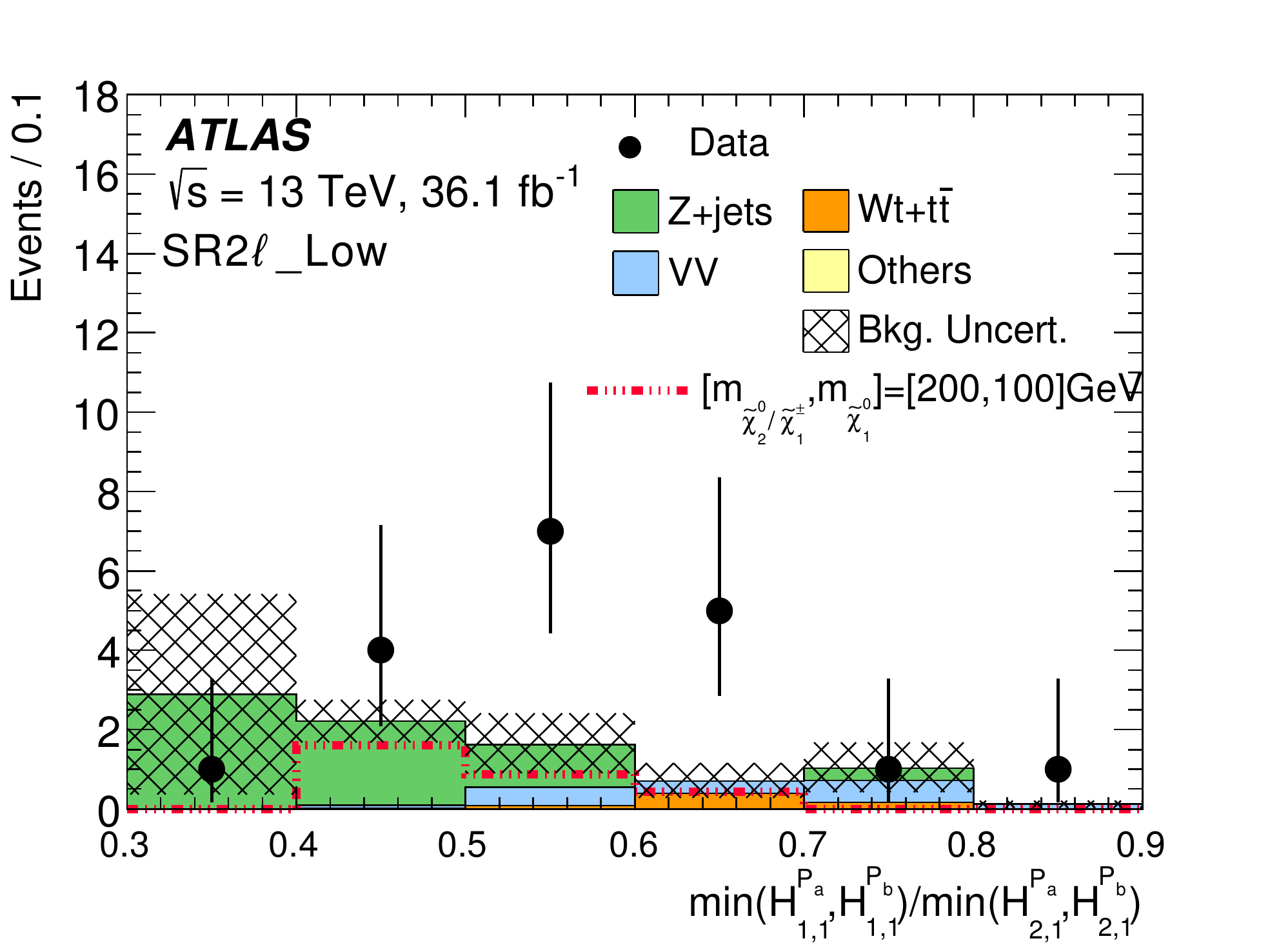}} \\ 
\vspace*{-0.005\textheight}\subfigure[]{\includegraphics[width=0.42\textwidth]{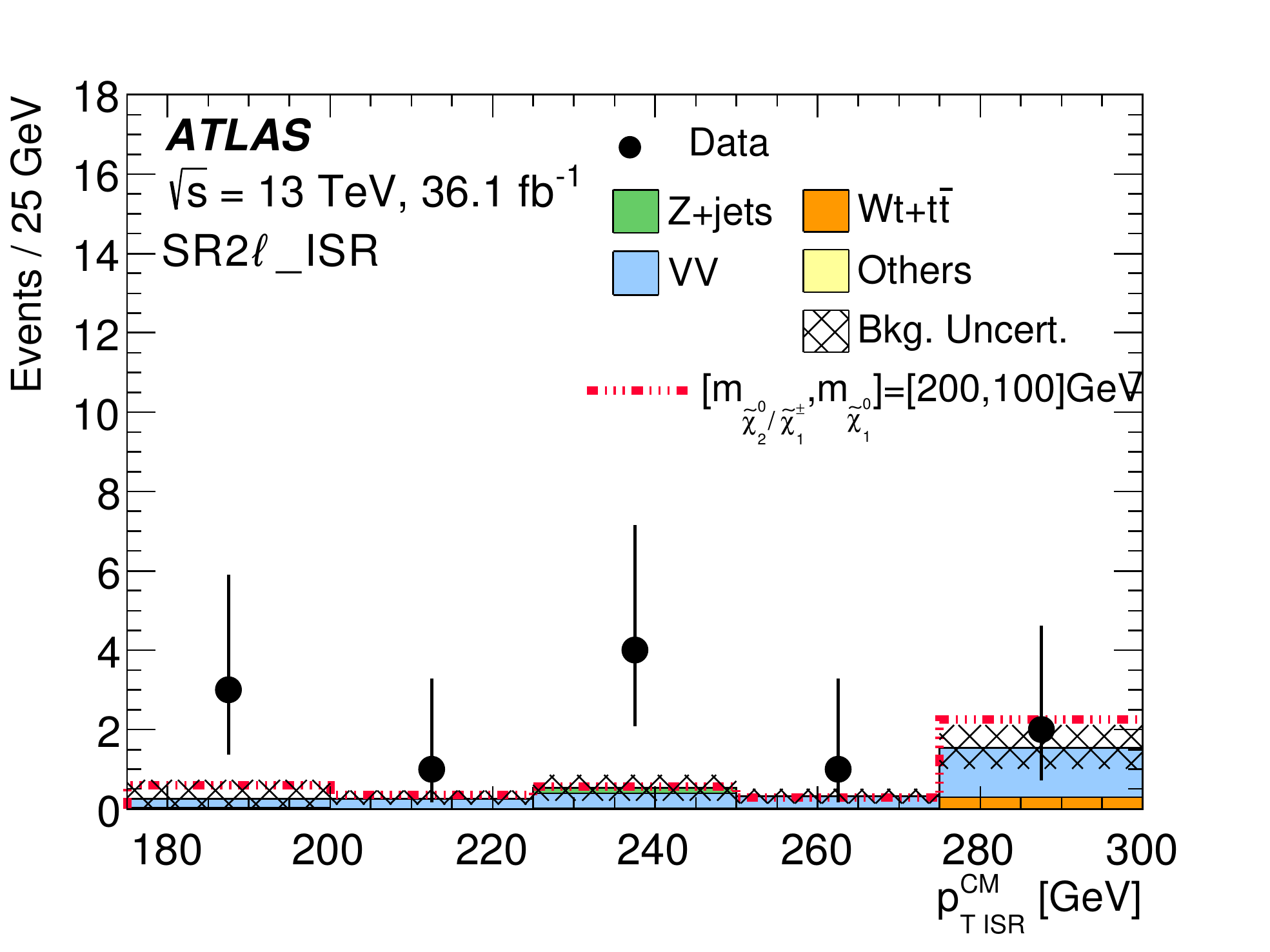}} 
\vspace*{-0.005\textheight}\subfigure[]{\includegraphics[width=0.42\textwidth]{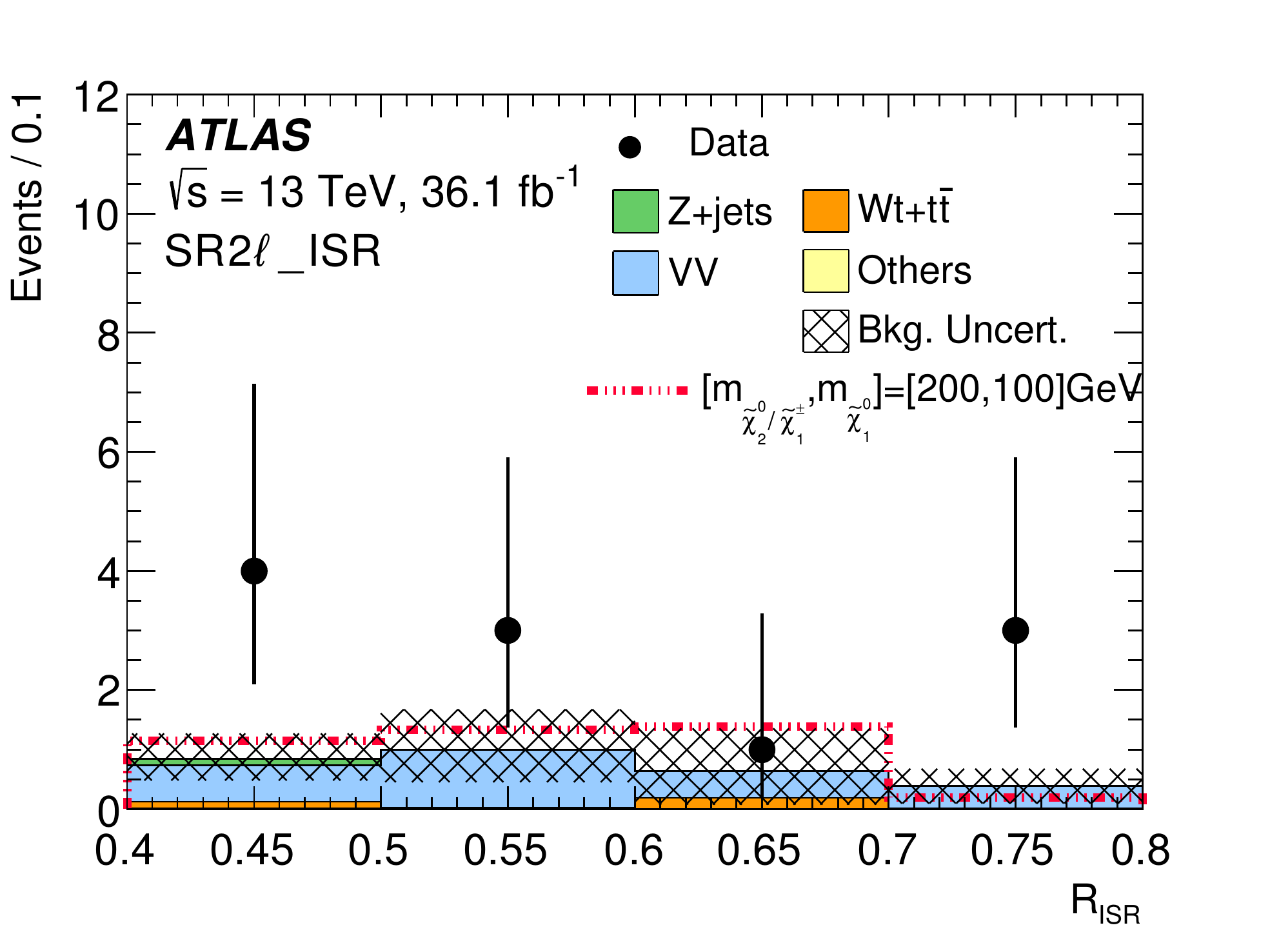}} 
\end{center}
\vspace*{-0.03\textheight}\caption{\label{fig:srRJR2L}
Distributions of kinematic variables in the signal regions for the $2\ell$ channels after applying all selection requirements. The histograms show the post-fit background predictions. 
The last bin includes the overflow. The FNP contribution is estimated from a data-driven technique and is included in the category ``Others''. Distributions for (a) $H_{4,1}^{\textrm{PP}}$ and (b) $\textrm{min}(H^{\textrm{P}_{\textrm{a}}}_{1,1},H^{\textrm{P}_{\textrm{b}}}_{1,1})/\textrm{min}(H^{\textrm{P}_{\textrm{a}}}_{2,1},H^{\textrm{P}_{\textrm{b}}}_{2,1})$ in SR$2\ell$\_Low, (c) $p_{\mathrm{T\ ISR}}^{~\textrm{CM}}$ and (d) $R_{\textrm{ISR}}$ in SR$2\ell$\_ISR are plotted. 
The hatched (black) error bands indicate the combined theoretical, experimental and MC statistical uncertainties.
The expected distribution for a benchmark signal model, normalized to the NLO+NLL cross-section (Section~\ref{sec:montecarlo}) times integrated luminosity, is also shown for comparison. 
}
\end{figure}

\begin{figure}[htbp]
\begin{center}  
\vspace*{-0.005\textheight}\subfigure[]{\includegraphics[width=0.42\textwidth]{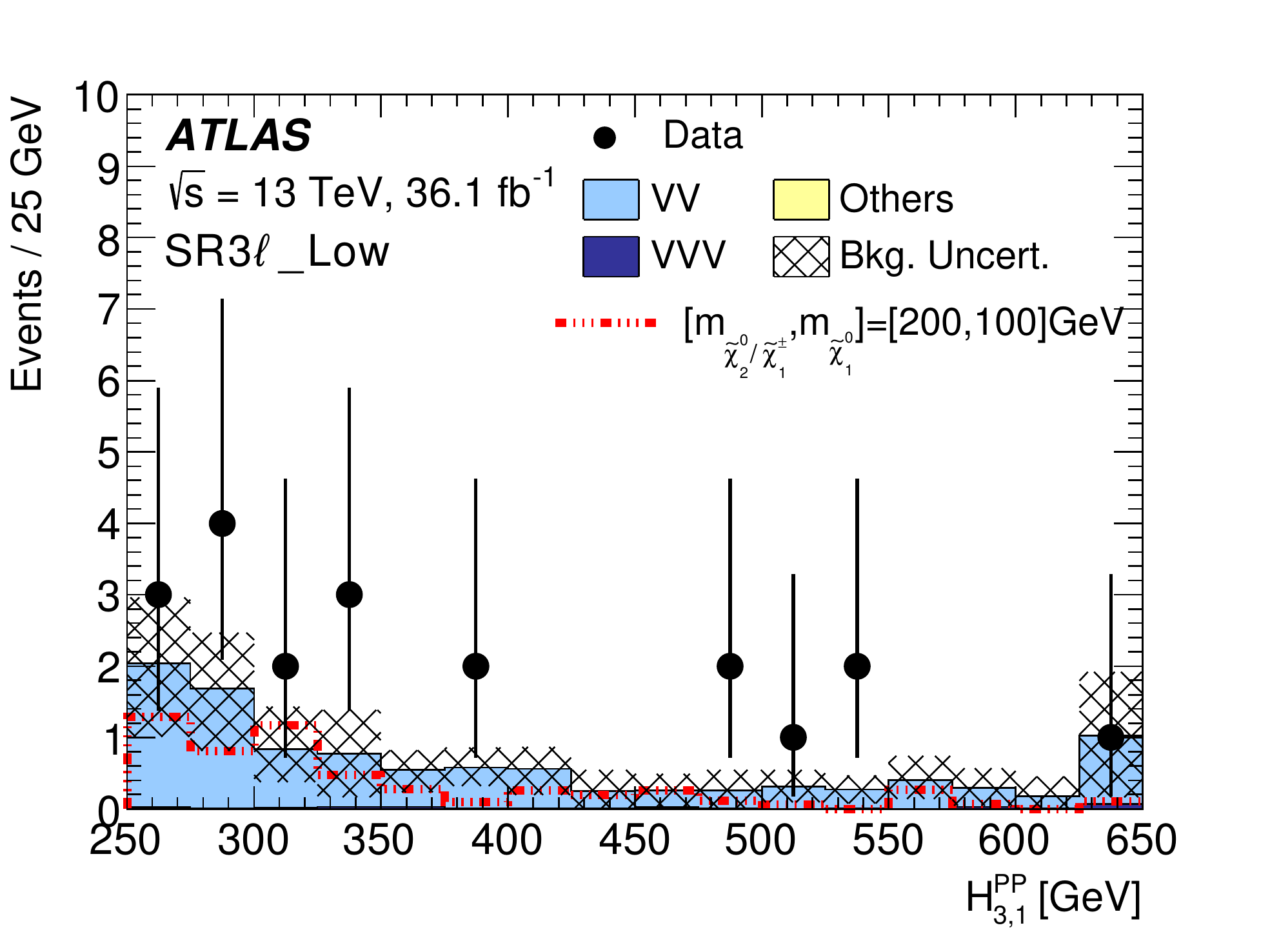}} 
\vspace*{-0.005\textheight}\subfigure[]{\includegraphics[width=0.42\textwidth]{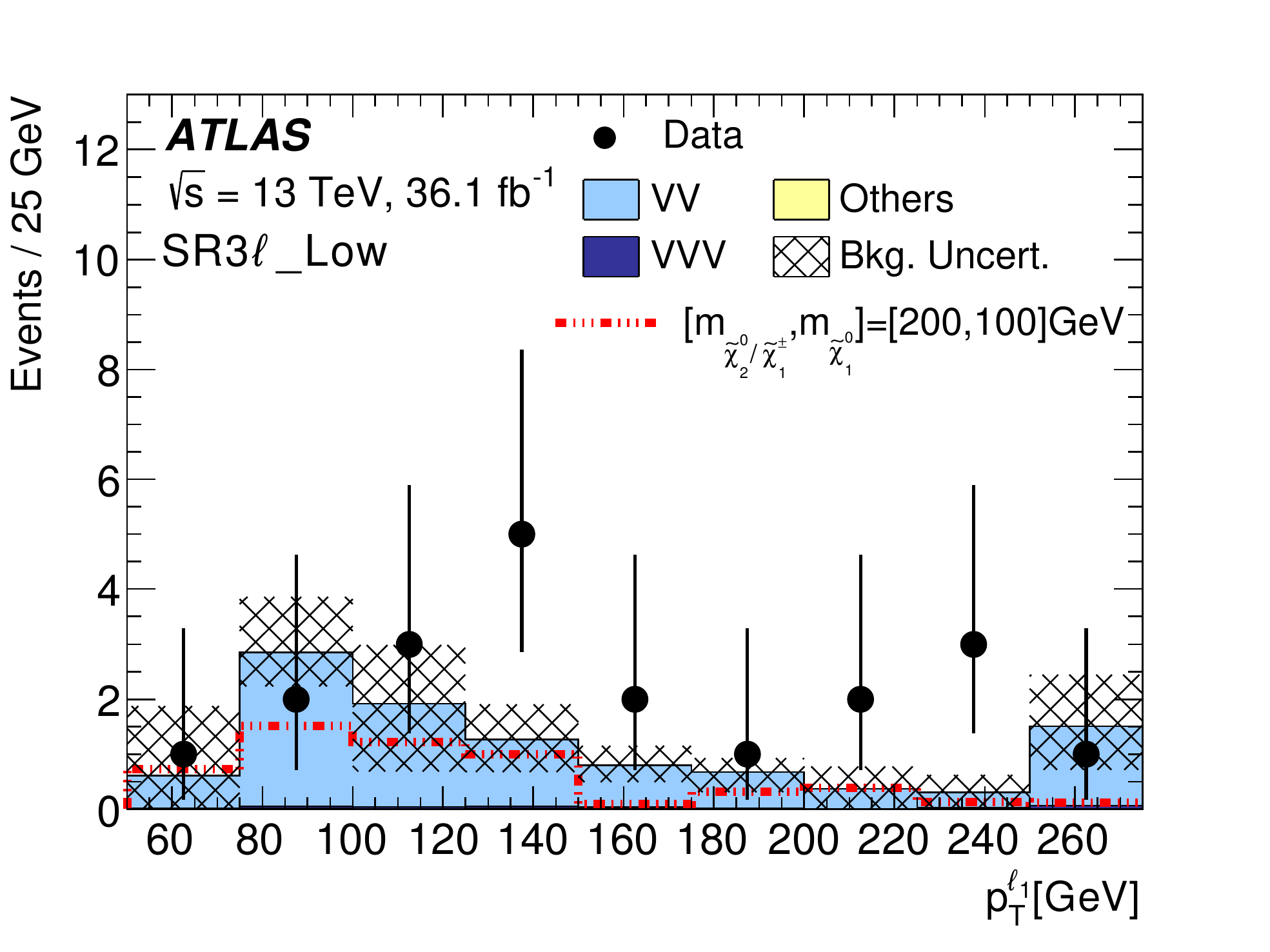}} \\ 
\vspace*{-0.005\textheight}\subfigure[]{\includegraphics[width=0.42\textwidth]{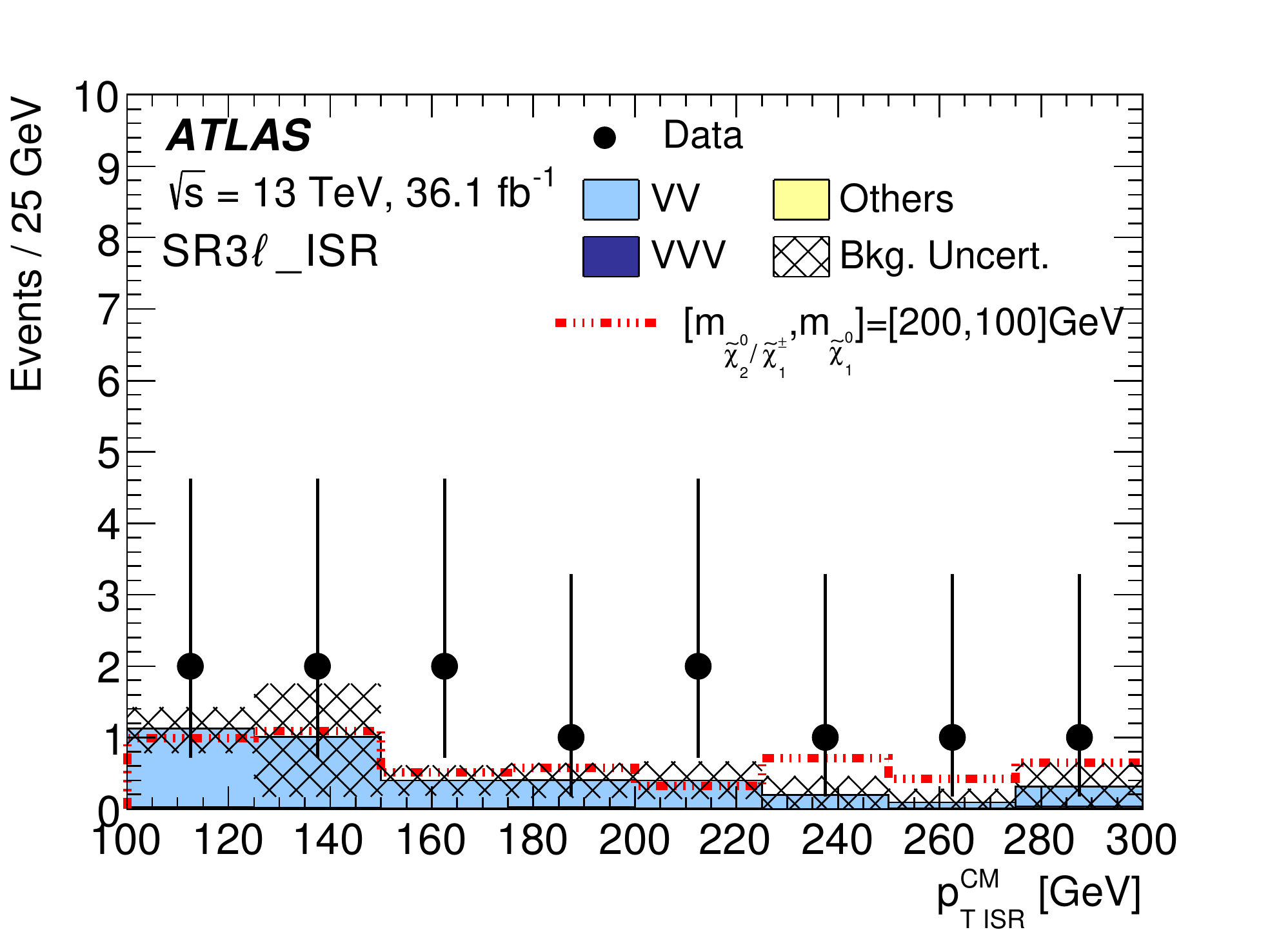}} 
\vspace*{-0.005\textheight}\subfigure[]{\includegraphics[width=0.42\textwidth]{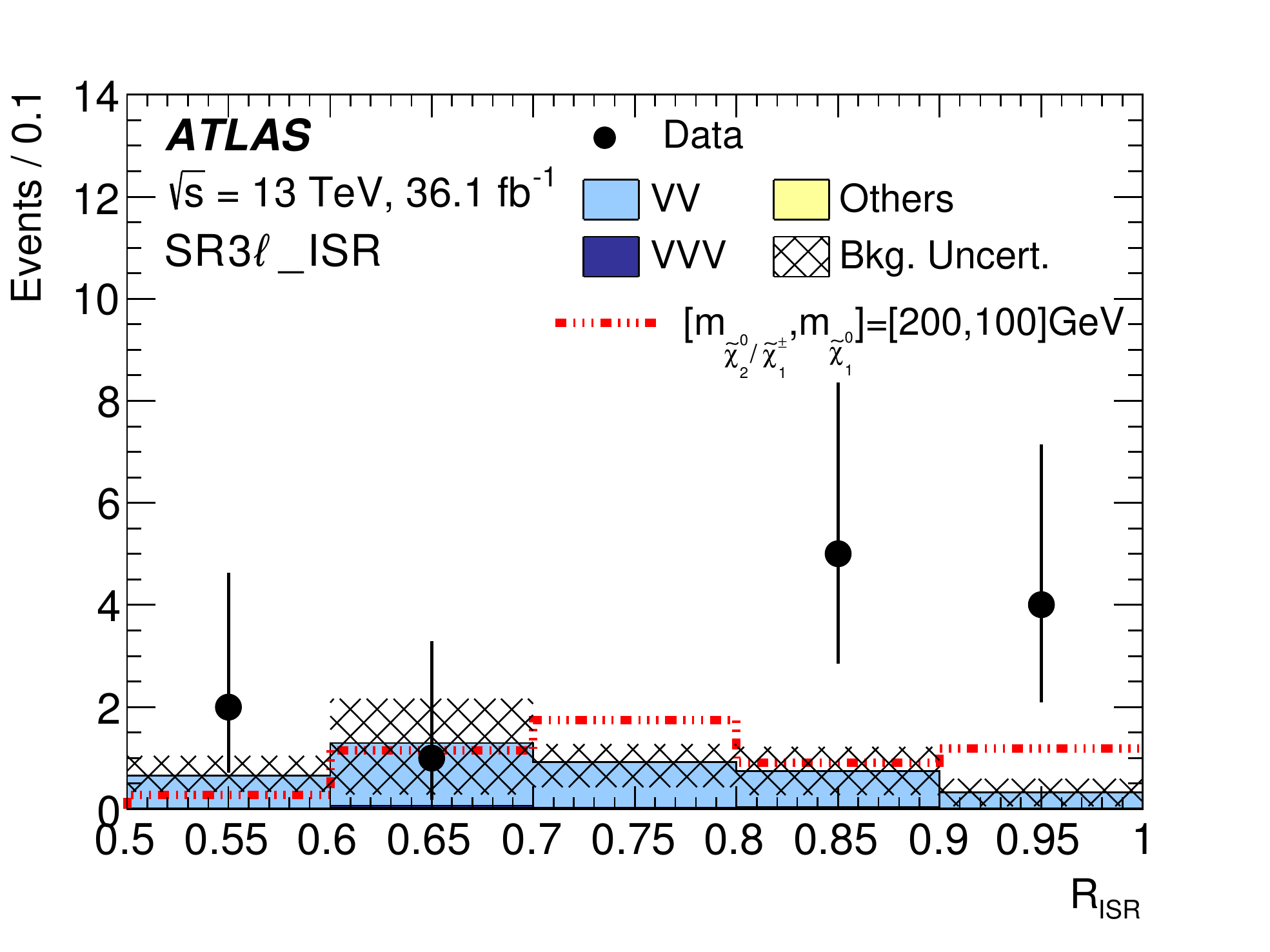}} 
\end{center}
\vspace*{-0.03\textheight}\caption{\label{fig:srRJR3L}
Distributions of kinematic variables in the signal regions for the $3\ell$ channels after applying all selection requirements. The histograms show the post-fit background predictions. The last bin includes the overflow. The FNP contribution is estimated from a data-driven technique and is included in the category ``Others''. Distributions for (a) $H_{3,1}^{\textrm{PP}}$ and (b) $p_{\textrm{T}}^{\ell_{1}}$ in SR$3\ell$\_Low, (c) $p_{\mathrm{T\ ISR}}^{~\textrm{CM}}$ and (d) $R_{\textrm{ISR}}$ in SR$3\ell$\_ISR are plotted. 
The hatched (black) error bands indicate the combined theoretical, experimental and MC statistical uncertainties.
The expected distribution for a benchmark signal model, normalized to the NLO+NLL cross-section (Section~\ref{sec:montecarlo}) times integrated luminosity, is also shown for comparison. 
}
\end{figure}

\begin{figure}[htbp]
\begin{center}
\vspace*{-0.005\textheight}\subfigure[]{\includegraphics[width=0.42\textwidth]{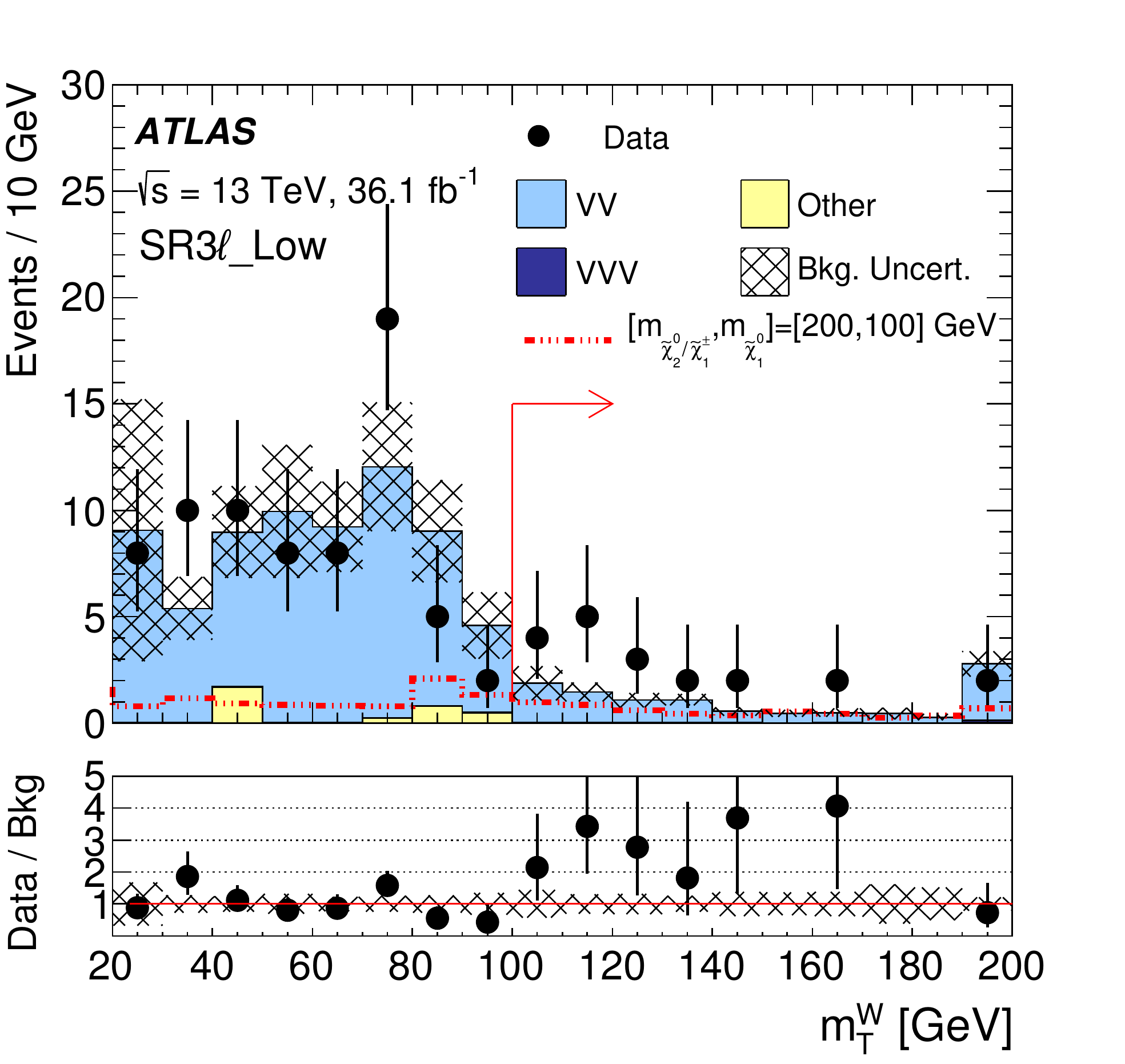}\label{fig:EitherOR3L_low}}
\vspace*{-0.005\textheight}\subfigure[]{\includegraphics[width=0.42\textwidth]{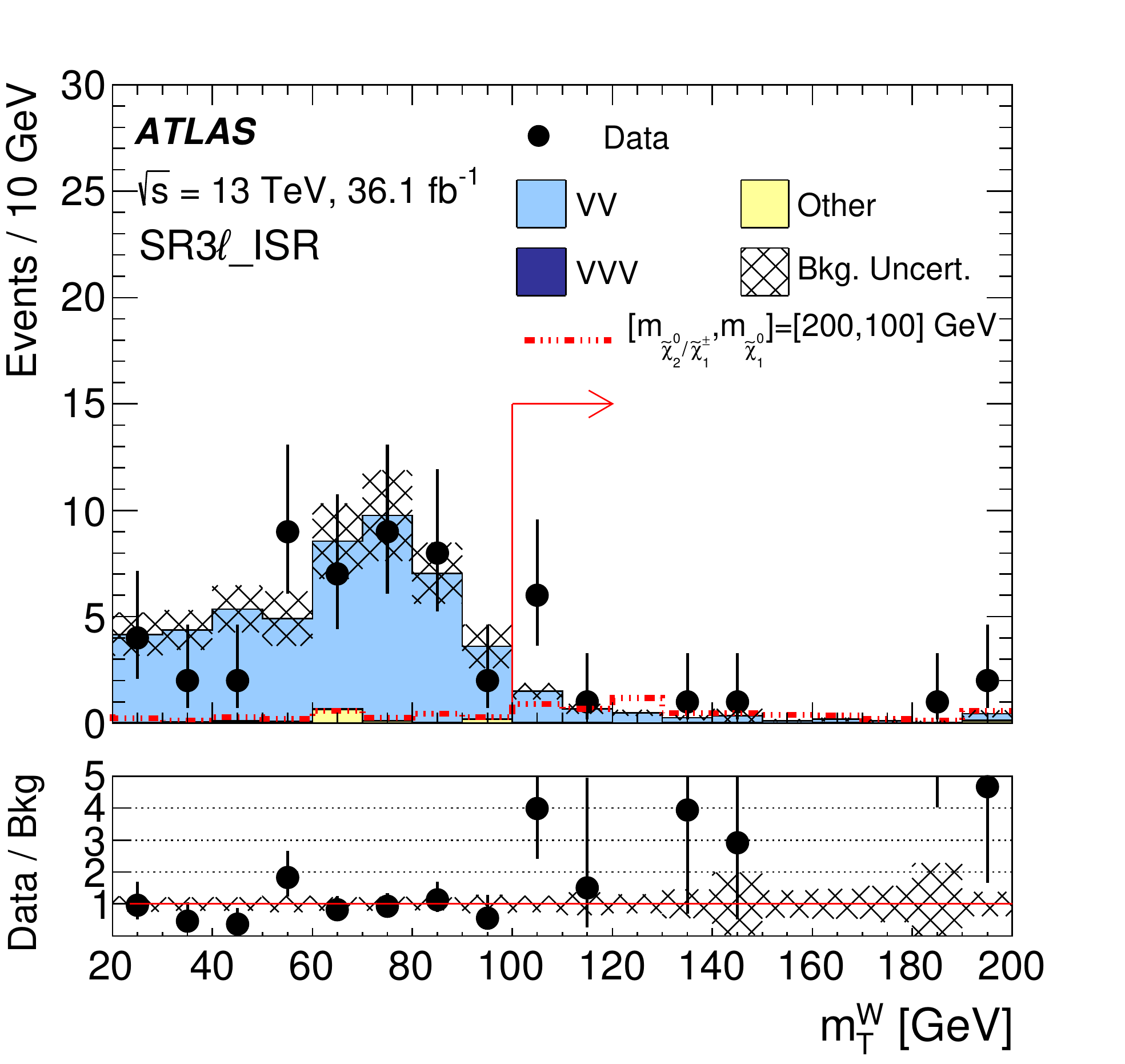}\label{fig:EitherOR3L_isr}}
\end{center}
\vspace*{-0.03\textheight}\caption{\label{fig:EitherOR3L}
The transverse mass of the unpaired lepton for events falling in either (a) SR$3\ell$\_Low or (b) SR$3\ell$\_ISR prior to the selection placed on this variable. The solid red line and arrow indicates the requirement defining these SRs. The last bin includes the overflow. The FNP contribution is estimated from a data-driven technique and is included in the category ``Others''. The hatched (black) error bands indicate the combined theoretical uncertainties on $VV$, experimental and MC statistical uncertainties. The expected distribution for a benchmark signal model, normalized to the NLO+NLL cross-section (Section~\ref{sec:montecarlo}) times integrated 
luminosity, is also shown for comparison.}
\end{figure}

\begin{figure}
\centering
	\subfigure[]{
	  \includegraphics[scale=0.38]{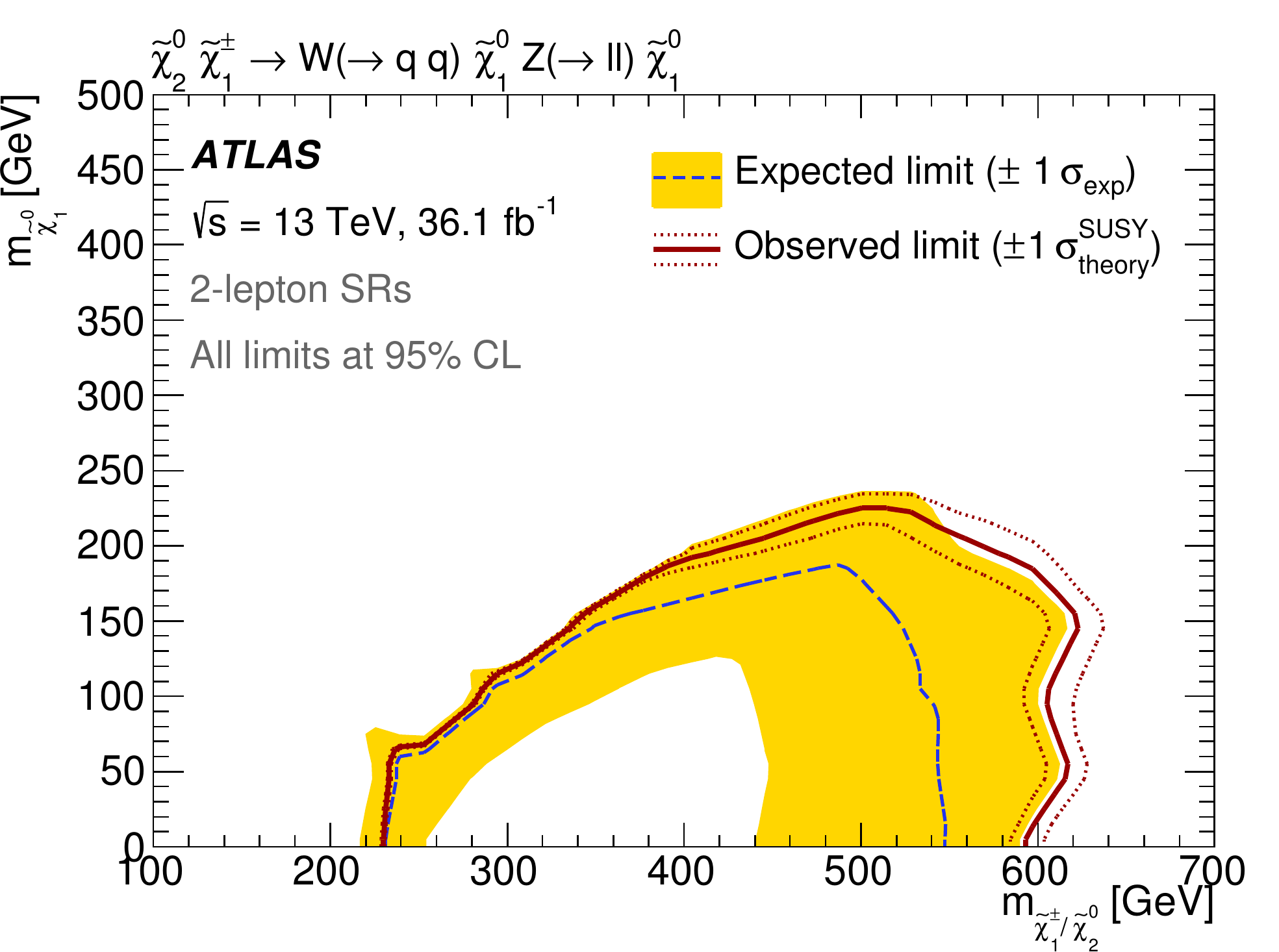}
          \label{fig:Money2L}
        }
	\subfigure[]{
	  \includegraphics[scale=0.38]{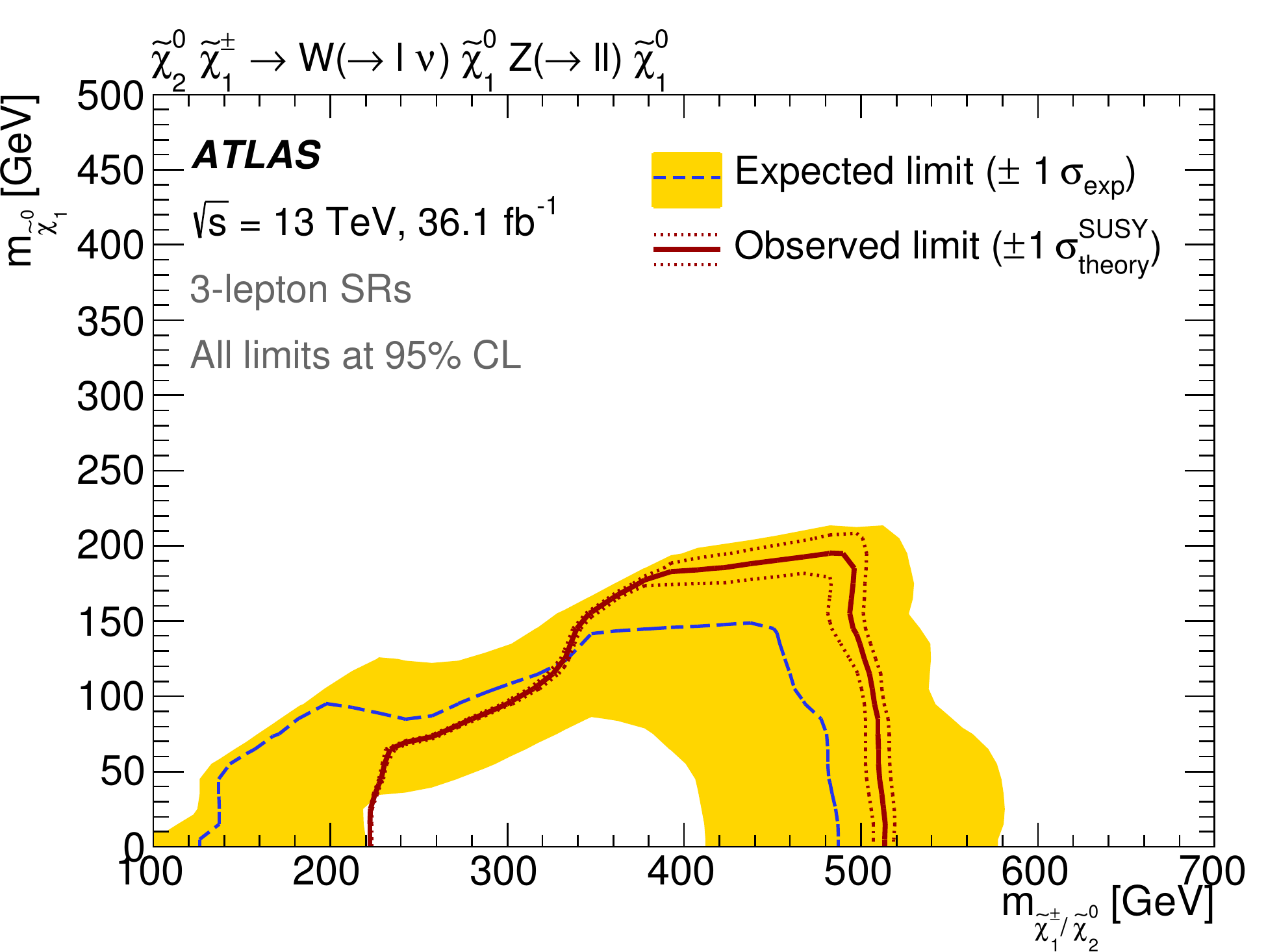}
          \label{fig:Money3L}
        }
	\subfigure[]{
	  \includegraphics[scale=0.38]{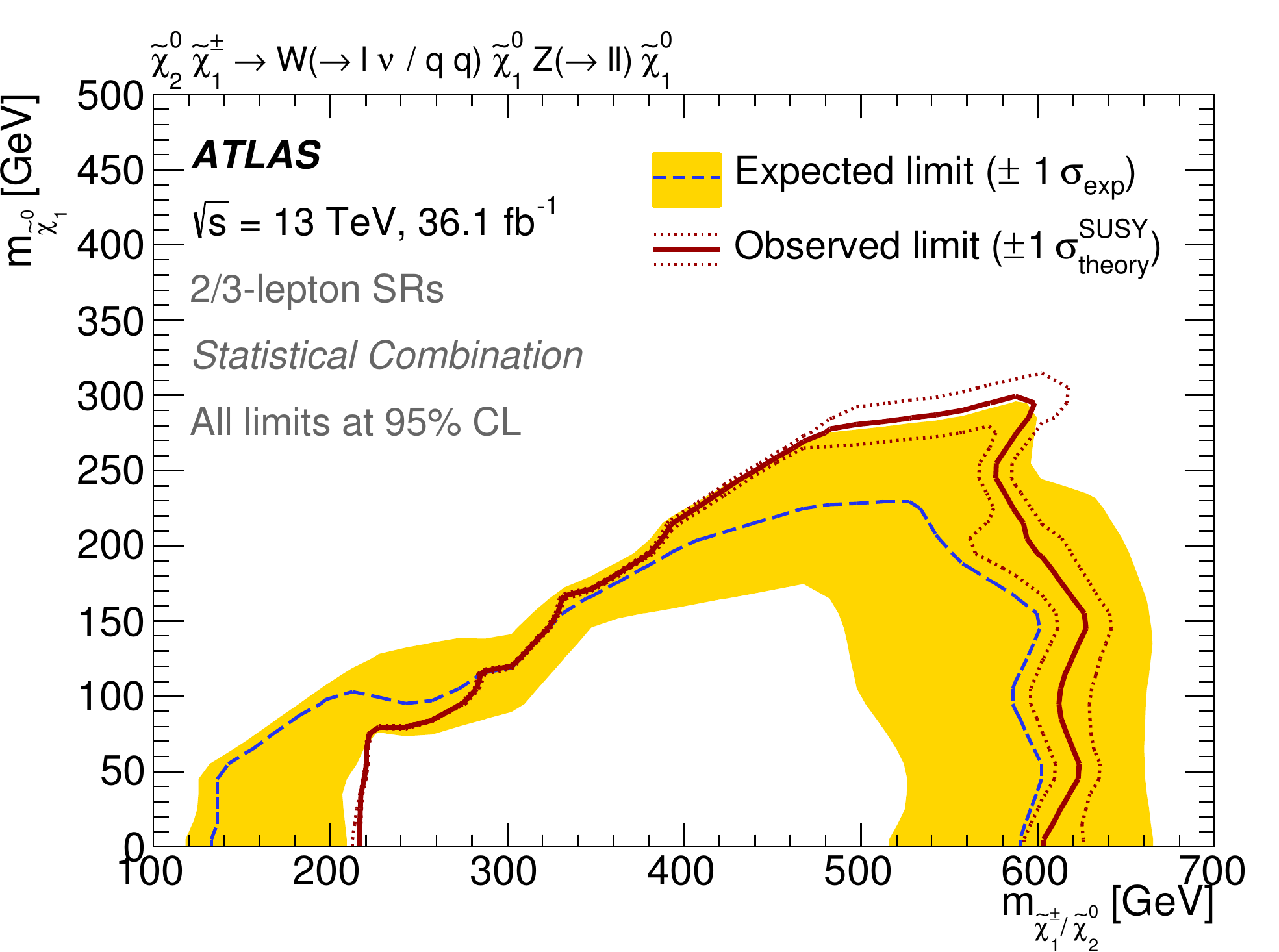}
          \label{fig:MoneyStatCombo}
        }
	\subfigure[]{
	  \includegraphics[scale=0.40]{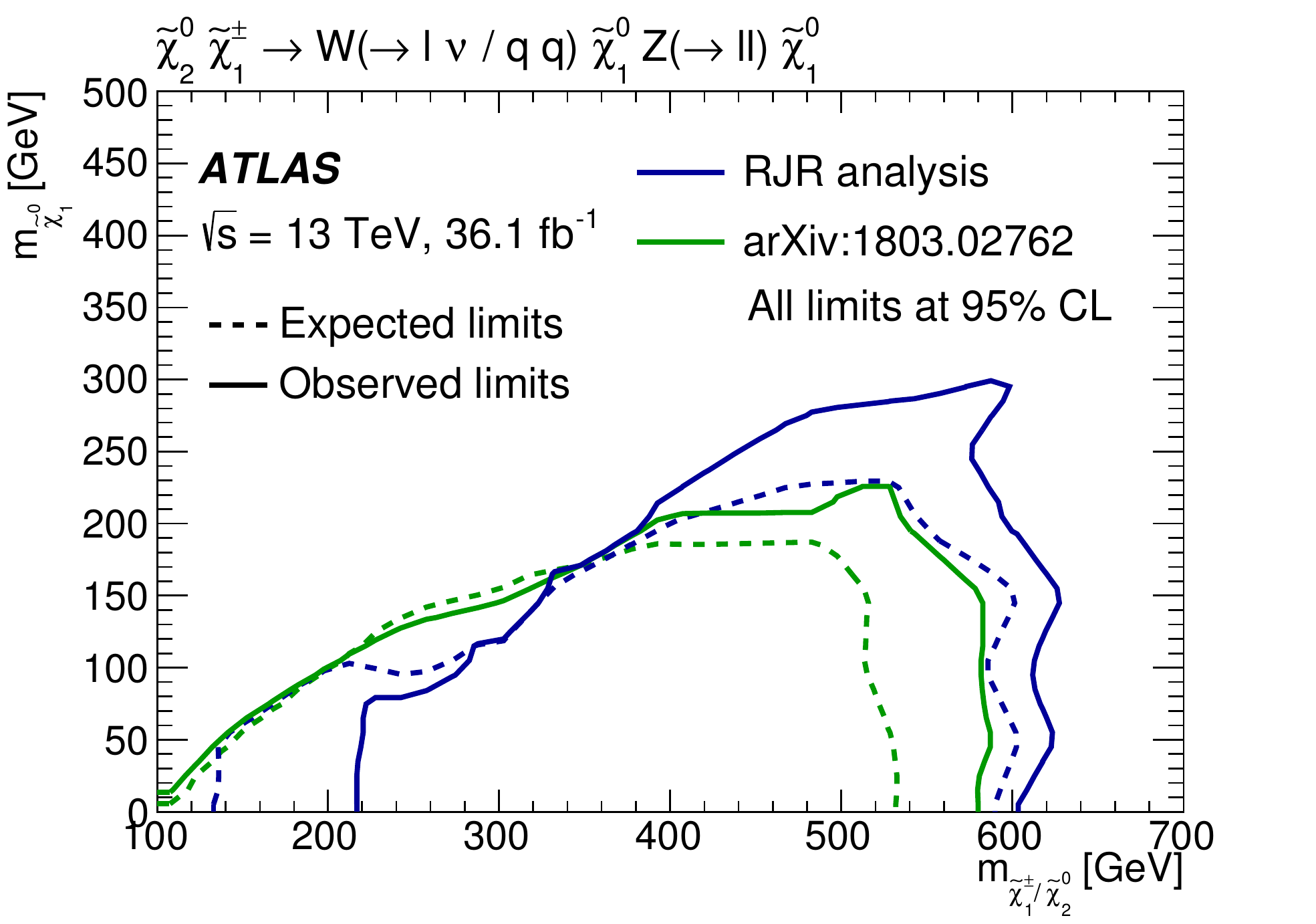}
          \label{fig:MoneyComparison}
	}
\caption{Exclusion limits at 95\% CL on the masses of $\chinoonepm/\ninotwo$ and $\ninoone$ from the analysis of 36.1~fb$^{-1}$ of 13~\TeV\ $pp$ collision data obtained from the (a) 2$\ell$ search, (b) the $3\ell$ search, (c) the statistical combination of the $2\ell$ and 3$\ell$ search channels, assuming 100\% branching ratio of the sparticles to decay to SM $W/Z$ bosons and $\ninoone$. The dashed line and the shaded band are the expected limit and its $\pm1\sigma$ uncertainty, respectively. The thick solid line is the observed limit for the central value of the signal cross-section. The dotted lines around the
observed limit illustrate the change in the observed limit as the nominal signal cross-section is scaled up and down by the theoretical uncertainty and (d) comparison between the exclusion limits from this analysis and Ref.~\cite{Aaboud:2018jiw}.
}
\label{fig:Money}		
\end{figure}

The current results extend the sensitivity and exclusion limits in the high- and intermediate-mass-splitting regions compared to those from Ref.~\cite{Aaboud:2018jiw}. However, the low-mass region where the mass splitting is $\approx$~100~\GeV~cannot be excluded due to the observed excess of events. The results in this region are of interest as they show an apparent disagreement with those quoted in Ref.~\cite{Aaboud:2018jiw} with similar sensitivity to this simplified model. The observed data excesses in SR3$\ell$\_ISR, SR3$\ell$\_Low, SR2$\ell$\_ISR and SR2$\ell$\_Low have associated significances of 3.0, 2.1, 2.0 and 1.4 standard deviations, respectively. As a result of these deviations from expectation the exclusion curves in Figure~\ref{fig:Money} demonstrate that there are regions where an exclusion would be expected but cannot be achieved with the data. A comparison with the analysis from Ref.~\cite{Aaboud:2018jiw} in Figure~\ref{fig:MoneyComparison} shows that there is a region of phase space in this simplified model, excluded at 95\% CL by that analysis, that cannot be excluded by the results of this analysis.

The RJR selection reduces background through testing how well the events exhibit properties anticipated for the topologies under investigation with a much looser requirement on the missing transverse momentum than in the analysis in Ref.~\cite{Aaboud:2018jiw}. The methods by which the analyses select the putative $Z$-boson candidate and define SRs with or without a system of jets consistent with ISR also differ. The overlap of the selected data events in the SRs between the two approaches is found to be smaller than 20\% and 30\% for the two-lepton and three-lepton channels, respectively. In the compressed regions the overlap percentage for the hypothetical signal $m_{\chinoonepm/\ninotwo}=200$~\GeV~and $m_{\ninoone}=100$~\GeV~for the 2$\ell$ (3$\ell$) search channel is found to be less than 5\% (15\%). 

In light of these results in the SR3$\ell$\_ISR, SR3$\ell$\_Low, SR2$\ell$\_ISR and SR2$\ell$\_Low regions, a variety of cross-checks were performed for both the $2\ell$ and $3\ell$ channels.

\begin{table}
  \caption{Breakdown of the observed and expected (in parentheses) number of events in terms of flavor composition in the SRs with an excess.}
  \begin{center}
    {\small
      \begin{tabular}{l r r}
        \toprule
        Signal region       &  SR$2\ell\_$Low     & SR2$\ell$\_ISR              \\[0.05cm]
        \hline\hline
        $ee$          & 9  (4.5$\pm$3.9)              &   3 (1.2$\pm$1.2)          \\
        $\mu\mu$      & 10 (3.9$\pm$2.6)              &   8 (1.5$\pm$1.5)          \\
        \toprule
        Signal region  &   SR3$\ell\_$Low          &   SR3$\ell$\_ISR             \\[0.05cm]
        \hline\hline
        $eee$           &  6 (3.5$\pm$0.7)         &  3 (1.1$\pm$0.3)             \\
        $ee\mu$         &  6 (2.0$\pm$0.4)         &  3 (0.9$\pm$0.3)             \\
        $\mu\mu\mu$     &  7 (2.7$\pm$0.6)         &  4 (1.5$\pm$0.4)             \\
        $\mu\mu e$      &  1 (1.9$\pm$0.4)         &  2 (0.4$\pm$0.1)            \\
         \toprule
      \end{tabular}
    }
  \end{center}
  \label{table:ExcessFlavorComposition}
\end{table}

Table~\ref{table:ExcessFlavorComposition} shows the breakdown of the composition of the lepton flavor for the events selected in the SR2$\ell\_$Low, SR2$\ell\_$ISR, SR3$\ell$\_Low and SR3$\ell$\_ISR regions, along with the expectation from the background estimation.

The validation-region distributions in Figure~\ref{fig:vrRJR2L} and Figure~\ref{fig:vrRJR3L} show that there is good agreement between the expectation from the background prediction and data in kinematic regions close to the SRs. For the SR3$\ell$\_ISR and SR3$\ell$\_Low regions, where the excesses are most significant, the composition of the events is studied in dedicated validation regions where the primary selection criteria in the signal region are inverted. In each of these distributions the observed events are in good agreement with the prediction, and the primary background from $WZ$ events in MC simulation describes the data in both shape and yield. These cross-checks do not indicate a significant mismodeling of any single component of the background. In all cases the main background components are studied with alternative generators and there is good agreement between these samples. Yields of events determined with data-driven methods are cross-checked with MC simulation samples and no significant discrepancies are observed.

\FloatBarrier

\section{Conclusion}
\label{sec:conclusion}

The paper presents a search for the electroweak production of neutralinos and charginos decaying into final states with exactly two or three electrons or muons and missing transverse momentum, performed using proton--proton collision data at $\sqrt{s}=13$~\TeV\ corresponding to an integrated luminosity of  36.1~fb$^{-1}$ recorded by the ATLAS detector at the LHC. Two distinct search channels based on recursive jigsaw reconstruction are considered where both the 2$\ell$ and 3$\ell$ channels target the same signal mode but with the $W$ boson decaying leptonically or to jets.

The statistical interpretation of the two search channels places exclusion limits on associated~$\chinoonepm\ninotwo$ production with gauge-boson-mediated decays. For a massless~$\ninoone$, $\chinoonepm$/$\ninotwo$ masses up to 600~\GeV\ are excluded. The results extend the region of supersymmetric parameter space previously excluded by LHC searches in the high- and intermediate-mass regions. In the low-mass and ISR signal regions an excess of events above the SM prediction is observed and the region of parameter space below $m_{\chinoonepm/\ninotwo}=220$~\GeV\ cannot be excluded.

The excesses observed in the 2$\ell$ and 3$\ell$ channels in the ISR (low-mass) signal regions correspond to local significances of 2.0 and 3.0 (1.4 and 2.1) standard deviations, respectively.

\section*{Acknowledgments}
We thank CERN for the very successful operation of the LHC, as well as the support staff from our institutions without whom ATLAS could not be operated efficiently.

We acknowledge the support of ANPCyT, Argentina; YerPhI, Armenia; ARC, Australia; BMWFW and FWF, Austria; ANAS, Azerbaijan; SSTC, Belarus; CNPq and FAPESP, Brazil; NSERC, NRC and CFI, Canada; CERN; CONICYT, Chile; CAS, MOST and NSFC, China; COLCIENCIAS, Colombia; MSMT CR, MPO CR and VSC CR, Czech Republic; DNRF and DNSRC, Denmark; IN2P3-CNRS, CEA-DRF/IRFU, France; SRNSFG, Georgia; BMBF, HGF, and MPG, Germany; GSRT, Greece; RGC, Hong Kong SAR, China; ISF, I-CORE and Benoziyo Center, Israel; INFN, Italy; MEXT and JSPS, Japan; CNRST, Morocco; NWO, Netherlands; RCN, Norway; MNiSW and NCN, Poland; FCT, Portugal; MNE/IFA, Romania; MES of Russia and NRC KI, Russian Federation; JINR; MESTD, Serbia; MSSR, Slovakia; ARRS and MIZ\v{S}, Slovenia; DST/NRF, South Africa; MINECO, Spain; SRC and Wallenberg Foundation, Sweden; SERI, SNSF and Cantons of Bern and Geneva, Switzerland; MOST, Taiwan; TAEK, Turkey; STFC, United Kingdom; DOE and NSF, United States of America. In addition, individual groups and members have received support from BCKDF, the Canada Council, CANARIE, CRC, Compute Canada, FQRNT, and the Ontario Innovation Trust, Canada; EPLANET, ERC, ERDF, FP7, Horizon 2020 and Marie Sk{\l}odowska-Curie Actions, European Union; Investissements d'Avenir Labex and Idex, ANR, R{\'e}gion Auvergne and Fondation Partager le Savoir, France; DFG and AvH Foundation, Germany; Herakleitos, Thales and Aristeia programmes co-financed by EU-ESF and the Greek NSRF; BSF, GIF and Minerva, Israel; BRF, Norway; CERCA Programme Generalitat de Catalunya, Generalitat Valenciana, Spain; the Royal Society and Leverhulme Trust, United Kingdom.

The crucial computing support from all WLCG partners is acknowledged gratefully, in particular from CERN, the ATLAS Tier-1 facilities at TRIUMF (Canada), NDGF (Denmark, Norway, Sweden), CC-IN2P3 (France), KIT/GridKA (Germany), INFN-CNAF (Italy), NL-T1 (Netherlands), PIC (Spain), ASGC (Taiwan), RAL (UK) and BNL (USA), the Tier-2 facilities worldwide and large non-WLCG resource providers. Major contributors of computing resources are listed in Ref.~\cite{ATL-GEN-PUB-2016-002}.

%

\printbibliography
\clearpage
 
\begin{flushleft}
{\Large The ATLAS Collaboration}

\bigskip

M.~Aaboud$^\textrm{\scriptsize 34d}$,    
G.~Aad$^\textrm{\scriptsize 99}$,    
B.~Abbott$^\textrm{\scriptsize 125}$,    
O.~Abdinov$^\textrm{\scriptsize 13,*}$,    
B.~Abeloos$^\textrm{\scriptsize 129}$,    
D.K.~Abhayasinghe$^\textrm{\scriptsize 91}$,    
S.H.~Abidi$^\textrm{\scriptsize 164}$,    
O.S.~AbouZeid$^\textrm{\scriptsize 143}$,    
N.L.~Abraham$^\textrm{\scriptsize 153}$,    
H.~Abramowicz$^\textrm{\scriptsize 158}$
% ...
\bigskip
\\

% ...
$^{13}$Institute of Physics, Azerbaijan Academy of Sciences, Baku; Azerbaijan.\\
% ...
$^{34}$$^{(a)}$Facult\'e des Sciences Ain Chock, R\'eseau Universitaire de Physique des Hautes Energies - Universit\'e Hassan II, Casablanca;$^{(b)}$Centre National de l'Energie des Sciences Techniques Nucleaires (CNESTEN), Rabat;$^{(c)}$Facult\'e des Sciences Semlalia, Universit\'e Cadi Ayyad, LPHEA-Marrakech;$^{(d)}$Facult\'e des Sciences, Universit\'e Mohamed Premier and LPTPM, Oujda;$^{(e)}$Facult\'e des sciences, Universit\'e Mohammed V, Rabat; Morocco.\\
% ...
$^{91}$Department of Physics, Royal Holloway University of London, Egham; United Kingdom.\\
% ...
$^{99}$CPPM, Aix-Marseille Universit\'e, CNRS/IN2P3, Marseille; France.\\
% ...
$^{125}$Homer L. Dodge Department of Physics and Astronomy, University of Oklahoma, Norman OK; United States of America.\\
% ...
$^{129}$LAL, Universit\'e Paris-Sud, CNRS/IN2P3, Universit\'e Paris-Saclay, Orsay; France.\\
% ...
$^{143}$Santa Cruz Institute for Particle Physics, University of California Santa Cruz, Santa Cruz CA; United States of America.\\
% ...
$^{153}$Department of Physics and Astronomy, University of Sussex, Brighton; United Kingdom.\\
% ...
$^{158}$Raymond and Beverly Sackler School of Physics and Astronomy, Tel Aviv University, Tel Aviv; Israel.\\
% ...
$^{164}$Department of Physics, University of Toronto, Toronto ON; Canada.\\
% ...
$^{a}$ Also at  Department of Physics, University of Malaya, Kuala Lumpur; Malaysia.\\
$^{b}$ Also at Borough of Manhattan Community College, City University of New York, NY; United States of America.\\
$^{c}$ Also at California State University, East Bay; United States of America.\\
$^{d}$ Also at Centre for High Performance Computing, CSIR Campus, Rosebank, Cape Town; South Africa.\\
$^{e}$ Also at CERN, Geneva; Switzerland.\\
% ...
$^{*}$ Deceased

\end{flushleft}


\FloatBarrier






\end{document}